\documentclass[11pt]{article}
\pdfoutput=1
\usepackage{jcapmod}
\usepackage{shorthand}

\usepackage{mathtools}
\usepackage{booktabs}
\usepackage[english]{babel}
\usepackage{amsmath,amssymb,amsbsy,amstext, amsthm, simplewick, amsfonts}
\usepackage{hyperref}
\usepackage{graphicx}
\usepackage[small]{caption}
\usepackage{siunitx}
\usepackage{upgreek}
\usepackage{framed}
\usepackage{wrapfig}
\usepackage{multirow}
\usepackage{bbm}
\usepackage[svgnames,dvipsnames,x11names]{xcolor}
\usepackage[utf8x]{inputenc}
\usepackage{selinput}
\usepackage{bm}
\usepackage{float}
\usepackage{geometry}
\usepackage{yfonts}
\usepackage{caption}
\usepackage{subcaption}
\usepackage{sidecap}
\usepackage{longtable}
\usepackage{anyfontsize}
\usepackage{dsfont}

\usepackage[framemethod=default]{mdframed}
\newmdenv[skipabove=7pt,
skipbelow=7pt,
rightline=false,
leftline=false,
topline=false,
bottomline=false,
backgroundcolor=gray!10,
linecolor=gray,
innerleftmargin=5pt,
innerrightmargin=5pt,
innertopmargin=5pt,
innerbottommargin=5pt,
leftmargin=0cm,
rightmargin=0cm,
linewidth=4pt]{eBox}
\newmdenv[skipabove=7pt,
skipbelow=7pt,
rightline=false,
leftline=false,
topline=false,
bottomline=false,
backgroundcolor=gray!10,
linecolor=gray,
innerleftmargin=5pt,
innerrightmargin=5pt,
innertopmargin=-5pt,
innerbottommargin=5pt,
leftmargin=0cm,
rightmargin=0cm,
linewidth=4pt]{eBox2}

\definecolor{blue3}{RGB}{31, 119, 180}
\definecolor{red3}{RGB}{	214, 39, 40}
\definecolor{orange3}{RGB}{255, 127, 14}
\definecolor{green3}{RGB}{44, 160, 44}
\definecolor{repBlue}{RGB}{31, 119, 180}
\definecolor{repRed}{RGB}{	214, 39, 40}
\definecolor{repGreen}{RGB}{44, 160, 44}

\renewcommand{\(}{\left(}
\renewcommand{\)}{\right)}
\renewcommand{\[}{\left[}
\renewcommand{\]}{\right]}

\def\be{\begin{equation}}
\def\ee{\end{equation}}
\newcommand{\bea}{\begin{eqnarray}}
\newcommand{\eea}{\end{eqnarray}}
\newcommand{\baa}{\begin{align}}
\newcommand{\eaa}{\end{align}}
\def\fnl{f_{\rm NL}}
\def\vp{\varphi}
\def\cG{\mathcal{K}}

\usepackage{colortbl}
\definecolor{lightgreen}{cmyk}{0.2, 0, 0.2, 0.2}
\definecolor{lightgray}{cmyk}{0.1,0.2,0,0.1}
\definecolor{lightgray2}{cmyk}{0.1,0.1,0,0.1}

\makeatletter
\newlength{\apb@width}
\newcommand{\autoparbox}[2][c]{\settowidth{\apb@width}{#2}\parbox[#1]{\apb@width}{#2}}

\makeatother


\def\beq{\begin{equation}}
\def\eeq{\end{equation}}

\allowdisplaybreaks[1]
\begin{document}


\newgeometry{top=2cm, bottom=2cm, left=2.9cm, right=2.9cm}

\begin{titlepage}
\setcounter{page}{1} \baselineskip=15.5pt 
\thispagestyle{empty}

\begin{center}
{\fontsize{20}{18} \bf Boostless Cosmological Collider Bootstrap}\\ [12pt]
\end{center}

\vskip 20pt

\begin{center}
\noindent
{\fontsize{12.5}{18}\selectfont 
Guilherme L.~Pimentel$^{1,2}$ and Dong-Gang Wang\hskip 1pt$^{3}$}
\end{center}

\begin{center}
  \vskip8pt
\textit{$^1$ Lorentz Institute for Theoretical Physics, Leiden University, \\Leiden, 2333 CA, The Netherlands}

  \vskip8pt
\textit{$^2$ Institute of Physics, University of Amsterdam,\\ Amsterdam, 1098 XH, The Netherlands}

  \vskip8pt
\textit{$^3$ Department of Applied Mathematics and Theoretical Physics, \\ University of Cambridge,
Wilberforce Road, Cambridge, CB3 0WA, UK}
\end{center}

\vspace{0.4cm}
 \begin{center}{\bf Abstract} 
 \end{center}
 \noindent

Cosmological correlation functions contain valuable information about the primordial Universe, with possible signatures of new massive particles at very high energies. 
Recent developments, including the cosmological bootstrap, bring new perspectives and powerful tools to study these observables.
In this paper, we systematically classify inflationary three-point correlators of scalar perturbations using the bootstrap method. 
For the first time, we derive a
complete set of single-exchange cosmological collider bispectra with new shapes and potentially detectable signals.
Specifically, we 
focus on the primordial scalar bispectra generated from the exchange of massive  particles with all possible boost-breaking interactions during inflation.
We introduce three-point ``seed'' functions, from which we bootstrap the inflationary bispectra of scalar and spinning exchanges using weight-shifting and spin-raising operators.  The computation of the seed function requires solving an ordinary differential equation in comoving momenta, a boundary version of the equation of motion satisfied by a propagator that linearly mixes a massive particle with the external light scalars.
 The resulting correlators are presented in analytic form, for any kinematics. These shapes are of interest for near-future cosmological surveys, as the primordial non-Gaussianity in boost-breaking theories can be large.
We also identify new features in these shapes, which are phenomenologically distinct from the de Sitter invariant cases. For example, the oscillatory shapes around the squeezed limit have different phases. Furthermore, when the massive particle has much lower speed of sound than the inflaton, oscillatory features appear around the equilateral configuration.

\noindent
\end{titlepage}

\newpage

\restoregeometry
\setcounter{tocdepth}{3}
\setcounter{page}{1}
\tableofcontents

\newpage

\section{Introduction}

The primordial Universe is a natural laboratory for fundamental physics, where the laws of the microscopic world can be tested via observations on cosmic scales.
In particular, since inflation is likely to have the highest energy densities accessible in nature, we expect that primordial correlations may provide the ultimate test of high energy physics~\cite{Meerburg:2019qqi, Achucarro:2022qrl}.
This idea is nicely manifested in ``cosmological collider physics" \cite{Arkani-Hamed:2015bza}, where the qualitative and quantitative features of inflation are recast in terms of a giant particle accelerator. Within this collider analogy, measuring correlation functions in the sky corresponds to measurements of interactions of the particles responsible for primordial fluctuations. These correlations could, for example, be mediated by new massive particles. This scenario leads to the natural question of studying these correlations and classifying their distinctive observational signatures. 
Specifically, new particles can mediate interactions among the curvature fluctuations, leaving their indirect imprints in the shapes of primordial non-Gaussianity (see~\cite{Chen:2009zp, Baumann:2011nk, Assassi:2012zq, Chen:2012ge, Pi:2012gf, Noumi:2012vr, Baumann:2012bc, Assassi:2013gxa, Gong:2013sma} for earlier studies, and \cite{Lee:2016vti, Flauger:2016idt, Chen:2016uwp, Chen:2016hrz, Kehagias:2017cym, Kumar:2017ecc, An:2017hlx, An:2017rwo, Baumann:2017jvh, Kumar:2018jxz, Bordin:2018pca, Goon:2018fyu, Anninos:2019nib, Kim:2019wjo, Alexander:2019vtb, Hook:2019zxa, Kumar:2019ebj, Liu:2019fag, Wang:2019gbi, Wang:2019gok, Maru:2021ezc, Lu:2021wxu, Wang:2021qez, Tong:2021wai, Pinol:2021aun, Cui:2021iie, Tong:2022cdz, Reece:2022soh} for recent ones). It is remarkable that there is the possibility to do particle spectroscopy in this extremely high energy environment, while having access only to the static pattern of density fluctuations at the end of inflation. To do so, precise predictions for the cosmological correlation functions are needed, as well as a detailed understanding of their analytic structure.

\vskip4pt
In recent years, our theoretical understanding of the statistics of primordial fluctuations has improved significantly. The  correlation functions at the end of inflation are now known in analytic form for a wide variety of processes. These advances come from a new perspective toward the investigation of cosmological correlators, following a ``bootstrap" philosophy \cite{Arkani-Hamed:2018kmz,Baumann:2019oyu,Baumann:2020dch, Arkani-Hamed:2017fdk,Arkani-Hamed:2018bjr,Benincasa:2018ssx, Sleight:2019mgd, Sleight:2019hfp,Sleight:2020obc, Pajer:2020wnj,Pajer:2020wxk, Jazayeri:2021fvk,Bonifacio:2021azc,Cabass:2021fnw,Hillman:2021bnk, Goodhew:2020hob, Cespedes:2020xqq, Melville:2021lst, Goodhew:2021oqg, Baumann:2021fxj, Meltzer:2021zin,Hogervorst:2021uvp,DiPietro:2021sjt}
(also see \cite{Baumann:2022jpr} for an up-to-date review of the subject).
In this new approach,  without reference to a specific model or Lagrangian,
the correlators are directly determined from a set of basic physical principles, such as locality, unitarity and symmetry. 

\vskip4pt
The cosmological bootstrap was first studied by exploiting the full de Sitter symmetries (the {\it de Sitter bootstrap}) \cite{Arkani-Hamed:2018kmz,Baumann:2019oyu,Baumann:2020dch}.  
From observations, we expect primordial fluctuations to be translation and rotation invariant, and dilatation covariant. For inflation, there is also the possibility of fluctuations being  invariant under de Sitter boosts. In this case, the constraints from all de Sitter isometries become very powerful. For example, it implies that the correlators have the same kinematical symmetries of Euclidean conformal field theories. From this perspective, we obtain analytic control of many correlators whose computation by conventional time evolution is rather intractable. 
Within the de Sitter bootstrap, it is possible to incorporate a mild breaking of boost symmetry, and thus compute primordial non-Gaussianity for slow-roll inflation. 

\vskip4pt
From a phenomenological perspective, however, it is interesting to drop the assumption of boost isometries. A simple way to break boosts is by giving a subluminal speed of propagation to the scalar fluctuations  \cite{Creminelli:2003iq,Silverstein:2003hf,Alishahiha:2004eh}. Generically, we expect the level of non-Gaussianity to be enhanced in this case, due to the smaller size of the sound horizon, compared to the Hubble radius during inflation \cite{Chen:2006nt,Creminelli:2006xe,Cheung:2007st}. Technically, this happens because the sound speed is controlled by an operator that induces strong self-interactions of the inflaton. Theories with small speed of sound and small non-Gaussianity typically require fine tuning. Therefore, if we detect primordial non-Gaussianity in the near future, it is  likely that primordial fluctuations break boost symmetries. In order to have the best of both worlds, we desire analytic control and understanding of shapes of non-Gaussianity, while encompassing scenarios in which de Sitter boosts are strongly broken.

\vskip4pt
In this paper, we make progress in developing the {\it boostless bootstrap} for the primordial bispectrum---the three-point function of density perturbations. 
 Despite having less symmetries at our disposal, the bispectrum is a simple observable, in which kinematics is tight enough that it is still possible to run the bootstrap. 
Recently, the boostless bootstrap was successfully applied  to classify three- and four- points of the massless fields in de Sitter by leveraging the remaining symmetries and locality constraints \cite{Pajer:2020wxk, Jazayeri:2021fvk, Bonifacio:2021azc,Cabass:2021fnw,Hillman:2021bnk}. 
 This general approach provides a complete set of correlators from single field inflation including all the boost-breaking interactions.
Our focus here is the bispectrum due to the presence of mediator massive particles during inflation. In other words, we are interested in developing a bootstrap for the  cosmological collider with potentially large non-Gaussianity, which is the most interesting case for upcoming observations. 
It is perhaps surprising that the bootstrap approach is applicable to the exchange bispectrum even in the boost-breaking scenario. We suspect that this is because the bispectrum is only sensitive to the longitudinal modes of the massive particles.

\begin{figure}[t]
   \centering
            \includegraphics[width=.75\textwidth]{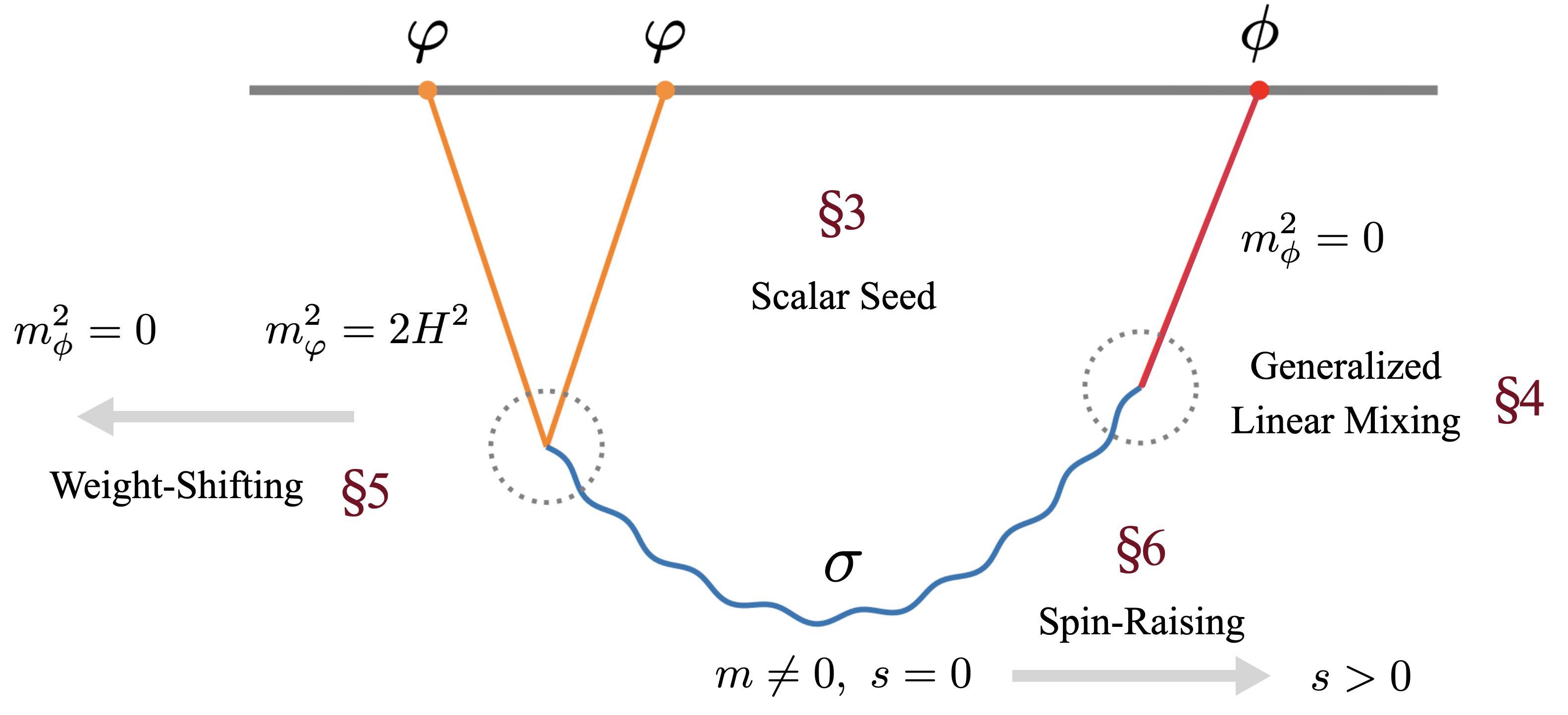}
   \caption{The structure of this paper. We compute the bispectrum of conformally coupled scalars with a massless scalar that linearly mixes with a massive scalar field. From this building block, and various ``weight-shifting" and ``spin-raising" operations, we generate a wide variety of inflationary bispectra  of phenomenological interest. As we relax the requirement of boost symmetry, the signals can be  large enough to be detectable in near-future cosmological surveys. The procedure to obtain the scalar seed, as well as the various weight-shifting moves, are presented in the sections indicated above.}
  \label{fig:roadmap}
\end{figure}

\vskip4pt
In the de Sitter bootstrap, we first compute de Sitter-invariant four-point functions, and then deform them to obtain a minimal level of boost breaking \cite{Arkani-Hamed:2018kmz}. In this paper, as boosts can be strongly broken, we compute the bispectrum using simpler building blocks, without reference to four-point functions (see \cite{Jazayeri:2022kjy} for an alternative approach).
The starting point is the correlator of two conformally coupled scalars and a massless scalar which linearly mixes with a scalar particle of arbitrary mass, as shown in Figure \ref{fig:roadmap}. The mixed propagator satisfies an interesting differential equation in time that internally ``collapses" the massive particle, producing the massless bulk-to-boundary propagator for a massless scalar. Then, we show that this ``scalar seed'' three-point correlator satisfies an inhomogeneous differential equation, and proceed to solve this equation analytically.
More surprisingly, from this scalar seed, we are able to bootstrap the inflaton correlators exchanging a particle of arbitrary mass {\it and} spin, as well as arbitrary  vertices (both for quadratic and cubic interactions).
Like in the de Sitter bootstrap, all possible boost-breaking interactions are derived from hitting the seed diagrams with ``weight-shifting" operators.
Similarly, we generate the spin-exchange bispectra from the scalar  seeds by using ``spin-raising" operators.

\vskip4pt
The resulting shapes share some similarities with their de Sitter symmetric counterparts, having features due to the mass and spin of the exchanged particles, but they also have new properties that are unique to the possibility of subluminal sound speeds\footnote{More precisely, the ratio of sound speeds between the inflaton fluctuations and the exchanged particles.}. The oscillatory phases are now different with the ones predicted by de Sitter invariant interactions. Moreover, the oscillations which are prominent around the squeezed limit in de Sitter invariant theories can also appear close to equilateral configurations. This is only possible in a scenario where the sound horizon of the mediator field is much smaller than the one of the inflaton. In that case, 
the exchange correlator can probe multiple sound horizon crossings for the massive particle
 before it decays into the inflaton. Even when all inflatons have similar wavelengths, the linear mixing leg provides a different clock during inflation, thus modulating the resulting non-Gaussian signal. 

\paragraph{Outline}
We start in Section \ref{sec:eft} reviewing the effective field theory  of inflation applied to the cosmological collider scenario. 
We highlight how large couplings are achieved for boost-breaking interactions in this setup, which illustrates why the boostless bootstrap is interesting.
The rest of the paper is organized as shown in Figure \ref{fig:roadmap}. 
In Section \ref{sec:seed}, we present the propagator of a massive scalar $\sigma$ linearly mixing with a massless field (the inflaton) $\phi$. Next, we apply this mixed propagator to compute the three-point function of two conformally coupled scalars $\vp$ with an inflaton. This correlator serves as a scalar seed of the bootstrap.
In Section \ref{sec:moremix}, we consider the most general quadratic interactions between $\phi$ and $\sigma$, and compute the resulting (generalized) scalar seeds.
The effects of different sound speeds between $\phi$ and $\sigma$ are taken into account here. These seed functions are related to the one in Section \ref{sec:seed} through recursive relations, whose explicit forms are presented in Appendix \ref{app:seeds}.
In Section \ref{sec:BBWS}, we introduce the boost-breaking weight-shifting operators, which map the seed functions with conformally coupled scalars to the three-point correlators of massless external fields. 
In Section \ref{sec:spin} we analyze spinning particle exchanges.
As only the longitudinal mode of the particle propagates, we find that relatively simple spin-raising operators relate
the spinning-exchange bispectra to the generalized scalar seeds.
We discuss the phenomenology of the new shapes in Section \ref{sec:pheno}. 

\vskip4pt
The appendices contain various technical details of the computations used throughout the main text.
In Appendix \ref{app:mix-p} we present the asymptotic behaviour of the mixed propagators. In Appendix \ref{app:seeds}
 we provide the solutions and singularity analysis of the generalized scalar seeds. In Appendix \ref{app:double}, we briefly comment on the double-exchange and triple-exchange diagrams.
 In Appendix \ref{app:spin}, we review the theory of free spinning particles in de Sitter space. Throughout
the paper we take the convention of natural units  $c=\hbar=1$,  the reduced Planck mass  $M_{\rm pl}^2 = 1/8\pi G$, and the  metric signature $(-,+,+,+)$.

\section{The EFT of Cosmological Colliders}
\label{sec:eft}
In this section, we briefly review the effective field theory (EFT) of  inflation, with a single clock picking a foliation of spacetime, and also additional massive fields beyond the inflaton. 
See the original papers \cite{Cheung:2007st, Lee:2016vti, Bordin:2018pca} for more details about the construction of the EFT.
  We illustrate the basic idea and collect the relevant results for the rest of the paper, showing the most relevant interaction vertices for large non-Gaussianities.

\vskip4pt
The key idea of the EFT is to separate the background dynamics from the dynamics of the quantum fluctuations. The background provides a natural foliation of spacetime, and dictates the allowed symmetries and interactions of the fluctuations. In this framework, many models of inflation lead to the same EFT. A specific model gives predictions for the EFT coefficients. More importantly, by being agnostic about the origin of the background dynamics,  the EFT provides a framework in which there is perturbative control of the fluctuations, even when the background dynamics is UV-sensitive. 

\begin{figure}[h]
   \centering
            \includegraphics[width=.95\textwidth]{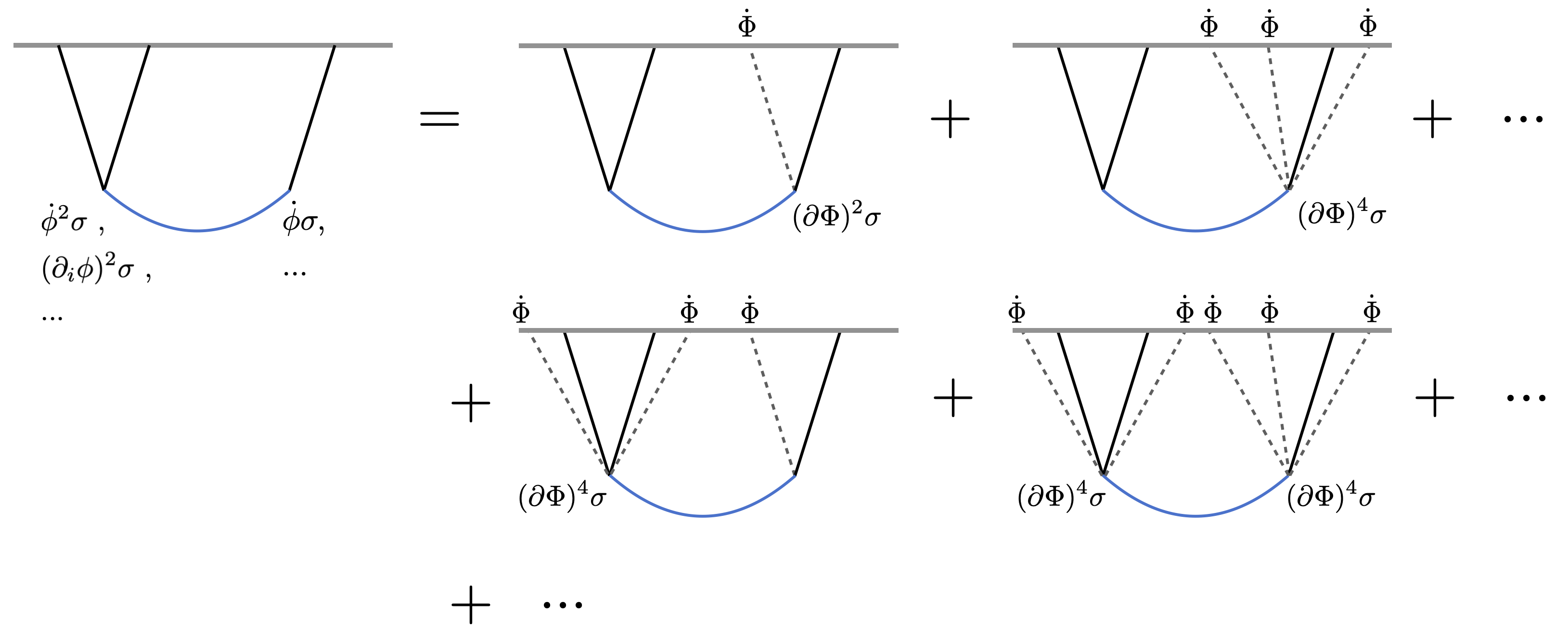}
   \caption{A diagrammatic illustration of how large boost-breaking interactions arise for the exchange bispectrum. In Feynman diagrams of cosmological correlators, the breaking of the boost symmetry is normally associated with evaluating external legs of the massless scalar to the background (the dashed lines with $\dot\Phi$).  The (boostless) EFT diagram on the left  encompasses all the higher-derivative contributions on the right-hand side systematically.
}
  \label{fig:eft}
\end{figure}

An example to keep in mind is that of a dynamical scalar field acting as the ``clock" of the background evolution (e.g. the inflaton). The time-dependence of this clock field $\Phi(t)$ is usually the source of the boost symmetry breaking in cosmology. If we look at its kinetic term $(\partial \Phi)^2$ and expand around the background solution, we obtain $\dot\Phi^2 g^{00}$. The metric component $g^{00}$ is the fluctuating degree of freedom in the EFT. If we look at higher-derivative operators involving $\Phi$, they would generate other operators for $g^{00}$, etc. The EFT packages all of these contributions in a background-agnostic fashion, meaning that operators that would be higher derivative, and thus suppressed, for the background dynamics, only appear as Wilson coefficients in the Lagrangian of the fluctuations, where the de Sitter boosts can be strongly broken.
This is  demonstrated by using   Feynman diagrams in Figure \ref{fig:eft}, where the EFT diagram with boost-breaking interactions is a sum of all the diagrams on the right hand side, with various legs being put to the background.

\vskip4pt
Within this framework, it is easy to show that non-Gaussian signals can be computed reliably as long as perturbation theory doesn't break down, namely $f_{\rm NL} \lesssim O(A_\zeta^{-1/2})$. For means of comparison, theories in which the background dynamics is weakly coupled typically predict a much lower $f_{\rm NL}\sim O(1)$. To test primordial non-Gaussianity in the near future, we'd like to have a framework that allows us to systematically classify non-Gaussianity for which $f_{\rm NL}\sim O(10)$, which are the current experimental bounds. As we will show below, the EFT provides us such a framework. Moreover, it strongly suggests that large non-Gaussianities are only achievable in models where the boost symmetries are broken \cite{Green:2020ebl}, both in the form of the interaction vertices, and in the dispersion relation of the fluctuations.

\vskip4pt
In order to study scalar fluctuations, it is often useful to consider the decoupling limit. In this limit, the metric has a scalar longitudinal mode, which is (by a gauge transformation) related to curvature perturbations. This is the Goldstone boson $\pi$ associated with the breaking of the time-translation invariance in an expanding spacetime. For most practical purposes, it can be treated as a massless field in the  quasi-de Sitter background of  inflation, but perhaps with a non-relativistic dispersion relation. 
In particular, we are  interested in its couplings to other massive fields during inflation. 
In general, these new particles can be massive scalars or spinning fields, and their interactions with the Goldstone might not be covariant. 

\vskip 6pt

First, let's briefly review the single-clock EFT focusing on self-interactions of the $\pi$ field.
Then we present the relevant results for its couplings with extra massive particles. 
At leading order in derivatives, the action of the single-clock EFT is 
\begin{align}
S = \int {\rm d}^4x\sqrt{-g}\left[\frac{1}{2}M_{\rm Pl}^2 R+M_{\rm Pl}^2\dot Hg^{00}-M_{\rm Pl}^2(3H^2+\dot H)+\sum_{n=2}^\infty \frac{M_n^4}{n!}(\delta g^{00})^n+\cdots\right] , \label{equ:action2}
\end{align}
where $\delta g^{00} \equiv g^{00} + 1$.
The coefficients of the operators $1$ and $g^{00}$ are adjusted to ensure that the background cosmology has Hubble rate $H$.  Then the action starts quadratic in fluctuations.  
For $M_n \to 0$ we recover slow-roll inflation. To see the fluctuating scalar degree of freedom, we introduce the ``pion" via a time reparametrization $t\to \tilde t = t+\pi(t,{\bf x})$.  The metric transforms as 
\begin{align}
g^{00} &\to g^{00}+2\partial_\mu \pi g^{0\mu} + \partial_\mu \pi \partial_\nu \pi g^{\mu \nu}\, . \label{equ:g00}
\end{align}
Substituting this into (\ref{equ:action2}) gives the action for the Goldstone boson. In general, this action contains a complicated mixing between the Goldstone mode and metric fluctuations.  We are interested in the decoupling limit, where the gravitational interactions are neglected~\cite{Cheung:2007st}. 
In this case, the transformation~(\ref{equ:g00}) is $g^{00} \to -1 -2 \dot \pi -\dot\pi^2 +a^{-2}(\partial_i \pi)^2$, and the Goldstone Lagrangian becomes
\beq
{\cal L}_\pi = M_{\rm Pl}^2 \dot H (\partial_\mu \pi)^2 + 2 M_2^4 \left[\dot \pi^2 - \frac{\dot \pi (\partial_i\pi)^2}{a^2} \right] +\left(2M_2^4-\frac{4}{3} M_3^4\right) \dot \pi^3 + \cdots\, . \label{Lpi}
\eeq
We see that $M_2 \ne 0$ induces a nontrivial sound speed for the Goldstone boson, 
\beq
c_s^2 \equiv \frac{M_{\rm Pl}^2 \dot H}{M_{\rm Pl}^2\dot H-2M_2^4}\, .
\eeq
A small value of $c_s$ (large value of $M_2$) is correlated with an enhanced cubic interaction $\dot \pi (\partial_i \pi)^2$ and large equilateral non-Gaussianity. This is partly why the boost-breaking scenario is phenomenologically important. The resulting bispectra and trispectra of the self-interacting pion were computed in \cite{Pajer:2020wxk, Jazayeri:2021fvk, Bonifacio:2021azc} via the ``boostless bootstrap'' approach.

\vskip4pt
Now we consider how the additional fields are coupled to the Goldstone in the EFT framework. We include both scalars and spinning fields in our discussion.\footnote{{For the EFT with spinning fields, here we follow the construction in \cite{Lee:2016vti} which assumes the full dS isometries for spinning particles. A different approach is presented in \cite{Bordin:2018pca} where  a dS-invariant UV completion is not required. However, as {\it only} the helicity-0 longitudinal mode contributes to the cosmological collider bispectrum, final results from these two approaches are expected to be the same. We leave more detailed discussion in Section \ref{sec:spin}.}}
For a spin-$s$ field $\sigma^{\mu_1\cdots \mu_s}$, the basic building blocks in the EFT are $\sigma^{0\cdots 0}$ and all Lorentz-invariant self-interactions, e.g.~$\sigma^{\mu_1\cdots\mu_s}\sigma_{\mu_1\cdots\mu_s}$. 
The latter are diff-invariant and will not induce couplings to the Goldstone in the decoupling limit. We may also have contractions with curvature tensors, which are higher order in derivatives. 
As this work focuses on single-exchange diagrams, we are interested in quadratic and cubic vertices with one massive field leg.
For this type of interactions, in order of increasing spin, we obtain:

\begin{itemize}

\item {\bf Spin-0}
Since the massive scalars  do not respect shift symmetry, the lowest derivative interactions with the Goldstone are simply given by
\begin{align}
{\cal L}_{\rm int}^{(0)} &= \omega_0^3 \hskip 1pt \delta g^{00}\sigma   + \tilde\omega_0^3  \hskip 1pt  (\delta g^{00})^2\sigma \,  .\label{spin0mixaction1}
\end{align}
In the decoupling limit, the mixing Lagrangian  becomes
\begin{align}
{\cal L}^{(0)}_{\rm int} = \rho_0\hskip 1pt\dot\pi_c\sigma + \frac{1}{\Lambda_0}\frac{(\partial_i\pi_c)^2\sigma}{a^2}  + \frac{1}{\tilde\Lambda_0}\dot\pi_c^2\sigma   ,\label{Lmix0}
\end{align}
where $\pi_c \equiv f_\pi^2 \pi$ is the canonically normalized Goldstone, with $f_\pi^4=2M_{pl}^2|\dot H|c_s$ being the symmetry breaking scale \cite{Baumann:2011su}.
The coupling constants are given by
\be
\rho_0 = -\frac{2\omega_0^3}{f_\pi^2}~,~~~~\Lambda_0 = -2\frac{f_\pi^2}{\rho_0}~,~~~~ \tilde\Lambda_0= \frac{f_\pi^4}{4\tilde\omega_0^3-\omega_0^3}~.
\ee
Here we see that, as a consequence of the nonlinearly realized time translation symmetry, the couplings $\rho_0$ and $\Lambda_0$ are correlated, though the $ \tilde\Lambda_0$ coupling is independent.
\item {\bf Spin-1}
For spin-1, the operators of the effective action involve $g^{00}$ and $\sigma^0$. Taking into account the tadpole constraints, the mixing Lagrangian at leading order in derivatives is  
\begin{align}
{\cal L}_{\pi \sigma}^{(1)} &= \omega_1^3 \hskip 1pt \delta g^{00}\sigma^0  + \tilde\omega_1^3  \hskip 1pt  (\delta g^{00})^2\sigma^0 \,  .\label{spin1mixaction1}
\end{align}
which, in the decoupling limit, gives 
\begin{align}
{\cal L}_{\pi \sigma}^{(1)} &= \omega_1^3  \hskip 1pt a^{-2}\big( 2\partial_i\pi\sigma_i -(\partial_i\pi)^2\sigma_0 - 2\dot\pi\partial_i\pi\sigma_i  \big) +(3\omega_1^3+4\tilde\omega_1^3)  \hskip 1pt \dot\pi^2\sigma_0 + \cdots\, .
\end{align}
Only the cubic mixing $\dot\pi\partial_i\pi\sigma_i$ will lead to the characteristic angular structure from spin exchange. Therefore, it is interesting to consider the bispectrum from the interaction vertices $\dot\pi\partial_i\pi\sigma_i$ and $\partial_i\pi\sigma_i$. Again, due to the nonlinearly realized symmetry, a single parameter $\omega_1$ controls the size of these two interactions. 
Combining the above, we can write
\begin{align}
{\cal L}^{(1)}_{\rm mix} =\frac{1}{a^2}\left(\rho_1\hskip 1pt\partial_i\pi_c\sigma_i + \frac{1}{\Lambda_1}\dot\pi_c\partial_i\pi_c\sigma_i \right) ,\label{Lmix1}
\end{align}
with two correlated couplings
\begin{align}
\rho_1\equiv \frac{2\omega_1^3}{f_\pi^2}\, , \quad \Lambda_1 \equiv -\frac{f_\pi^2}{\rho_1}\, 
\, . \label{equ:coup}
\end{align}

\item {\bf Spin-2 and higher}
For the interactions between a massive spin-2 field and the Goldstone boson, the steps are similar. Focusing on the cubic operator which produces a characteristic angular dependence from spin exchange, we obtain
\begin{align}
{\cal L}^{(2)}_{\rm mix} =\frac{1}{a^4}\left(\rho_2\hskip 1pt\partial_i\partial_j\pi_c\hat\sigma_{ij} + \frac{1}{\Lambda^2_2}\, \dot\pi_c\partial_i\partial_j\pi_c\hat\sigma_{ij} \right) ,\label{Lmix2} 
\end{align}
This time the $\rho_2$ and $\Lambda_2$ parameters are independent. A similar structure persists at higher spin $s> 2$; we find the following mixing Lagrangian 
\begin{align}
{\cal L}^{(s)}_{\rm mix} = \frac{1}{a^{2s}}\left(\rho_s \hskip 1pt\partial_{i_1\cdots i_s}\pi_c\hat\sigma_{i_1\cdots i_s} + \frac{1}{\Lambda_s^s}\dot\pi_c\partial_{i_1\cdots i_s}\pi_c\hat\sigma_{i_1\cdots i_s}\right) ,\label{spinsLmix} 
\end{align}
where $\partial_{i_1\cdots i_s}\equiv \partial_{i_1}\cdots\partial_{i_s}$ and $\rho_s$, $\Lambda_s$ are independent parameters. 

\end{itemize}

The couplings are free parameters, but must satisfy some bounds to keep the effective theory under theoretical control. The necessary requirements are that the interactions must be treated perturbatively, that the fluctuations propagate subluminally, and that the couplings are technically natural, in the sense of being robust to radiative corrections. A detailed analysis implies the following bounds \cite{Lee:2016vti}
\begin{align}
\frac{\rho^2_s}{m^2}\, \le\, \,\frac{1-c_s^2}{c_s^3}\, ,~~~~~~~\left(\frac{H}{\Lambda_s}\right)^s\ \lesssim\ \[ \frac{\big(2\pi A_\zeta^{1/2}\big)^{s+1}}{c_s^{5}}\]^{1/2}\,  ,\label{couplingbounds}
\end{align}
where $m$ is the mass of the additional field, and $A_\zeta$ is the  amplitude of the curvature perturbation power spectrum.
While sizes of some interactions can be strongly constrained, in general  large non-Gaussianity signals are still allowed.

\paragraph{From EFT to Bootstrap}
The discussion above shows that the couplings can be large within the EFT, in particular when de Sitter boosts are broken.
We may take another look at Figure \ref{fig:eft}.
For a weakly coupled theory,
the bispectrum is dominated by the first diagram on the right, which is the case studied in the de Sitter bootstrap \cite{Arkani-Hamed:2018kmz}. There the symmetry breaking is mild because of the slow-roll condition. 
However, for theories with strongly coupled dynamics, all the diagrams on the right may contribute, and thus it is possible to have sizable breaking of the boost symmetry.
The upshot of the EFT analysis is that for small sound speed and boost-breaking interactions, the primordial bispectra for masses $m\sim H$ are potentially detectable, with strengths that could be as large as the currently allowed bounds for equilateral non-Gaussianity.

\vskip4pt
With this general picture and motivation in mind, in the next sections we will develop the boostless bootstrap of cosmological colliders. 
Our goal is a precise determination of all primordial bispectrum shapes due to the exchange of a massive, scalar or spinning particle, in theories that break boost symmetry. 
In practice, we ignore all the coupling constants above, but keep the forms of the boost-breaking interactions as specific examples. 
As we will show, from the bootstrap, we will systematically obtain a complete set of these correlators, with complete analytic understanding of their shape functions.

\section{The Three-Point Scalar Seed}
\label{sec:seed}

Cosmological correlators at the reheating surface are evaluated at late times in a quasi-de Sitter space, well approximated by its asymptotic spacelike boundary. 
A key insight of the cosmological bootstrap is that the time evolution and interactions of particles during inflation are encoded in the momentum dependence of cosmological correlators on the boundary. 
The local, causal and unitary bulk evolution imply that the cosmological correlators satisfy an interesting set of differential equations.
While this idea was first realized by exploiting all the de Sitter isometries \cite{Arkani-Hamed:2018kmz}, 
we expect that similar equations exist for theories in which de Sitter boosts are broken. Indeed, we will derive the differential equations below, from the known bulk time integrals for the correlators in the case of broken boosts.

\vskip4pt
In this section, we will derive and solve the differential equations for a ``seed" cosmological correlator. We begin by introducing a linear mixing bulk-to-boundary propagator, show that it satisfies a differential equation of its own, and use that observation to construct the primary scalar seed of three-point functions. This is the correlator of two conformally coupled scalars and one inflaton exchanging a scalar particle of arbitrary mass. The seed function derived here provides a benchmark example of a  cosmological collider correlator with broken boosts, and will serve as the building block for the general bispectra of inflation.

\subsection{Free Propagators in de Sitter}
\label{sec:free}
We begin with a brief review of free propagators of scalar fields during inflation and Feynman rules for computing cosmological correlators.
Expert readers may skip this part and move on to Section \ref{sec:mix} directly.

\vskip4pt
The background geometry of the inflationary universe can be well approximated by de Sitter (dS) space, with line element
\be
ds^2=a(\eta)^2{(-d\eta^2 + d{\bf x}^2)}~,~~~~~~a(\eta)=- \frac{1}{H\eta}
\ee
where $H$ is the Hubble scale and $\eta$ is the conformal time. In the following we consider quantum fields propagating on this fixed background. Instead of restricting to de Sitter invariant theories, our analysis shall incorporate the cases with broken boost symmetries, while keeping the dilations, spatial translations and rotations intact. This means that not all interactions are built out of contractions of the background metric with spacetime derivatives. Sometimes they will involve contractions with the space components or the time component of the metric only.
In general, free scalars are described by the action
\be
S_2 = \int d \eta d^3{\bf x} \frac{1}{H^2\eta^2}\[ \frac{1}{2} (\partial_\eta\sigma)^2-\frac{1}{2} c_\sigma^2(\partial_i\sigma)^2 - \frac{1}{2} m^2 \sigma^2 \] ~,
\ee
where $m$ and $c_\sigma$ are the mass and the sound speed of the field $\sigma$ respectively. At this level, the breaking of the dS boosts is associated with $c_\s\neq 1$. In the Fourier space, we decompose the field operator as $\s({\bf k},\eta)=\sigma_k(\eta) a({\bf k})+h.c.$. Since we may absorb $c_\s$ in the momentum ${\bf k}$ by redefining $c_\s{\bf k}\rightarrow{\bf k}$, without losing generality we shall set $c_\sigma=1$ in the following analysis.  Then the mode function $\sigma_k(\eta)$ satisfies
the equation of motion  
\be \label{sigmaeom}
\(\mathcal{O}_\eta +{m^2}/{H^2} \) \sigma_k ( \eta) = 0 ,
~~~~~~{\rm with}~~
\mathcal{O}_\eta \equiv  \eta^2 \partial^2_\eta - 2\eta \partial_\eta + k^2\eta^2~.
\ee
Assuming Bunch-Davies vacuum at early times, the $\s$ mode function is explicitly given by
\be \label{sigmak}
 \sigma_k(\eta) = -i \frac{H\sqrt{\pi}}{2} e^{i\pi/4}e^{-\pi\mu/2}
 (-\eta)^{3/2}H^{(1)}_{i\mu}(-k\eta) ~ \xrightarrow{\eta\rightarrow-\infty} ~ i H\eta\frac{e^{-ik\eta}}{\sqrt{2k}}
\ee
where 
\be
\mu=\sqrt{\frac{m^2}{H^2}-\frac{9}{4}}
\ee
On the late-time boundary $\eta\rightarrow 0$, the massive scalar behaves as 
\be \label{scaling}
\lim_{\eta\rightarrow 0} \s({\bf k},\eta)= O^+({\bf k})\eta^{\Delta^+} + O^-({\bf k})\eta^{\Delta^-}
\ee
where the conformal dimensions are
$\Delta^\pm = \frac{3}{2}\pm i \mu$.
Two particular cases that we will be interested in are the massless scalar $\phi$ (with $m^2=0$) and the conformally coupled scalar $\vp$ (with $m^2=2H^2$).
For later convenience, we restore the sound speeds of these two fields and set them as $c_s$, and then their mode functions and the corresponding scaling dimensions $\Delta$ (in the sense of the power law behaviour of the decaying mode) are given by
\bea 
\phi_{k}(\eta) &=&\frac{H}{\sqrt{2c_s^3k^3}}(1+i c_s k\eta)e^{-ic_sk\eta}
~,~~~~~~~~\Delta=3 \\ 
\vp_{k}(\eta) &=&i \frac{H\eta}{\sqrt{2c_sk}}e^{-ic_sk\eta}~, ~~~~~~~~~~~~~~~~~~~~~~~\Delta=2
\eea
The scalar curvature fluctuations are well approximated by those of the massless scalar $\phi$.\footnote{Explicitly,  $\phi$ is related to the canonically normalized Goldstone $\pi_c$ in the previous section via $\phi=\pi_c/\sqrt{c_s^3}$.}
In this paper, we shall focus on the three-point functions of these two types of scalars. They will be external lines in Feynman diagrams, while the $\sigma$ field with general mass corresponds to the exchanged massive particle.

\vskip4pt
From the bulk perspective, the standard approach to compute cosmological correlators is the Schwinger-Keldysh or in-in formalism~\cite{Maldacena:2002vr, Weinberg:2005vy}. Here we give a very brief introduction to the method, and  refer to recent reviews in~\cite{Chen:2010xka, Baumann:2014nda} for more details. 
First, it is convenient to introduce two types of propagators for the quantum fields, {\it bulk-to-bulk} and {\it bulk-to-boundary} propagators.
As the massive field $\sigma$ appears in the internal lines, we are interested in its {\it bulk-to-bulk propagators}
\bea
G_{++}^\sigma(k,\eta, \eta') &=& \sigma_k(\eta) \sigma^*_k(\eta') \Theta(\eta - \eta') +
\sigma^*_k(\eta) \sigma_k(\eta') \Theta(\eta' - \eta) \nn\\ 
G_{+-}^\sigma(k,\eta, \eta') &=&  
\sigma^*_k(\eta) \sigma_k(\eta')  \nn\\ 
G_{-+}^\sigma(k,\eta, \eta') &=&  
\sigma_k(\eta) \sigma^*_k(\eta') \nn\\ 
G_{--}^\sigma(k,\eta, \eta') &=& \sigma_k(\eta) \sigma^*_k(\eta') \Theta(\eta' - \eta) +
\sigma^*_k(\eta) \sigma_k(\eta') \Theta(\eta - \eta')~,
\eea
where $\Theta$ is the Heaviside function and we use $+$ and $-$ to represent the time-ordered and anti-time-ordered pieces in the in-in integration contour. These propagators describe the motion of a comoving mode $\sigma_k$ from one bulk time $\eta$ to another time $\eta'$.
By using the equation of motion \eqref{sigmaeom}, the $G_{\pm\pm}^\sigma(k,\eta, \eta') $ propagators satisfy the following inhomogeneous equation
\be \label{G-diff-eq}
\(\mathcal{O}_\eta +{m^2}/{H^2}\)G_{\pm\pm}^\sigma(k,\eta, \eta') = \mp i H^2 \eta^2\eta'^2 \delta(\eta-\eta') ~, 
\ee
while $ G_{\pm\mp}^\sigma(k,\eta, \eta')$ satisfies the homogeneous equation correspondingly.
These propagators are related to each other by complex conjugation, $G_{++}=G_{--}^*$ and $G_{+-}=G_{-+}^*$.

\vskip4pt
For fields associated with the external lines, we introduce the {\it bulk-to-boundary propagators}, which describe the propagation from some bulk time $\eta$ to the late-time boundary of de Sitter $\eta_0\rightarrow0$.
For the massless scalar $\phi$, they are given by
\be
K_+(k,\eta) =  \phi_{k}(\eta_0) \phi^*_{k}(\eta)~, ~~~~~~~~
K_-(k,\eta) =  \phi^*_{k}(\eta_0) \phi_{k}(\eta)~,
\ee
while the ones for the conformally coupled scalar $K_\pm^\vp(k,\eta) $ are given by the same form but with the $\vp_k(\eta)$ mode function. 
It is clear that the bulk-to-boundary propagators satisfy the homogeneous equation $\(\mathcal{O}_\eta +{m^2}/{H^2}\right)K=0$, and they have $K_+=K^*_-$.

\vskip4pt
Using this set of propagators, it becomes straightforward to derive the Feynman rules for computing boundary correlators in interaction theories. 
For contact diagrams, only the bulk-to-boundary propagators are needed, and we have one time integral from $\eta=-\infty$ to $\eta\rightarrow0$ to capture the field interactions in the bulk.
The computation becomes more complicated when we study exchange diagrams from this bulk perspective. The internal lines are associated with bulk-to-bulk propagators which lead to multiple nested time integrals in the in-in formalism. 
In general, it is very difficult to find analytical expressions, and thus one has to resort to numerical methods for solving these integrals.

\subsection{A Mixed Propagator}
\label{sec:mix}

To simplify the computation of three-point correlators from exchange processes, we first consider the two-point function $\cG_\pm\equiv \langle\s_\pm(\eta)\phi(\eta_0)\rangle$. This object is a bulk-to-boundary propagator generated by quadratic interactions. Physically, it describes the conversion process from a massive field $\sigma$ at the bulk time $\eta$ to an inflaton, which then freely propagates to the boundary.
In this section let us focus on the simplest quadratic interaction $\dot\phi\sigma$.
Setting the coupling constant to unity, we express this mixed propagator as
\be \label{mix0}
\cG_\pm (k, \eta) =\pm i   \int_{-\infty}^{0} d\eta' a(\eta')^3 \[ G_{\pm\pm}^\sigma(k,\eta, \eta') \partial_{\eta'} K_{\pm}(k, \eta')
- G_{\pm\mp}^\sigma(k,\eta, \eta') \partial_{\eta'} K_{\mp}(k, \eta')\]~.
\ee
Note that in \eqref{mix0}, we make the simplifying assumption that $\phi$ and $\sigma$ have the same sound speed, which is not generally true for boost-breaking theories. We consider the case of different sound speeds and other mixing interactions in Section \ref{sec:moremix}.

\vskip4pt
This linear mixing is ubiquitous in cosmological backgrounds when multiple fields are present. In maximally symmetric spacetimes, one can always diagonalize the field basis and remove such mixings. 
During inflation, the de Sitter boosts are broken by the time-dependent inflaton profile $\Phi(t)$, which generically leads to quadratic interactions.\footnote{One simple example is to consider the shift-symmetric coupling $(\partial \Phi)^2\sigma$. As shown in Figure \ref{fig:eft}, expanding $\Phi(t,{\bf x})=\Phi(t)+\phi(t,{\bf x})$  with $\dot\Phi(t)\neq0$, we obtain the linear mixing $\sim\dot\Phi \dot\phi\sigma$.
In the EFT analysis, it is given by the first term in \eqref{Lmix0}.
 In general, the $\dot\phi\sigma$ interaction will arise when the inflaton trajectory deviates from the geodesics in the multi-dimensional field manifold \cite{Achucarro:2010da}.}
The mixed propagator generates new shapes of cosmological correlators, beyond those of self-interactions of the inflaton, as we shall see.

\vskip4pt
The explicit form of this simplest mixed propagator has been studied in \cite{Chen:2017ryl}.
Here instead of computing the $\eta'$ integral directly, we are interested in deriving
a differential equation for $\cG$. From the equation for $G$-propagators in \eqref{G-diff-eq},
it is easy to see that the evolution of the mixed propagator satisfies the following inhomogeneous equation
\be \label{eqGm}
\(\mathcal{O}_\eta +{m^2}/{H^2}\) \cG_\pm (k, \eta) = - {H} \frac{\eta^2}{2k} e^{\pm ik\eta} .
\ee
Notice that the free bulk-to-boundary propagator $K_\pm$ satisfies a similar equation, but without the source term.
As a consequence of the $\dot\phi\s$ interaction, this nonzero source marks the main feature of the mixed propagator.
Next we introduce the dimensionless mixed propagator $\hat\cG_\pm (k \eta) \equiv (k^3/ H) \cG_\pm(k, \eta)  $, which is a function of the combination $k\eta$ only.
Therefore, we are allowed to trade $\eta$-derivatives with $k$-derivatives on $\hat \cG$, and \eqref{eqGm} is equivalent to
\be \label{eqGkm}
\( \mathcal{O}_k +{m^2}/{H^2} \)\hat\cG_\pm (k \eta)  =
-\frac{1}{2} k^2 \eta^2  e^{\pm ik\eta}~,~~~~~~ {\rm with}~\mathcal{O}_k\equiv k^2\partial_{k}^2 - 2k\partial_k + k^2\eta^2  .
\ee
To better understand the evolution behaviour of the mixed propagator, it is useful to look at its early-time and late-time limits, where the time integrals can be performed. At early times, or equivalently the short-wavelength limit $k\gg -1/\eta$, we have
\bea \label{tauinf}
\lim_{\eta\rightarrow-\infty}\hat\cG_+ (k\eta) &=& \frac{i}{4} k\eta e^{ik\eta} \log(-2k \eta)~.
\eea
Thus the free Bunch-Davies vacuum has been dressed by the mixing, but still only positive-frequency mode appears.
The late-time behaviour corresponds to the soft limit
\be   \label{softK}
\lim_{k\rightarrow 0}\hat\cG_+ (k\eta) = 
  \sum_\pm A_\pm\(\frac{-k\eta}{2}\)^{\frac{3}{2}\pm i\mu} 
~,~~~~~~{\rm with}~A_\pm =   \frac{\pi^{3/2} e^{\frac{\pi\mu}{2}\mp\frac{i\pi}{4}}(1\mp ie^{-\pi\mu})}{\sinh(2\pi \mu)\Gamma(1\pm i\mu)}~,
\ee
which encodes the scaling of $\sigma$ on the boundary in \eqref{scaling}. This interesting behaviour of the mixed propagator is crucial for the new features in the cosmological correlators.

\subsection{The Primary Scalar Seed}

In this section, we compute the three point function of two conformally coupled scalars exchanging a massive particle with a massless scalar. We do the computation for the cubic vertex $\vp^2\s$ and set the coupling constants to be unity in our analysis. This correlator will serve as a seed to build all the cosmological collider bispectra  later. 

\vskip4pt
With the help of the mixed propagator, the exchange interaction can be simplified into a ``contact-like'' form. Explicitly the three-point function is given by
\bea \label{ccphi}
\langle \vp_{{\bf k}_1} \vp_{{\bf k}_2} \phi_{{\bf k}_3} \rangle'  &=& i \int d\eta a(\eta)^4 \[ K^\vp_+(k_1,\eta) K^\vp_+(k_2,\eta) \cG
_+(k_3,\eta) -c.c.\] + {\rm perms.}\nn\\
&=&  \frac{iH\eta_0^2}{4k_1k_2k_3^2} \mathcal{\hat I} (k_{12}, k_3) + {\rm perms.}~,
\eea
where $k_{12}\equiv k_1+k_2$ and the prime on the correlator means that the momentum-conserving delta function
has been stripped.
Meanwhile, we have defined the {\it primary scalar seed} \footnote{The name {\it primary} is used because later we will need other scalar seeds.}
\be \label{singleM}
\mathcal{\hat I} (k_{12}, k_3) 
\equiv \frac{1}{k_3} \int_{-\infty}^{0} \frac{d\eta}{\eta^2} \[
e^{i k_{12}\eta}
\hat\cG_+ (k_3\eta) -
e^{-i k_{12}\eta}
\hat\cG_- (k_3\eta)\] ,
\ee
which will be used as a building block to construct more general three-point functions. Notice that $\mathcal{\hat I}$ is dimensionless and depends only on the ratio $k_3/k_{12}$.
The direct integration of \eqref{singleM} is difficult.
Instead, using  \eqref{eqGkm}, we find 
\be
\[ (k_{12}^2-k_3^2)\partial_{k_{12}}^2 + 2 k_{12} \partial_{k_{12}} + \frac{m^2	}{H^2}-2 \] \mathcal{\hat I} (k_{12}, k_3)
= i\frac{k_3}{k_T}~,
\ee
where $k_T\equiv k_1+k_2+k_3$.
In terms of $u\equiv {k_3}/{k_{12}}$, this equation becomes
\be \label{seedeq}
\[ \Delta_u +\(\mu^2+ \frac{1}{4}\) \] \mathcal{\hat I}  (u)
= {i}\frac{u}{1+u} ,
\ee
where we have defined the differential operator
\be
\Delta_u\equiv u^2(1-u^2)\partial_{u}^2 -2u^3\partial_{u}~.
\ee
This equation has a hidden conformal symmetry, which is closely related to the differential equation of the four-point scalar seed in de Sitter bootstrap.
Meanwhile the boost-breaking effect of the linear mixing is manifested in the source term. Next, we will first derive its analytical solution explicitly, and then discuss the  connection with the scalar seed in de Sitter bootstrap.

\paragraph{Analytical solution}
The equation \eqref{seedeq} is a second order ordinary differential equation with three singular points $u=0,\pm1$.
To find its solution, we separate $\mathcal{\hat I} (u)$ into a homogeneous part $\mathcal{\hat H} (u)$ and a particular part $\mathcal{\hat S} (u)$.
For the particular solution with $u\in [0,1]$, we use the following series expansion around the regular singular point $u=0$
\be \label{partisol1}
\mathcal{\hat S}(u) = i  \sum_{n=0}^{\infty}c_n u^{n+1} .
\ee
Substituting this ansatz into \eqref{seedeq}, we find the recursive relation of the series coefficients
\be
c_0 = \frac{1}{\frac{1}{4}+\mu^2}~,~~~~~~~~c_1=\frac{-1}{\frac{9}{4}+\mu^2}~,~~~~~~~~c_n=\frac{(-1)^n+n(n-1)c_{n-2}}{(n+\frac{1}{2})^2+\mu^2}~,
\ee
which can be solved as
\be
c_n = \sum_{m=0}^{\lfloor n/2\rfloor}
\frac{(-1)^n n! / (n-2m)!}{\[\(n+\frac{1}{2}\)^2 + \mu^2\]\[\(n-\frac{3}{2}\)^2 + \mu^2\]...\[\(n+\frac{1}{2}-2m\)^2 + \mu^2\]} .
\ee
This series solution is regular at $u=0$, but has singular behaviour when $u\rightarrow\pm 1$.

\vskip4pt
Next, we derive the general solution $\mathcal{\hat H} (u)$, which can be written as
\be \label{homo-0}
\mathcal{\hat H} (u) = -\frac{i }{2}\sum_{\pm}
C_{\pm} \( \frac{iu}{2\mu} \)^{\frac{1}{2}\pm i\mu} {}_{2}F_1 
\[ \frac{1}{4}\pm \frac{i\mu}{2}, \frac{3}{4}\pm \frac{i\mu}{2}; 1 \pm {i\mu} ; u^2 \]~,
\ee
with two free coefficients $C_\pm$. To fix them, we impose the non-analytic  behaviour of the primary scalar seed at $u\rightarrow 0$.
By using the soft limit of the mixed propagator in \eqref{softK}, the integral in $\mathcal{\hat I}$ can be explicitly solved as 
\be
\mathcal{\hat I} (u\rightarrow0)
= - \frac{i}{2}\sum_{\pm}B_\pm\(\frac{u}{2}\)^{\frac{1}{2}\pm i\mu},~~~~{\rm with} ~~ B_\pm= \frac{\pi^{3/2}}{\cosh\pi\mu} \(1\mp\frac{i}{\sinh\pi\mu}\)
 \frac{\Gamma(\frac{1}{2}\pm i\mu)}{\Gamma(1\pm i\mu)}~.
\ee
This non-analytic  soft limit can only be present in the homogeneous solution, as $\mathcal{\hat S}(u\rightarrow0)$ is a rational function. Thus by matching the coefficients in this limit, we fix $C_\pm = (-i\mu)^{\frac{1}{2}\pm i\mu} B_\pm$. It is interesting that this limit automatically fixes the two free coefficients of the differential equation, so no additional boundary condition is necessary.

\begin{figure}[t]
\centering
      \includegraphics[scale=0.8]{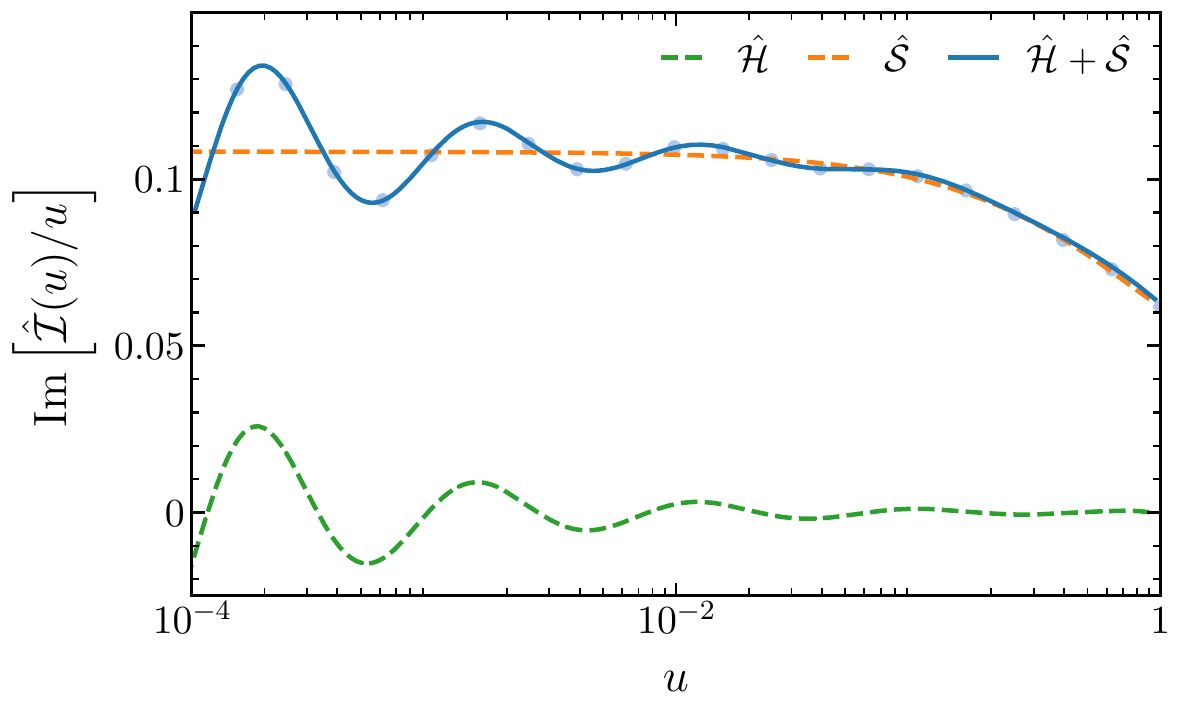}
           \caption{The analytical solution of the primary scalar seed with $\mu=3$. For comparison, the blue dots are numerical results from the direct integration of \eqref{singleM}.} 
    \label{fig:seed}
\end{figure}

\vskip4pt
The form of the solution is demonstrated in Figure \ref{fig:seed}. As we can see, the homogeneous part contains the non-analytic oscillations in the $u\rightarrow0$ limit; while the series solution is convergent as long as $u$ is not too close to 1. 
The full expression $\mathcal{\hat I} (u) = \mathcal{\hat S} (u)+\mathcal{\hat H} (u)$ matches well with the numerical computation. Thus without integrating \eqref{singleM} directly, we find the exact and practical solution of the primary scalar seed, which will be extensively used in our following analysis.

\paragraph{Singularity structure} In the derivation above, the singular behaviour of $\mathcal{\hat S}(u)$ and $\mathcal{\hat H}(u)$ at $u=0$ is manifest. Now we look at $u\rightarrow\pm1$. For the homogeneous solution \eqref{homo-0}, we can see from the hypergeometric function that there are logarithmic singularities as $u\rightarrow \pm1$.
When $u\rightarrow 1$, this corresponds to the folded limit $k_{12}=k_3$, and the homogeneous solution goes to 
\be \label{homo-folded}
\lim_{u\rightarrow 1}\mathcal{\hat H}  (u) = \frac{i}{2}\frac{\pi}{\cosh\pi\mu}\log(1-u)  .
\ee
For vacuum with only positive-frequency mode, we don't expect singular behaviour in this limit. Therefore, the homogeneous and particular parts must cancel their logarithmic singularities against each other.
For $u\rightarrow-1$, this limit corresponds to the situation where the total energy involved in the process vanishes, $k_T\equiv k_{12}+k_3\rightarrow0$.  The singularity of the three-point function is allowed in this limit, which is known as the  $k_T$-pole \cite{Maldacena:2011nz, Raju:2012zr}. From the homogeneous solution, we have
\be
\lim_{u\rightarrow -1}\mathcal{\hat H} (u)= \frac{1}{2} \( \lim_{u\rightarrow e^{i\pi}} + \lim_{u\rightarrow e^{-i\pi}} \)\mathcal{\hat H} (u) = -\frac{i}{2}
\frac{\pi}{\cosh\pi\mu}\log(1+u)~.
\ee

\vskip4pt
The singularity of the series solution is not so straightforward to obtain. The detailed derivation is in Appendix \ref{app:seeds}; here we show the final result. For the folded limit $u\rightarrow 1$
\be \label{cs1u1}
\lim_{u\rightarrow 1}\mathcal{\hat S}  (u) =-\frac{i}{2}\frac{\pi}{\cosh(\pi\mu)}\log(1-u)~,
\ee
which precisely cancels the logarithmic singularity of the homogeneous solution in \eqref{homo-folded}. Meanwhile, there is a physical $k_T$-pole in the series solution
\be \label{series0-kT}
\lim_{u\rightarrow -1}\mathcal{\hat S}  (u)  = -\frac{i}{4} \log^2(1+u)~,
\ee
which dominates over the one from the homogenous solution.
Thus the three-point function has a total-energy pole of $\log^2(k_T)$, and the residue of the pole is independent of the mass of the exchanged field. This result can be understood by considering the early-time limit of the mixed propagator in \eqref{tauinf}. Substituting it in the primary scalar seed, we find 
\be \label{I0-kT}
\lim_{k_T\rightarrow0}\mathcal{\hat I} 
= \frac{i}{4}\int_{-\infty}^{0} \frac{d\eta}{\eta} \[
e^{i k_{T}\eta}
 \log(-2k_3 \eta) +c.c.\] \rightarrow -\frac{i}{4} \log^2 k_T~.
\ee
Thus this $k_T$-pole is a feature of the deformed vacuum state of the mixed propagator.
Notice that the total energy singularity here is also a partial energy one, where  the energy flowing
into the cubic vertex vanishes. We will see in the next section that these two poles become distinguishable when a nontrivial sound speed is involved.

\paragraph{Comparison with the dS bootstrap} 
In the cosmological bootstrap with full de Sitter isometries, the building block is a four-point function $\hat F$ of conformally coupled scalars exchanging a massive scalar \cite{Arkani-Hamed:2018kmz}. It was shown that as a consequence of conformal symmetry, this four-point scalar seed satisfies
\be
\[ \Delta_u +\(\mu^2+ \frac{1}{4}\) \] \hat{F}(u,v) = \frac{uv}{u+v}\ee
with $u\equiv \frac{|{\bf k}_1+{\bf k}_2|}{k_{12}}$ and $v\equiv \frac{|{\bf k}_1+{\bf k}_2|}{k_{34}}$. 
In the $v\to 1$ limit, the differential equation for the four-point function is the same as \eqref{seedeq}.  This connection between the four-point and three-point seed functions can be made manifest in the bulk picture. Let us look at the time integral from the right $\varphi^2\s$ vertex in the four-point exchange diagram
\be
\sum_{\pm}(\pm i)\int d\eta' a^4 K^\vp_{\pm} (k_3,\eta')K^\vp_+ (k_4,\eta') G_{\pm+}(|{\bf k}_1+{\bf k}_2|,\eta',\eta) ~,
\ee
where we use the $+$ propagator of the $k_4$ leg as an example.
Next, we take the $k_4\to 0$ and ${|{\bf k}_1+{\bf k}_2|\rightarrow k_3}$ limit (i.e. $v\rightarrow1$), this vertex becomes
\be
\sum_{\pm}(\pm i)\int \frac{d\eta'}{\eta'^2} e^{ik_3\eta'} G_{\pm+}(k_3,\eta',\eta) \sim \mathcal{K}_+(k_3,\eta)~,
\ee
which reduces to the mixed propagator \eqref{mix0} from the $\dot\phi\s$ interaction. Therefore, the three-point function $\langle\varphi\varphi\phi\rangle$ has the same form with the soft limit $k_4\rightarrow0$ of the $\langle\varphi\varphi\varphi\varphi\rangle$ correlator.
In \cite{Arkani-Hamed:2018kmz} it was shown that this is the ``weight shifting" procedure to turn the de Sitter symmetric four-point function into the slow-roll suppressed three-point function of conformally coupled scalars and a massless scalar. 
We can also check that the four-point seed solution  reproduces the $\hat{\mathcal{I}}$ solution here by taking $v\rightarrow1$. The shape function is identical to $\hat{F}(u,1)$, as expected.

\vskip4pt
However, it is important to comment on the distinction between these two seed functions. 
In the three-point seed equation \eqref{seedeq}, we make no assumption of weakly broken boosts, and thus  in general, the resulting correlators such as $\langle\varphi\varphi\phi\rangle$ are {\it not} slow-roll suppressed.
This is manifest from the schematic in Figure \ref{fig:eft}: while the dS bootstrap focuses on the first diagram on the right-hand side, the boostless bootstrap analyzes the diagram on the left-hand side directly.
Another advantage of the three-point scalar seed is that at the technical level it has a much simpler solution, considering that
the four-point seed solution is given by a
two-variable generalization of the hypergeometric series. 
As we are mainly interested in the bispectra, it becomes more straightforward to work on the three-point functions from the beginning.
Furthermore, the primary seed function can be simply extended to describe more complicated boost-breaking interactions, as we will see in the following sections.

\paragraph{Unitarity and Locality} 
We close this section by making some remarks about how unitarity and locality are manifested in the exchange bispectra.
As two fundamental principles, unitarity and locality have played crucial roles in the modern studies of scattering amplitudes. For cosmological correlators, some of the consequences of unitarity come from the {cosmological optical theorem} (COT) \cite{Goodhew:2020hob}, while a consequence of locality is in the {manifestly local test} (MLT)  \cite{Jazayeri:2021fvk}. 
These tests, which are based on the assumption of {\it free} bulk-to-boundary propagators in both contact and exchange diagrams,  previously did not take into account the linear mixing with the external fields. 
Thus it is interesting to check whether COT and MLT are still satisfied when mixed propagators are involved.
Since $\langle\vp\vp\phi\rangle$ in \eqref{ccphi} provides the simplest three-point function with a mixed propagator, 
we use this result as a demonstration. The analysis can be easily extended to more complicated single-exchange processes with arbitrary quadratic and cubic interactions.

\vskip4pt
For the analysis of the COT, it is convenient to use the dimensionless bulk-to-boundary propagators $\hat{K}(k,\eta)= k^{3} K(k,\eta)$. One key step for deriving the COT is to notice that these free propagators are Hermitian analytic
$\[\hat{K}(-k^*,\eta)\right]^* = \hat{K}(k,\eta)$,
which follows from the choice of the Bunch-Davies vacuum.
For the mixed propagator, although the expression becomes more complicated because of the quadratic interaction, we find that this property of Hermitian analyticity is nicely inherited by $\hat \cG$, namely
\be
\[\hat{\cG}_\pm(-k^*,\eta)\]^* = \hat{\cG}_\pm(k,\eta)~.
\ee
These identities of the analytic continuation of propagators can be commuted with the time integral. As a result, the primary scalar seed \eqref{singleM} satisfies
\be
 \mathcal{\hat I} (k_{12}, k_3) + \[ \mathcal{\hat I} (-k_{12}, -k_3) \]^* = 0 ~,
\ee
which corresponds to the COT for contact diagrams \cite{Goodhew:2020hob}. This condition basically means that $\mathcal{\hat I}$ is imaginary.
Interestingly, the exchange bispectrum with a mixed propagator looks ``contact"-like as far as unitarity is concerned. 

\vskip4pt
For the MLT, one may start with the observation that the free bulk-to-boundary propagators of massless fields in de Sitter space satisfy
\be \label{mlt}
\partial_k \hat{K}^{(n)}(k,\eta) |_{k=0} = 0~,
\ee
where $n\geq0$ corresponds to the number of time derivatives on $\hat{K}$.
This condition gives powerful boundary constraints for both contact and exchange correlators with external free massless fields. 
In particular it has been applied to bootstrap all  three- and four-point functions from  tree-level boost-breaking interactions in single field inflation \cite{Jazayeri:2021fvk, Bonifacio:2021azc, Cabass:2021fnw}.
The corresponding relation for the mixed propagator becomes more complicated than the one in \eqref{mlt}. As shown in \eqref{softK}, $\hat{\cG}$ encodes the non-analytic scaling of the massive field on the boundary.
Intuitively, this is because the mixed propagator describes the non-local conversion from $\s$ to $\phi$ in the bulk evolution. 
Thus the MLT in its current form is not applicable to these correlators.
It remains an interesting question about how a generalized version could incorporate the behaviour of the mixed propagators.

\section{More Mixed Propagators}
\label{sec:moremix}

In the previous section, we bootstrapped the $\langle\vp\vp\phi \rangle$ bispectrum with the simplest $\dot\phi\s$ linear mixing. In this section, we consider more general mixing vertices, incorporating all the possible quadratic interactions between $\phi$ and $\sigma$.
We will construct the general form of the mixed propagators by first introducing their building blocks in Section \ref{sec:buildmix}.
Next, in Section \ref{sec:gen-seeds} we will propose the generalized scalar seeds for the three-point functions with higher derivative quadratic interactions.
We show that they can be obtained from the primary scalar seed with $\dot\phi\s$ mixing through recursive relations. 
While the exchanged field is still a massive scalar in the analysis here, later on we will show that this  generalization of the seed functions will be crucial for bootstrapping the three-point function of spin exchanges.

\subsection{Mixed Propagators from General Interactions}
\label{sec:buildmix}

To extend our analysis of boost breaking bispectra, we must take into account the effects of different sound speeds for the interacting fields, as well as general quadratic interactions between them. Let us first consider effects due to reduced sound speeds, which is typical in theories with broken boosts. 
We assume that the inflaton $\phi$ and the conformally coupled scalar $\vp$ have a sound speed $c_s$, while the one for the exchanged field $\sigma$ is $c_\sigma$. 
Without loss of generality, we rescale $k\rightarrow c_\sigma k$ and $c_s\rightarrow c_s/c_\sigma$, thus removing the $c_\sigma$-dependence. After the rescaling, ``$c_s$" can take any positive value, being a {\it ratio} of sound speeds. Therefore we consider general (either sub or superluminal) $c_s$ for the external scalars below, where the free propagators are $K^{\phi,\vp}_\pm(c_sk,\eta)$ and $G^\s(k,\eta,\eta')$. We will restore the $c_\s$-dependence in the phenomenological analysis in Section \ref{sec:pheno}.

\vskip4pt
Next, we consider the boost-breaking quadratic interactions between $\sigma$ and $\phi$. The general form of the linear mixing vertex can be written as
\be
\mathcal{L}^{\phi\sigma} = a^{-n_{\partial_i}} \partial_{i}^{n_{\partial_i}} \( \partial_t^{n_{\partial_t}}\phi \) \s
, \label{generalmix}
\ee
where $n_{\partial_i}$ and $n_{\partial_t}$ are the number of spatial and time derivatives on $\phi$. 
Notice that we have moved all spatial and time derivatives to $\phi$ via integration by parts. Naively $n_{\partial_i}$ should be even, but we also consider $n_{\partial_i}$ odd, to account for the possible contractions between internal polarization vectors $\epsilon_{i}$ and $\partial_i$. This is the case for the linear mixing with spinning fields, which we analyze in Section \ref{sec:spin}.  
The $\dot\phi\s$ mixing discussed in the previous section corresponds to $n_{\partial_i}= 0$ and $n_{\partial_t}=1$.

\vskip4pt
Motivated by the analysis above, we will first propose the building blocks of mixed propagators, and then consider how to construct the ones for all possible mixings.

\subsubsection{Building Blocks of the Mixed Propagators}

To capture the higher-derivative interactions and effects of different sound speeds, 
we introduce 
\bea \label{mix-ncs}
\hat\cG_\pm^{(n)} (k \eta ; c_s) 
&\equiv & \pm i \frac{c_s^3 k^{3+n}}{H^{n+1}} \int_{-\infty}^{0} d\eta' a(\eta')^{3-n} \Big[ G_{\pm\pm}^\sigma(k,\eta, \eta') \partial_{\eta'} K_{\pm}(c_s k, \eta')
\nn\\
&&~~~~~~~~~~~~~~~~~~~~~~~~~~~~~~~~~~
- (-1)^n G_{\pm\mp}^\sigma(k,\eta, \eta') \partial_{\eta'} K_{\mp}(c_s k, \eta')
\Big]
\eea
as the dimensionless building blocks for more general mixed propagators.
We have chosen the $k$-dependent prefactor to make $\hat\cG^{(n)}$ a function of the combination $k\eta$.
For $n=0$ and $c_s=1$ we retrieve the dimensionless mixed propagator for the $\dot\phi\s$ interaction. 
The index $n$ counts the number of spatial derivatives in these two-point vertices---for general $n$, this is the propagator coming from the quadratic interaction $a^{-n}\hat{k}_{i_1}...\hat{k}_{i_n} \partial_{i_1...i_n}\dot\phi\s$. 

\vskip4pt
Following the strategy of Section \ref{sec:mix}, by using the inhomogeneous equation of the $G^\s$ propagator \eqref{G-diff-eq} and then trading $\eta$ derivatives with $k$ derivatives, we find the differential equation of  $\hat\cG^{(n)}$
\be \label{eqGkmn}
\(\mathcal{O}_k +{m^2}/{H^2}\)
 \hat\cG_\pm^{(n)} (k \eta; c_s) = - \frac{1}{2}  c_s^2(-k\eta)^{n+2} e^{\pm ic_sk\eta} .
\ee
While the left-hand side of the equation remains the same as \eqref{eqGkm}, the generalization to arbitrary $n$ and $c_s$ is manifested in the source term, which has the form of a free bulk-to-boundary propagator $\partial_\eta K$.
We leave detailed discussions of $\hat\cG^{(n)}$  and their asymptotic behaviours to Appendix \ref{app:mix-p}. Below we show that $\hat\cG^{(n)}$ can be reduced to the simplest mixed propagators plus a sum of free propagators.

\paragraph{Recursive relations} We focus on $\hat\cG^{(n)}_+$, as $\hat\cG^{(n)}_-$ can be easily obtained from complex conjugation.
Consider the source term of the generalized equation in \eqref{eqGkmn}: $\hat{S}^{(n)}=(-k\eta)^{n+2} e^{ ic_sk\eta} $, and notice that
\be \label{Ksource-re}
\mathcal{O}_k\hat{S}^{(n)}=  (n+2)(n-1) \hat{S}^{(n)} - 2i(n+1)c_s\hat{S}^{(n+1)}+(1-c_s^2)\hat{S}^{(n+2)} ~.
\ee
This relation connects the source terms with different orders of $n$.
From this relation, there are two different results depending on the sound speed.
\begin{itemize}
\item For $c_s=1$, the last term in \eqref{Ksource-re} vanishes. By using the differential equations of  $\hat\cG_+^{(n)} (k \eta) $ and  $\hat\cG_+^{(n-1)} (k \eta) $,  we find the recursive form
\be \label{Krecur}
\hat\cG_+^{(n)} = -\frac{i}{2n} \left( \[ \mu^2+\(n-\frac{1}{2}\)^2\] \hat\cG_+^{(n-1)} + \frac{1}{2} \hat{S}^{(n-1)} \right).
\ee
Thus applying this relation iteratively, the $n$-th order building block of the mixed propagator can be expressed in terms of the simplest one with $n=0$ and a sum of source terms.
\item For $c_s\neq1$, by using \eqref{Ksource-re} and \eqref{eqGkmn} with different $n$,  
we find
\be \label{Krecurcs}
\hat\cG_+^{(n)}  =\frac{1}{1-c_s^2}\left(
2i(n-1)c_s \hat\cG_+^{(n-1)} 
-\[\(n-\frac{3}{2}\)^2+\mu^2\]\hat\cG_+^{(n-2)}  -\frac{c_s^2}{2}\hat{S}^{(n-2)}\right) .
\ee
Thus in addition to $\hat\cG^{(0)}$, we also need $\hat\cG^{(1)}$ to obtain $\hat\cG^{(n)}$ for arbitrary $n$.  For $c_s\ll 1$, since $\hat\cG^{(n)}\sim c_s^2$,
the relation simplifies to $\hat\cG^{(n)} \simeq - \[\(n-{3}/{2}\right)^2+\mu^2\right] \hat\cG^{(n-2)} -{c_s^2}\hat{S}^{(n-2)}/{2}$,
and thus up to a sum of source terms, we get $\hat\cG^{(n)} \propto \hat\cG^{(0)}$ for even $n$, or $\hat\cG^{(n)} \propto \hat\cG^{(1)}$ for odd $n$.
\end{itemize}
These two recursive relations simplify the discussion of higher-derivative quadratic interactions.
As we shall show next, using them we can reduce the mixed propagators from complicated interactions to simple ones with some constant prefactors.

\subsubsection{Building Mixed Propagators}

Now we build the propagator for generic mixing, as in \eqref{generalmix}. We are mainly interested in the ones with higher derivatives, and thus mixings like $\phi\sigma$ and $\partial_i\phi\sigma$, as well as nonlocal interactions with inverse Laplacians will not be included in the discussion here.
In addition, for interactions of two scalars, the number of spatial derivatives ${n}_{\partial_i}$ is supposed to be even, as required by the rotational invariance. But we may have an odd number of spatial derivatives when the massive field has spin. We leave the discussion on mixed propagators with spinning particles to Section \ref{sec:spin}, and assume that ${n}_{\partial_i}$ is positive and even  in this section.
Explicitly, the mixed propagator from the general quadratic interaction in \eqref{generalmix} takes the following  form
\be \label{mix-general}
\cG_\pm^{(n_{\partial_t},n_{\partial_i})} ( k, \eta) =\pm i (\mp ik)^{n_{\partial_i}} \int_{-\infty}^{0} d\eta' a(\eta')^{4-n_{\partial_i}-n_{\partial_t}} \[ G_{\pm\pm}^\sigma \partial_{\eta'}^{{n}_{\partial_t}} K_{\pm} 
- G_{\pm\mp}^\sigma  \partial_{\eta'}^{{n}_{\partial_t}} K_{\mp} \]~,
\ee
where we use the upper indices $(n_{\partial_t},n_{\partial_i})$ to denote the number of time and spatial derivatives respectively.
Let us analyze various cases separately:
\begin{itemize}
\item For ${n}_{\partial_t}=1$, the quadratic interaction can be
brought to the  form  
$
a^{-n_{\partial_i}} \partial_{i}^{n_{\partial_i}} \dot\phi  \sigma
$,
and the corresponding mixed propagator is simply given by
\be
\cG_\pm^{(1,n_{\partial_i})}  (k , \eta) = (\mp i)^{n_{\partial_i}}  \frac{H^{n_{\partial_i}+1}}{c_s^3k^{3}}\hat\cG_\pm^{\(n_{\partial_i}\)} ~.
\ee
\item For ${n}_{\partial_t}=0$, the interaction vertex is $a^{-n_{\partial_i}} \partial_{i}^{n_{\partial_i}} \phi  \sigma$. To express its mixed propagator in terms of the building blocks, we resort to the mode function of the massless scalar. Then we find
\be
\cG_\pm^{(0,n_{\partial_i})}  (k , \tau) = (\mp i)^{n_{\partial_i}+1} \frac{H^{n_{\partial_i}}}{c_s^3k^{3}}\[\hat\cG_\pm^{\(n_{\partial_i}-1\)} \mp i \hat\cG_\pm^{\(n_{\partial_i}-2\)} \]~.
\ee
We may further rewrite the result by using the recursive relations of the building blocks. In particular, for $c_s=1$ we find
\be
\cG_\pm^{(0,n_{\partial_i})}  (k , \eta) = \frac{-(\mp i)^{n_{\partial_i}}H^{n_{\partial_i}}}{2(n_{\partial_i}-1)c_s^3k^3}  \[\(\mu^2+(n_{\partial_i}-\frac{1}{2})^2\)\hat\cG_\pm^{\(n_{\partial_i}-2\)} -\frac{1}{2} (k\eta)^{n_{\partial_i}} e^{\pm ik\eta}\]~.
\ee
\item For ${n}_{\partial_t}>1$, we can use the equation of motion of $\phi$ to reduce its number of time derivatives to be ${n}_{\partial_t}=0,1$. Or, equivalently we notice that from the mode function of the massless scalar the time derivatives of the bulk-to-boundary propagator satisfy
\be
\partial^{{n}_{\partial_t}}_\eta K_\pm =  (\pm ic_s k)^{n_{\partial_t}}\[ \frac{1-{n}_{\partial_t}}{c_s^2k^2\eta}\mp\frac{i}{c_sk}\]\partial_\eta K_\pm~,~~~~~~{n}_{\partial_t}\geq0~.
\ee
By using this relation in the mixed propagators, the quadratic interaction with arbitrary numbers of spatial and time derivatives leads to
\be \label{mix-g2}
\cG_\pm^{(n_{\partial_t},n_{\partial_i})}  (k , \eta) = (\pm i)^{n_T}\frac{H^{n_{T}}}{c_s^3 k^3}\[({n}_{\partial_t}-1)\hat\cG_\pm^{\(n_{T}-2\)} \mp i \hat\cG_\pm^{\(n_{T}-1\)} \]~,
\ee
where we have set $n_T={n}_{\partial_t}+{n}_{\partial_i}$ as the total number of derivatives on $\phi$.
\end{itemize}

Therefore, as we have seen, the mixed propagators from higher derivative quadratic interactions in general can be written in terms of linear combinations of building blocks $\cG^{(n)}$. By using the recursive relations in \eqref{Krecur} and \eqref{Krecurcs}, they can be further reduced to the lower-$n$ mixed propagators with some constant prefactors and a sum of source terms.
At last, let us comment on the time derivatives on $\s$ in the quadratic interaction, which can be moved onto $\phi$  by repeated use of integration by parts. This procedure also generates additional source terms in the mixed propagators, but since they  have the form of free bulk-to-boundary propagators and lead to simple contact interactions in three-point functions, we neglect their contribution in our analysis.

\vskip4pt
To summarize, in this section we derived the mixed propagators for arbitrary boost-breaking quadratic interactions. For a ``cosmological collider" diagram with massive scalar exchange, the most relevant mixing is the one with lowest derivative $\dot\phi\s$, thus we are mainly interested in $\cG_\pm  (k , \eta) =H\hat\cG_\pm^{(0)}/(c_sk)^3  $.
Nonetheless, the general discussion of higher derivative mixings will be useful for the bootstrap of spinning exchanges in Section \ref{sec:spin}.

\subsection{Generalized Scalar Seeds}
\label{sec:gen-seeds}

Having determined the propagator for arbitrary linear mixing, we proceed to generalize the three-point seed function. 
What we have in mind is the bispectrum between two conformally coupled scalars and one massless scalar, as in \eqref{ccphi}.
Replacing the linear mixing propagator $\hat\cG$ with the general building block $\hat\cG^{(n)}$, defined in \eqref{mix-ncs}, we propose
\be \label{seedcs}
\mathcal{\hat I}^{(n)} (k_{12}, k_3;c_s) 
= \frac{1}{k_3} \int_{-\infty}^{0}  \frac{d\eta}{\eta^2} \[
e^{i k_{12}c_s\eta}
 \hat\cG_+^{(n)} (k_3 \eta;c_s)-(-1)^n
e^{-i k_{12}c_s\eta} \hat\cG_-^{(n)} (k_3 \eta;c_s)\]~,
\ee
as the {\it generalized scalar seeds}. When $n=0$ and $c_s=1$, we recover the primary scalar seed \eqref{singleM}. Physically,
$\mathcal{\hat I}^{(n)}$ are associated with the three-point functions with higher derivative quadratic interactions and different sound speeds. 
The relative sign  $(-1)^n$ of the second term in the integrand is introduced such that these correlators are real.
We will discuss how to derive cosmological correlators from these seeds functions in the following sections, while here we look at the analytical form of their shapes.
By definition $\mathcal{\hat I}^{(n)} $ are dimensionless, being functions of
\be \label{ucs}
u\equiv \frac{k_3}{c_s k_{12}} .
\ee
Using \eqref{eqGkmn}, we find the differential equation
\be \label{seedneqcs}
\[ \Delta_{u} +\(\mu^2+ \frac{1}{4}\) \] \mathcal{\hat I}^{(n)} (u; c_s)
= (-i)^{n-1} n!c_s^2\(\frac{u}{1+c_s u}\)^{n+1} .
\ee
Comparing to the equation satisfied by the primary scalar seed, \eqref{seedeq}, both the effects of higher derivative interactions and the sound speed are encoded in the source term, while the left-hand side of the equation remaining unchanged.
Similarly this generalized equation can be solved  by imposing boundary conditions in the soft limit.
We leave the explicit derivation of its solutions and their singularity analysis to Appendix \ref{app:seeds}. Below, we first show how to obtain the answer recursively, and then qualitatively discuss the general solution. Depending on the value of $c_s$, there are two different cases to consider for the recursive relations:

\begin{itemize}
\item {For $c_s=1$,} the inhomogeneous source in \eqref{seedneqcs}, which is a higher-order contact term
$\hat{\mathcal{C}}^{(n)}= \[{u}/({1+  u})\right]^{n+1} $, satisfies
\be \label{Crecur-re}
\Delta_u \hat{\mathcal{C}}^{(n)}  =n(n+1) \hat{\mathcal{C}}^{(n)} -2  (n+1)^2\hat{\mathcal{C}}^{(n+1)}    
~.
\ee 
Using the differential equations of $ \mathcal{\hat I}^{(n)}$ and $ \mathcal{\hat I}^{(n-1)}$, 
 we find 
\be
\mathcal{\hat I}^{(n)}  = -\frac{i}{2n} \[ \mu^2+\(n-\frac{1}{2}\)^2\] \mathcal{\hat I}^{(n-1)}  - \frac{(-i)^{n-1} n!}{2n^2} \hat{\mathcal{C}}^{(n-1)} ~.
\ee
Applying this relation iteratively, the $n$-th order solution can be written as a sum of $\mathcal{\hat I}^{(0)}  $ and contact terms
\be \label{In}
\mathcal{\hat I}^{(n)}   = (-i)^n n! \[s_n \mathcal{\hat I}^{(0)}  +i \sum_{m=0}^{n-1} s_{nm} \hat{\mathcal{C}}^{(n-m-1)} \]    ,
\ee
where the $0$-th order function $\mathcal{\hat I}^{(0)}  $ is the primary scalar seed we have derived in Section \ref{sec:seed}. The coefficients $s_n$ and $s_{nm}$ are
\bea
s_{nm} &=& -\frac{\[\(n-\frac{1}{2}\)^2 + \mu^2\]\[\(n-\frac{3}{2}\)^2 + \mu^2\]...\[\(n+\frac{1}{2}-m\)^2 + \mu^2\]}{2^{m+1}\[n!/(n-m-1)!\]^2}~, 0\leq m< n  \\
s_n &=&  \frac{1}{2^n(n!)^2}
 \[\(n-\frac{1}{2}\)^2 + \mu^2\]\[\(n-\frac{3}{2}\)^2 + \mu^2\]...\[\frac{1}{4} + \mu^2\]~,~~~~~~n>0~.
\eea
The contact terms are rational polynomials of the momenta with $k_T$ poles. They are degenerate with higher-derivative contact interactions of the primordial fluctuations. For cosmological colliders, only the first term in \eqref{In} is relevant, capturing the effects of massive particle production. In this sense, for $c_s=1$ the generalized scalar seeds can be simply reduced to the primary one, and increasing the number of derivatives in the quadratic interaction will not change the shape of the seed function.
\item {For $c_s\neq 1$,} the $n$-th order contact term becomes $\hat{\mathcal{C}}^{(n)}= \[{u}/({1+c_s u})\right]^{n+1} $, which satisfies 
\be \label{Crecur-recs}
\Delta_u \hat{\mathcal{C}}^{(n)}  =n(n+1) \hat{\mathcal{C}}^{(n)} -2 c_s (n+1)^2\hat{\mathcal{C}}^{(n+1)}   +
(n+1)(n+2)(c_s^2-1) \hat{\mathcal{C}}^{(n+2)} 
~.
\ee
Using this and the differential equations \eqref{seedneqcs}, we obtain  
\begin{small}
\be
\mathcal{\hat I}^{(n)}  = \frac{1}{1-c_s^2} \( 2i(n-1)c_s \mathcal{\hat I}^{(n-1)}  
- {\[ \(n-\frac{3}{2}\)^2+ \mu^2 \]} \mathcal{\hat I}^{(n-2)} -(-i)^{n-1}(n-2)!  {c_s^2} \hat{\mathcal{C}}^{(n-2)}  \)~,
\ee
\end{small}where the $n$-th order seed is expressed as a sum of the $(n-1)$-th and $(n-2)$-th order solutions  and the $(n-2)$-th order contact term.
Therefore, to obtain $\mathcal{\hat I}^{(n)}$ by using recursion relations,
two of the generalized seeds are needed as input, $n=0$ and $n=1$. 
\end{itemize}

\begin{figure}[t!]
   \centering
      \includegraphics[height =5.3cm]{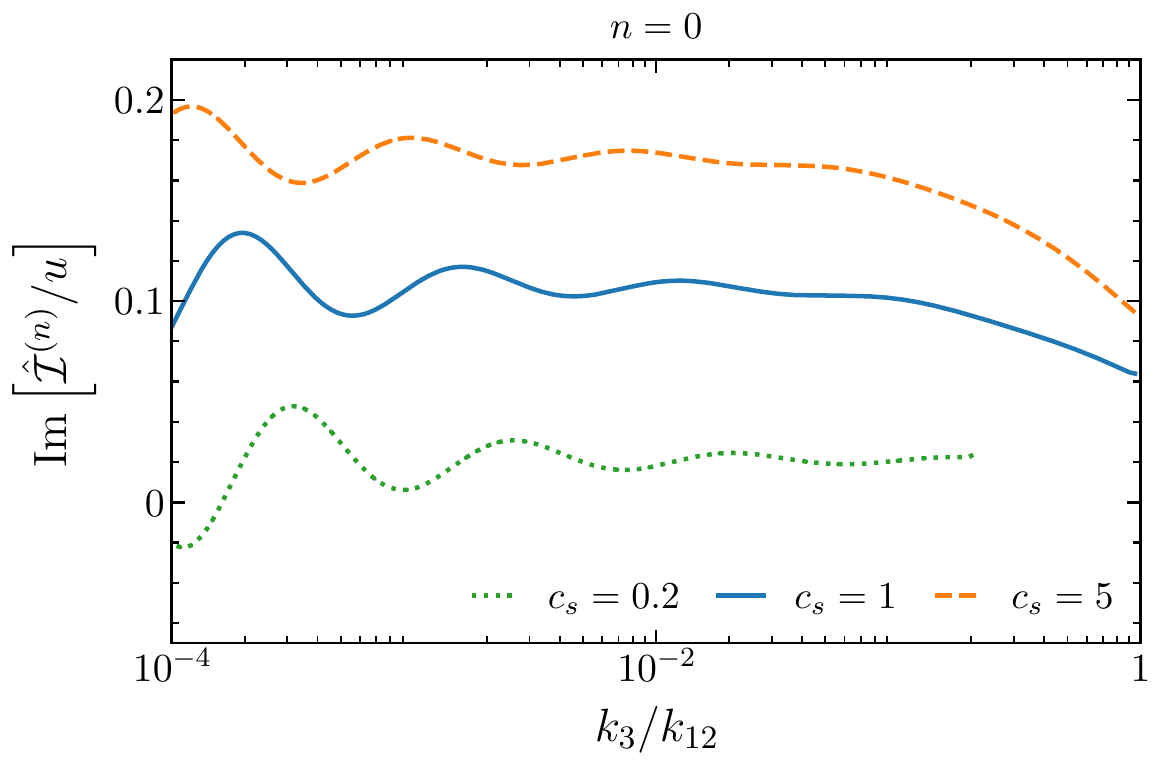}  \hspace{0.1cm}
\includegraphics[height =5.3cm]{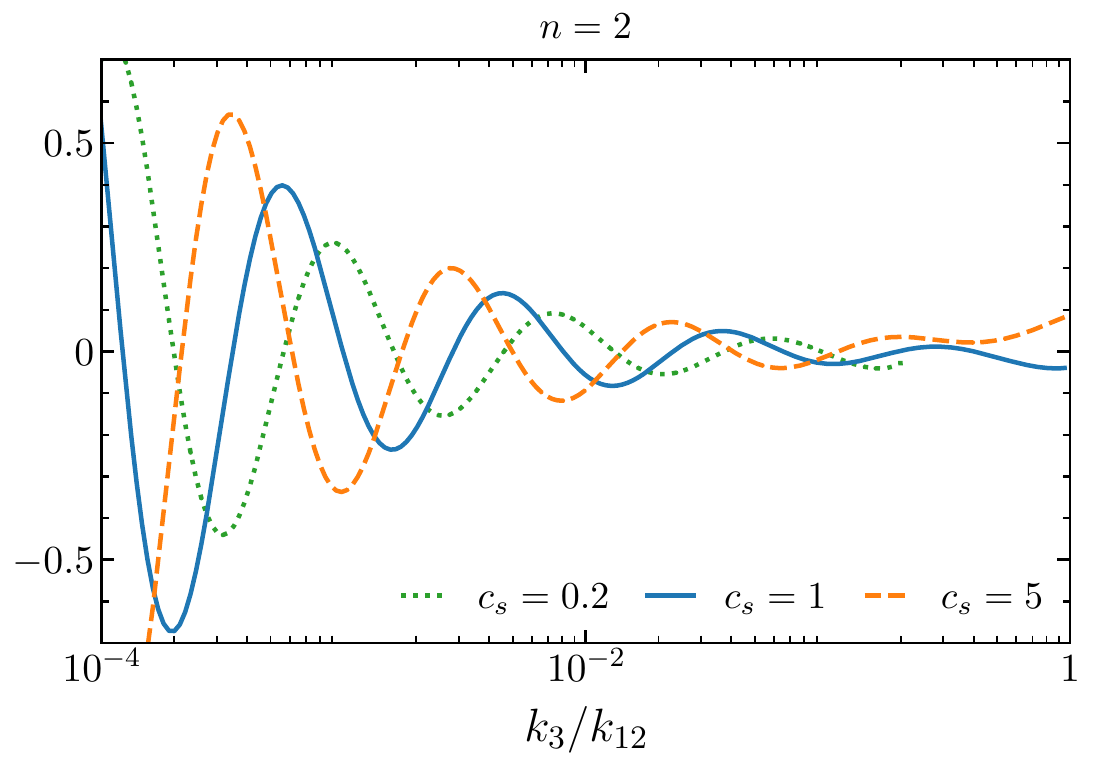} 
      \caption{Analytical solutions of the generalized scalar seeds  for $n=0$ ({\it left}) and $n=2$ ({\it right}), with $\mu=3$ and the sound speed ratio $c_s=0.2,~1,~5$. Since we work on the solutions for $u\in[0,1]$, the plots are stopped at $k_3/k_{12}<c_s$ when $c_s<1$. } 
      \label{fig:seedcs}
\end{figure}

Let us comment on the situation where the sound speed ratio $c_s$ is not 1. 
It is more informative to take a look at the explicit solutions of the differential equation \eqref{seedneqcs} derived in Appendix \ref{app:Insol}. 
Figure \ref{fig:seedcs} shows the generalized scalar seeds with different values of $n$ and $c_s$.
Like the primary scalar seed, the solutions here are expressed in terms of the homogeneous part \eqref{homo-n} and the particular part \eqref{series-ncs}.
The particular solutions are rational series expansions which are analytic around $u=0$. Thus they do not contribute to the oscillatory features of cosmological colliders, although the presence of $c_s$ leads to significant modification on the series coefficients.\footnote{As an example, when $c_s\ll1$,  new non-Gaussianity shapes are expected to arise in the series solutions. This is recently discussed in detail by \cite{Jazayeri:2022kjy}. As in this work we mainly focus on the oscillatory signals of  cosmological colliders, this particular regime is not included in our analysis.}
We are mainly interested in the homogeneous solutions  which are non-analytic at $u\rightarrow0$. 
This non-analyticity in the soft limit corresponds to the productions of massive particles, as we have discussed in the primary scalar seed. 
For $c_s\neq1$,
 in addition to $c_s$-dependent overall prefactors, 
the homogeneous solutions are  affected through the $c_s$ in the definition of $u$ in \eqref{ucs}. 
In particular we see from Figure \ref{fig:seedcs} that,
 the oscillations in the squeezed limit $k_3/k_{12}\ll 1$, which are in terms of $u$, are shifted away from the $c_s=1$ shapes. This feature leads to interesting phenomenology in the cosmological correlator. We discuss it in detail in Section \ref{sec:pheno}.
 
\vskip4pt
The appearance of $c_s$ also affects the singularity structure of the seed functions. 
For general $c_s$, the folded  configuration now corresponds to $u\rightarrow 1/c_s$, while $u\rightarrow -1/c_s$ reflects the limit of vanishing total energies $E_{\rm tot}\equiv c_s (k_{12}+k_3)\rightarrow 0$.
Meanwhile, there is another singularity of $\mathcal{\hat I}^{(n)} $ at $u\rightarrow-1$.
This is the partial energy pole when the sum of energies in the cubic vertex goes to zero $E_L\equiv k_3 + c_s k_{12} \rightarrow 0 $.\footnote{One may expect another partial energy pole at $E_R=k_3+c_sk_3\rightarrow0$ where the energies in the quadratic vertex vanish. This is simply the non-analytic soft limit which we have used as boundary conditions for solving the differential equation \eqref{seedneqcs}.}
When $c_s=1$, the $E_{\rm tot}$- and $E_L$-poles coincide with each other, as we have seen in the primary scalar seed. But in general they are two physical singularities in the exchange bispectra.
The detailed analysis is presented in Appendix \ref{app:Insin}.

\paragraph{Summary} In this section, by incorporating all possible linear mixings between $\phi$ and $\s$, we extended the three-point scalar seed to the most general one $\mathcal{\hat I}^{(n)} $ with arbitrary sound speed and high-derivative quadratic interactions. 
Their analytical expressions are obtained from recursive relations and also explicit solutions of the differential equation \eqref{seedneqcs}.
These generalized seed functions provide building blocks for bootstrapping inflationary bispectra from both scalar and spin exchanges, as we will show in Section \ref{sec:BBWS} and Section \ref{sec:spin} respectively.

\section{Boost-Breaking Weight-Shifting Operators}
\label{sec:BBWS}

We have derived the three-point seed functions that are associated with the bispectra of two conformally coupled scalars and one inflaton. 
In this section, we will apply these results to compute the scalar exchange bispectra of three massless external scalars. We will introduce a set of weight-shifting operators to map the  conformally coupled scalars $\vp$ to the  massless fields $\phi$. In particular, we generalize the  $\vp^2\s$ cubic vertex to the ones of $\phi\phi\s$-type with any boost-breaking interactions.
As the inflaton fluctuations are directly related to the primordial curvature perturbations, these correlators are most relevant for observations.

\vskip4pt
In Section \ref{sec:bbws}, we derive the weight-shifting operators for generic  boost-breaking  cubic interactions.
In Section \ref{sec:scalar}, we apply these operators to generate the phenomenologically interesting bispectra from massive scalar exchanges.

\subsection{From $\Delta=2$ to $\Delta=3$}
\label{sec:bbws}

For two cases of interest, the conformal weights $\Delta$ of the two boundary operators in the three-point correlators are given by: $\Delta=2$, corresponding to
conformally coupled scalars; and $\Delta=3$ which are the ones of massless fields. We perform the analysis one by one.

\paragraph{The $\langle\vp\vp\phi\rangle$ bispectrum}
Let us first explicitly compute the three-point function with two $\Delta=2$ boundary operators.
With the generalized scalar seeds, it becomes straightforward to obtain $\langle\vp\vp\phi\rangle$ from  the simplest cubic coupling $\vp^2\s$. We use the general version of the mixed propagator in \eqref{mix-g2}, and the three-point function in \eqref{ccphi} becomes
\bea \label{ccphi2}
\langle \vp_{{\bf k}_1} \vp_{{\bf k}_2} \phi_{{\bf k}_3} \rangle'  
&=&  \frac{i\eta_0^2}{4k_1k_2c_s^2} I_{\vp\vp\phi} + {\rm perms}~,
\eea
where
\be
 I_{\vp\vp\phi} = \int \frac{d\eta}{\eta^2}\[ e^{ic_sk_{12}\eta }\cG
_+(k_3,\eta) - c.c. \]=  i^{n_T} \frac{H^{n_{T}}}{c_s^3 k_3^2} 
 \[ ({n}_{\partial_t}-1) \mathcal{\hat I}^{(n_T-2)} - i   \mathcal{\hat I}^{(n_T-1)} \]~.
\ee
For the  linear mixing $\dot\phi\s$, this integral is simply given by  $ I_{\vp\vp\phi} =  \({H}/{c_s^3k_3^2}\right) \mathcal{\hat I}^{(0)} $. Thus the analysis of the generalized scalar seeds in Section \ref{sec:gen-seeds} can be directly applied for the $\langle\vp\vp\phi\rangle$ correlator.
It is interesting to consider an extremal case where
 the exchanged field is heavy, $m^2\gg H^2$. Then $\s$ can be integrated out and we expect the correlator becomes the one from the contact interaction $\vp^2\dot\phi$. 
 In our formalism, this bispectrum can be obtained by looking at the differential equation \eqref{seedneqcs} in the $\mu\rightarrow\infty$ limit, where we may drop the differential operator $\Delta_u$. Thus the scalar seed is simply given by $\mathcal{\hat I}^{(0)} \simeq ic_s^2 \mu^{-2} {u}/\(1+c_s u\right)$, and we find
\be
\langle \vp_{\bf k_1} \vp_{\bf k_2} \phi_{\bf k_3} \rangle'  
\sim  \frac{1}{k_1k_2k_3^2k_T}+ {\rm perms}~,
\ee
which matches the bispectrum shape from the contact interaction $\vp^2\dot\phi$ as expected.

\vskip4pt
For the later convenience, here let us also take a look at  cubic interactions with arbitrary time derivatives on the massive field $\s$
\be
\vp^2\partial^{n_\s}_{t}\s = \vp^2(-H\eta\partial_{\eta})^{n_\s}\s +... ~,
\ee
where we have only kept the highest derivatives term when we change to the conformal time. 
These time derivatives on $\s$ lead to modifications on the mixed propagators.  
Since the building block $\hat{\cG}^{(n)}$ is a function of the combination $k\eta$, we are able to trade $\eta$-derivatives on it with $k$-derivatives and get a differential operator 
$(k\partial_k)^{n_\s}$. 
As a result, the mixed propagator in \eqref{mix-g2} is changed to
\be \label{mix-g3}
\cG_\pm  (k , \eta)=(\pm i)^{n_T}   \frac{H^{n_{T} +{n_\s}}}{c_s^3k^3} (-k\partial_k)^{n_\s}\[({n}_{\partial_t}-1)  \hat\cG_\pm^{\(n_{T}-2\)} \mp i   \hat\cG_\pm^{\(n_{T}-1\)} \] + ...
\ee
where the ellipses denote  terms proportional to the free bulk-to-boundary propagators.
Accordingly, the bulk integral $ I_{\vp\vp\phi}$  of the $\langle\vp\vp\phi\rangle$ correlator now becomes
\bea \label{Ivpvpphi}
 I_{\vp\vp\phi} = i^{n_T}  \frac{H^{n_{T} +{n_\s}}}{c_s^3k_3^3} (-k_3\partial_{k_3})^{n_\s} k_3\[ ({n}_{\partial_t}-1) \mathcal{\hat I}^{(n_T-2)} - i   \mathcal{\hat I}^{(n_T-1)} \]~.
\eea
While these $k_3\partial_{k_3}$ operators are not frequently used for scalar exchange, a similar procedure plays an important role for deriving the spin-raising operator in Section \ref{sec:spin}.

\paragraph{The $\langle\phi\phi\phi\rangle$ bispectrum}
Now we consider the three-point function where the three external fields are massless scalars.
This corresponds to the inflaton bispectra which are most relevant for observations. To compute these correlators, one way to proceed is to repeat what we did for the $\langle\vp\vp\phi\rangle$ correlator: introduce a bulk integral $I_{\phi\phi\phi}$ based on the cubic vertex,  derive its differential equation  using the mixed propagator, and solve for its solution with proper boundary conditions.
Although in principle it can be done, this procedure may become rather complicated, since the boost-breaking cubic vertices may take various forms, and for each of them we need to solve an inhomogeneous equation correspondingly.
Here we take a more efficient `bootstrap''  approach by making the use of the scalar seeds. The key insight is that the $\langle\phi\phi\phi\rangle$ bispectra can be 
 generated  by acting on the seeds with various differential operators. This is similar in spirit to the ``weight-shifting'' approach of \cite{Arkani-Hamed:2018kmz, Baumann:2019oyu,Baumann:2020dch}.\footnote{{Similar differential operators that change the weights (masses) of scalar fields were first introduced in the context of the conformal bootstrap \cite{Karateev:2017jgd}.}} Despite the absence of boost symmetry, interestingly, such operators still exist, as shown in \cite{Hillman:2021bnk} for correlators in single field inflation.

\vskip4pt
Here our goal is to systematically derive the $\langle\phi\phi\phi\rangle$ correlator from the scalar seeds.
Concretely, we aim to map the $\vp^2\s$ vertex to a general cubic vertex of  $\phi \phi \s$-type with boost-breaking interactions.
That is to say, 
 the scaling dimension of two boundary operators needs to be shifted from $\Delta=2$ to $\Delta=3$, and at the same time we should take into account the time and spatial derivatives on them.
Let us begin by proposing the generic form of the cubic vertex  as
\be
\partial_i^{n_{s}} \(  \partial_t^{n_1}\phi\partial_t^{n_2}\phi\partial_t^{n_3}\s\)~,
\ee
where $n_{s}$ is the total number of spatial derivatives, $n_1$, $n_2$ and $n_3$ are the numbers of time derivatives for the two massless field $\phi$ and the massive scalar $\s$ respectively.
Notice that the spatial derivatives can act on any field in the vertex, and $n_s$ is an even number for scalar interactions.
In Fourier space, the above cubic vertex leads to (at highest derivatives)
\be
(- {\bf k}_a\cdot {\bf k}_b)^{n_s/2}a^{-\tilde{n}_T-n_3}  \partial^{n_1}_\eta \phi_{k_1} \partial^{n_2}_\eta \phi_{k_2} \partial^{n_3}_\eta \s_{k_3}~,
\ee
where $\tilde{n}_T=n_1+n_2+n_s$ is the total number of derivatives on the inflaton $\phi$, and ${\bf k}_a\cdot {\bf k}_b$ ($a,b=1,2,3$) corresponds to the possible  contractions of momenta. 
Then the three-point function of the massless scalar is given by the following bulk integral
\begin{small}
\bea \label{phi30}
\langle \phi_{\bf k_1} \phi_{\bf k_2} \phi_{\bf k_3} \rangle' = i (- {\bf k}_a\cdot {\bf k}_b)^{n_s/2} \int d\eta a(\eta)^{4-\tilde{n}_T} \[ \partial^{n_1}_\eta K_+(k_1,\eta)\partial^{n_2}_\eta K_+(k_2,\eta)   \cG
_+(k_3,\eta) - c.c. \]+ {\rm perms}~.
\eea
\end{small}Next we need to take care of the time derivatives inside the integral. 
For the mixed propagator, we have absorbed its time derivatives of $a^{-n_3}\partial^{n_3}_\eta\s$ into its expression, and trade them with $k$-derivatives. This leads to a differential operator $k^{n_3}\partial_{k}^{n_3}$ as we discussed around \eqref{mix-g3}.
For the free propagator $K(k,\eta)$ of massless field $\phi$, in general its $n$-th order time derivative takes a simple form
\be \label{dnK}
\partial^{n}_\eta K_\pm(k,\eta) 
= \frac{H^2}{2c_s^3k^3}(\pm ic_sk)^{n}\( 1-n -k\partial_k\)e^{\pm ic_sk\eta}~.
\ee
This is a key observation that helps us convert the $\vp$ propagators into the ones of the massless field $\phi$ through differential operations.
Substituting it into \eqref{phi30}, and taking the $k$-derivatives outside of the bulk integral, we find
\bea \label{phi3}
\langle \phi_{\bf k_1} \phi_{\bf k_2} \phi_{\bf k_3} \rangle' =(-1)^{\tilde{n}_T}\frac{iH^{\tilde{n}_T}}{4c_s^6k_1^2k_2^2}  \mathcal{W}_{12} I_{\vp\vp\phi}(k_{12},k_3)+ {\rm perms}~,
\eea
where $I_{\vp\vp\phi}(k_{12},k_3)$ is the bulk integral associated with the  $\langle\vp\vp\phi\rangle$ correlator in \eqref{Ivpvpphi}.
Meanwhile, we have introduced a dimensionless differential operator 
\begin{eBox}
\be \label{bbws}
\mathcal{W}_{12}\equiv-c_s^{2-n_s}({\bf k}_a\cdot {\bf k}_b)^{n_s/2}k_1^{n_1-1}k_2^{n_2-1}\(1-n_1-k_1\partial_{k_1}\)\(1-n_2-k_2\partial_{k_2}\)\partial_{k_{12}}^{\tilde{n}_T-2}~,
\ee
\end{eBox}
which is the {\it boost-breaking weight-shifting operator}\footnote{We wish to thank Enrico Pajer for helping with its derivation.}.
From the intuition of bulk calculation, this operator exactly maps $\vp\vp$ to $\partial_i^{n_s}\(\partial_t^{n_1}\phi\partial_t^{n_2}\phi\right)$ in time integral of the cubic vertex.
As we will show shortly in the examples, the form of $\mathcal{W}_{12}$ returns to the one in de Sitter bootstrap when we consider the dS-invariant cubic interaction. In general, this operator is capable of generating all the boost-breaking $\phi\phi\s$-type vertices from the $\vp^2\s$ one.

\vskip4pt
The result in \eqref{phi3} directly works on the boundary correlators. This  general expression  provides  all the possible inflaton bispectra from the exchange of one massive scalar field. 
Starting with the generalized scalar seeds, we can map them to arbitrary boost-breaking cubic interactions by performing differential operators $k_3\partial_{k_3}$  (for higher derivatives on $\s$) and $\mathcal{W}_{12}$ (for higher derivatives on $\phi$). 
In the following we shall consider simple but observationally relevant interactions, and then 
 apply the operator \eqref{bbws} to  bootstrap examples of the inflaton bispectra from massive scalar exchange.

\subsection{Scalar Exchange Bispectra}
\label{sec:scalar}

For scalar exchange, we  focus on the simplest quadratic interaction $\dot\phi\s$, which gives the mixed propagator $\cG_{\pm}(k,\eta;c_s)= H/(c_s^3k^3)\hat\cG^{(0)}_{\pm}(k\eta;c_s)$.
The most relevant cubic interactions are the ones with lowest derivatives
\be
\dot\phi^2 \s~, ~~~~~~ (\partial_i\phi)^2\s~.
\ee
When the de Sitter boosts are not broken, they are restricted to take the particular combination $(\partial_\mu\phi)^2\s$. This cubic coupling automatically induces a small linear mixing, which is slow-roll suppressed.
In boost-breaking theories, these two cubic vertices can appear independently with large interactions, as we have shown in the EFT analysis around \eqref{Lmix0}.
For these low-derivative interactions, the exchange three-point function can be expressed as
\bea \label{phi3b}
\langle \phi_{\bf k_1} \phi_{\bf k_2} \phi_{\bf k_3} \rangle' =  \frac{iH^3}{4 c_s^9 k_1^2k_2^2k_3^2} \mathcal{W}_{12}
\mathcal{\hat I}^{(0)}+ {\rm perms}~.
\eea
Then the bispectra shapes are fully determined once we specify the weight-shifting operators based on cubic vertices. Explicitly, they are given by:
\begin{itemize}
\item The $\dot\phi^2\sigma$  vertex. The weight-shifting operator
 is given by
 \be
\mathcal{W}^{\dot\phi^2\sigma}_{12} = -c_s^2 k_1k_2 \partial_{k_{12}}^2~.
\ee
In the squeezed limit $k_3\rightarrow0$ and $k_1=k_2$,  the bispectrum \eqref{phi3b} becomes
\be \label{sq3pt-0-bb}
\lim_{k_3\rightarrow 0}\langle \phi_{\bf k_1} \phi_{\bf k_2} \phi_{\bf k_3} \rangle' =  \frac{i A_0H^3}{4c_s^4 k_1^3k_3^3} 
\[ \frac{\Gamma\(\frac{5}{2}+i\mu\)}{\Gamma\(1+i\mu\)}(1+i \sinh\pi\mu )\(\frac{k_3}{4c_s k_1}\)^{\frac{3}{2}+i\mu}-c.c\]~,
\ee
where $A_0 = {\sqrt{\pi}\Xi_0(\mu, c_s)}/{ \sinh\pi\mu}$ with the $\Xi_0$ function defined in \eqref{Xixi}.

\item The $(\partial_i\phi)^2\sigma$  vertex. It leads to a different  operator
\be
\mathcal{W}^{(\partial_i\phi)^2\sigma}_{12} = -\frac{1}{2k_1k_2} (k_3^2 - k_1^2 - k_2^2) {\(1-k_1\partial_{k_1}\)\(1-k_2\partial_{k_2}\)}~,
\ee
where we have used the momentum conservation to rewrite ${\bf k}_1\cdot {\bf k}_2 = (k_3^2 - k_1^2 - k_2^2)/2$.
For this bispectrum, the squeezed limit is 
\bea \label{sq3pt-0-bb2}
\lim_{k_3\rightarrow 0}\langle \phi_{\bf k_1} \phi_{\bf k_2} \phi_{\bf k_3} \rangle' &=&  \frac{-iA_0H^3}{4c_s^6 k_1^3k_3^3} 
\[ \frac{\Gamma\(\frac{1}{2}+i\mu\)}{\Gamma\(1+i\mu\)}\(\frac{3}{2}+i\mu \)\(\frac{9}{2}+i\mu \) \right.\nn\\
&& ~~~~~~~~~~~~~~\left. \times(1+i \sinh\pi\mu )
 \(\frac{k_3}{4c_s k_1}\)^{\frac{3}{2}+i\mu}-c.c\]~.
\eea

\item 
As a nontrivial check, we also reconstruct the bispectrum from de Sitter invariant interaction $(\partial_\mu\phi)^2\sigma$ from the two results above. Setting $c_s=1$, we get
 the dS-invariant weight-shifting operator 
\be \label{ws-dS}
\mathcal{W}_{12}^{\rm dS} = - \mathcal{W}^{\dot\phi^2\sigma}_{12}  + \mathcal{W}^{(\partial_i\phi)^2\sigma}_{12}= \frac{1}{2} \(k_{12}^2-k_3^2\) \partial^2_{k_{12}} - \frac{1}{2 k_1k_2}\( k_3^2 - k_1^2 -k_2^2 \) \(1-k_{12}\partial_{k_{12}}\)~,
\ee
which reproduces the one in the de Sitter bootstrap \cite{Arkani-Hamed:2018kmz}.
We also check the soft limit of this bispectrum
\be \label{sq3pt-0-dS}
\lim_{k_3\rightarrow 0}\langle \phi_{\bf k_1} \phi_{\bf k_2} \phi_{\bf k_3} \rangle' =  \frac{iA_0H^3}{4k_1^3k_3^3} 
\[ \frac{\Gamma\(\frac{7}{2}+i\mu\)}{\Gamma\(1+i\mu\)}\frac{(1+i \sinh\pi\mu )}{\frac{1}{2}+i\mu}\(\frac{k_3}{4 k_1}\)^{\frac{3}{2}+i\mu}-c.c\]~,
\ee
and find that it  is in agreement with (6.130) in \cite{Arkani-Hamed:2015bza}.
\end{itemize} 

Through these simple examples, we find the boost-breaking interactions generate new bispectrum shapes for the cosmological collider physics, which differ from the one with all the de Sitter symmetries. Furthermore, their sizes are not supposed to be slow-roll suppressed.
We leave the detailed discussion on the phenomenological implications in Section \ref{sec:pheno}.

\vskip4pt
Finally, let us look at the bispectrum shapes that arise
from integrating out a heavy scalar with $m\gg H$.
The boost-breaking weight-shifting operators can also help us derive these contact three-point correlators with any number of derivatives.
Since in this situation the solutions of the generalized scalar seeds are well approximated by the contact term $\mathcal{\hat I}^{(n)}\sim u^{n+1}/(1+c_s u)^{n+1}$, schematically the inflaton bispectrum can be written as
\be \label{scalar-contact}
\langle \phi_{\bf k_1} \phi_{\bf k_2} \phi_{\bf k_3} \rangle' 
\supset  \frac{1}{k_1^2 k_2^2 k_3^3} \mathcal{W}_{12} \(k_3 \partial_{k_3}\)^{m}
\[k_3\(\frac{k_3}{k_T} \)^{n+1}\] + {\rm perms}~.
\ee
These shape functions are rational polynomials of the absolute values of the momenta. They can be interpreted as coming from higher-derivative inflaton self-interactions. 
A complete set of boostless contact bispectra from single-field inflation have been derived from symmetries and locality constraints in \cite{Pajer:2020wxk, Jazayeri:2021fvk, Bonifacio:2021azc}, while our approach \eqref{scalar-contact} provides a consistency check for the computation of these shapes.
For illustration, we take the lowest-derivative vertices again
as examples. 
Consider the $\dot\phi\s$ mixing, and the scalar seed is given by $\mathcal{\hat I}^{(0)}\sim u/(1+c_s u)$.
Then the cubic vertex $\dot\phi^2\s$ leads to the following shape of the three-point function
\be
\langle \phi_{\bf k_1} \phi_{\bf k_2} \phi_{\bf k_3} \rangle' \sim \frac{1}{k_1k_2k_3^2}\partial_{k_{12}}^2
\(\frac{k_3}{k_T} \)+{\rm perms}  \sim \frac{1}{k_1k_2k_3k_T^3}~,
\ee
which is the one from the contact interaction $\dot\phi^3$.
Similarly integrating out the $\s$ field in the $(\partial_i\phi)^2\s$ vertex leads to the bispectrum from $(\partial_i\phi)^2\dot\phi$
\bea
\langle \phi_{\bf k_1} \phi_{\bf k_2} \phi_{\bf k_3} \rangle' &\sim &  \frac{ k_3^2 - k_1^2 - k_2^2 }{ k_1^3k_2^3k_3^2} \(1-k_1\partial_{k_1}\)\(1-k_2\partial_{k_2}\) \(\frac{k_3}{k_T} \) + {\rm perms} \nn\\
&=&  \(k_3^2 - k_1^2 - k_2^2\)  \frac{2k_1k_2+k_{12}k_T+k_T^2}{k_T^3k_1^3k_2^3k_3} + {\rm perms}~.
\eea
Both of the above bispectra have the equilateral type scaling in the soft limit, unlike the non-analytical behaviour of massive field exchange.
If we consider higher derivative vertices, they correspond to more derivatives in the weight-shifting operators and/or higher order seed functions, both of which lead to higher powers of $k_T$ in the denominator of the shape function. This is in agreement with the analysis in \cite{Pajer:2020wxk, Jazayeri:2021fvk, Bonifacio:2021azc}.

\section{Exchange of Spinning Particles}
\label{sec:spin}

Massive spinning particles leave unique imprints in primordial non-Gaussianity. In particular, they modulate the bispectrum with a spin-dependent envelope.
In this section, we extend to compute the boostless bispectra from the single exchange of spinning particles during inflation.
We make the use of the generalized scalar seeds in Section \ref{sec:gen-seeds}, and map them to spin-exchange correlators by ``spin-raising" operators.

\vskip4pt
In Section \ref{sec:spin1} we discuss the exchange of spin-1 fields in detail, using it as a case study for the general strategy. 
In Section \ref{sec:mix-s}, we introduce the basics of free and mixed propagators of fields with arbitrary spins during inflation, while the free theory of spinning fields in de Sitter space is summarized in Appendix \ref{app:spin}.
In Section \ref{sec:spin-s} we 
derive the three-point functions from spinning exchanges with generic boost-breaking interactions.

\subsection{Spin-1 Exchange}
\label{sec:spin1}

In this section, we describe in detail the derivation of the bispectrum from spin-1 exchange. We introduce the necessary ``spin-raising" operator to obtain it from scalar exchange seeds. It is helpful to study this simplest case to gain insight about how to bootstrap the generic spinning exchange bispectra. However, from a phenomenological perspective, this case is not the most interesting one, as the signals in the squeezed limit from odd spin exchange is more suppressed than the one with even spins.

\subsubsection{Spin-1 Propagators}

First, let us derive the mixed propagators with spin-1 fields. We will focus on the longitudinal mode of the massive spin-1 particle (the only component that contributes to the exchange diagram) and establish its connection with the massive scalars. 

\paragraph{Free Theory in de Sitter}

{The notion of spin  is less unambiguous in the EFT of inflation. With the full dS isometries, we can have a dS-invariant description for the spinning fields, which leads to the EFT in \cite{Lee:2016vti}. While the dS boosts are broken,  it becomes possible to construct another type of  EFT \cite{Bordin:2018pca}, where more general theories of spinning fields are allowed  but it remains unclear about how to embed them in a UV-complete theory. Meanwhile, for the interest of this work, we notice that only the helicity-0 longitudinal mode will contribute to  the scalar bispectra of cosmological colliders. As long as this single component of the spinning field is concerned, there is no much difference in these two EFT approaches. Thus in this paper  we shall take the first approach with dS isometries, where the UV completion is much better understood.}

For a  massive spin-1 field $\s_\mu$ in de Sitter space, its quadratic action is given by
\be \label{actions1}
S_2 = \int d^4x \sqrt{-g} \[ - \frac{1}{2}\nabla_\mu\s_\nu \nabla^\mu\s^\nu +  \frac{1}{2}(\nabla^\mu\s_\mu)^2- \frac{1}{2}(m^2+3H^2)\s^\mu\s_\mu\]~,
\ee
which is equivalent to the Proca action up to integration by parts. 
Here we have chosen the mass definition such that $m$ becomes the mass of $\s_\mu$ in the flat space limit, and we can check that the action becomes gauge invariant when $m\rightarrow0$.
From this action, the equation of motion of the $\s_\mu$ field is
\be \label{eoms1}
\[\nabla^\nu\nabla_\nu- \(m^2+3H^2\)\] \s_\mu = 0~,
\ee
and we also find the transverse condition
\be \label{s1constr}
\nabla^\mu\s_\mu = 0~.
\ee
The spinning field  $\s_\mu$ may also have a nontrivial sound speed $c_\s$, even though it 
appears in dS-invariant forms in the above expressions. 
Like in the massive scalar case, in the free theory we can always rescale this sound speed into the spatial coordinates $c_\s\partial_i\rightarrow\partial_i$ (or, equivalently $c_\s k\rightarrow k$ in the Fourier space), such that it disappears in the final expression.
As long as all the components of $\s_\mu$ have the same sound speed, this rescaling can lead us back to \eqref{actions1} -- \eqref{s1constr}. 
 The spinning fields with different sound speeds for each component were discussed in \cite{Bordin:2018pca}, {but since only one component makes contribution to the final scalar three-point function, the theories with multiple sound  speeds  do not lead to additional bispectrum shapes. Therefore for convenience, our strategy is to focus on the dS-invariant theories but allow one uniform sound speed for the spinning field.}

\vskip4pt
To discuss the mode functions, it is more convenient to expand the spinning fields into their helicity basis.
For spin-1 fields, this decomposition becomes
\be
\s_\mu=\sum_{\lambda=-1}^1 \s_\mu^{(\lambda)}~,
\ee
where $\lambda=0$ gives the longitudinal ($l$) mode and the transverse ($t$) ones correspond to $\lambda=\pm1$. The transverse modes have only the spatial components $\s^{(\pm1)}_i= \s_t^{\pm1}\epsilon^{(\pm1)}_i$, with the polarization vectors satisfying $k_i\epsilon^{(\pm1)}_i({\bf k})=0$.
The temporal ($T$) and spatial ($S$) components of the longitudinal mode with $\lambda=0$ can be further expressed as 
\be
\sigma_\eta^{(0)}= \sigma^T_l~, ~~~~~~~~ \sigma_i^{(0)}= \sigma^S_l \epsilon_i^{(0)}~,
\ee
with the longitudinal polarization vector $\epsilon_i^{(0)}({\bf k})=\hat k_i$.
As we will show very soon, only the longitudinal mode contributes to the spin-1 mixed propagator and thus leads to nonzero exchange bispectra. In the following we will focus on the $\lambda=0$ modes, and drop the upper index $^{(\lambda)}$ in the mode functions and polarization vector. 

\paragraph{Longitudinal modes} For future convenience, we introduce the new notation for $\lambda=0$ longitudinal modes
\be
\sigma^{(0)}_1 = \sigma^T_l~, ~~~~~~~~ \sigma^{(1)}_1 = \sigma^S_l~.
\ee
For a general expression $\sigma^{(n)}_s$, the lower index represents the spin\footnote{Note that for the spinning mode functions we do not use the lower index to represent the momentum $k$, which differs from the notation of scalar fields.} and the upper index $n\leq s$ is used for the number of spatial polarization directions.
For the Fourier mode $\s_1^{(0)}$, the equation \eqref{eoms1} becomes
\be
\(\mathcal{O}_\eta +\mu_1^2  + \frac{9}{4} \)\sigma^{(0)}_0 = 0~,~~~~{\rm with}~~\mu_1=\sqrt{\frac{m^2}{H^2}-\frac{1}{4}}
\ee
which is the same as the one of massive scalars in \eqref{sigmaeom}.
Imposing the Bunch-Davies initial condition and the correct normalization \cite{Lee:2016vti}, we obtain
\bea
 \sigma_{1}^{(0)}(k,\eta) =   \frac{H}{2m}\sqrt{\frac{\pi}{k}}e^{i\pi/4}e^{-\pi\mu_1/2}
 (-k \eta)^{3/2}H^{(1)}_{i\mu_1}(-k\eta) ~,
\eea
which is related to the scalar mode function $\sigma_k(\eta)$ in \eqref{sigmak} by 
\be
 \sigma_{1}^{(0)}(k,\eta) =  k N_1  \sigma_k(\eta)~,~~~~~~{\rm with}~~N_1=\frac{i}{m} ~.
\ee
The equation and mode function solution of the $\s_1^{(1)}$ mode are more complicated, but we notice that the two longitudinal modes are related through the constraint \eqref{s1constr}
\be \label{s1-s0}
\sigma^{(1)}_1  = -i U^{(1)}_\eta \sigma^{(0)}_1~,~~~~~~ {\rm with}~ ~U^{(1)}_{\eta} \equiv \frac{1}{k} \(\partial_\eta-\frac{2}{\eta}\) .
\ee
Therefore we are able to connect the $\lambda=0$ longitudinal mode of $\s_\mu$ with the solution of the massive scalar field discussed in Section \ref{sec:free}.
As we will show below, this is the key observation that allows us to map the results of mixed propagators and seed functions with massive scalars to the spinning case.
Specifically, the bulk-to-bulk propagators of the $\s_\eta=\sigma^{(0)}_1$ mode are simply given by
\be
G^{(0)}_{\pm\pm} (k,\eta, \eta')  = \frac{k^2}{m^2} G^\sigma_{\pm\pm} (k,\eta, \eta')~, ~~~~ G^{(0)}_{\pm\mp} (k,\eta, \eta')  = \frac{k^2}{m^2} G^\sigma_{\pm\mp} (k,\eta, \eta')~.
\ee
For the $\s_i=\sigma^{(1)}_1 \epsilon_i$ mode, its propagators are a bit more complicated, but can also be expressed in terms of $G^\s$ as
\bea \label{s1b2b}
G_{ij,\pm\mp}^{(1)}(k,\eta, \eta') &=& 
|N_1|^2 k^2 U^{(1)}_{\eta} U^{(1)}_{\eta'} G_{\pm\mp}^\sigma(k,\eta, \eta')\epsilon_i\epsilon_j~, \nn\\
G_{ij,\pm\pm}^{(1)}(k,\eta, \eta') &=& 
 |N_1|^2 k^2 U^{(1)}_{\eta} U^{(1)}_{\eta'} G_{\pm\pm}^\sigma(k,\eta, \eta') \epsilon_i\epsilon_j
 \mp i\eta^2 \delta(\eta-\eta')\frac{H^2}{m^2}\epsilon_i\epsilon_j ~.
\eea
The $\delta$-function term in $G_{ij,\pm\pm}^{(1)}$ is to cancel the ones generated when the time derivatives hit the $\Theta$-functions in the bulk-to-bulk propagator.

\paragraph{Spin-1 Mixed Propagator}
Next, we consider the linear mixing between the massive spin-1 field and the inflaton. At leading order in derivatives, there are two quadratic interactions: $\partial_\eta\phi\s_\eta$ and $\partial_i\phi\s_i$. Since they are related from the constraint equation in \eqref{s1constr}, we can mainly focus our analysis on the $\partial_i\phi\s_i$ mixing \cite{Lee:2016vti}.
From this quadratic vertex, we find the contraction from the transverse mode $k_i\epsilon_i^{(\pm1)}$ vanishes, and thus only the spatial component of the longitudinal mode $\s_i=\sigma^{(1)}_1 \epsilon_i$ gives the nonzero contribution to 
the spin-1 mixed propagator.
Explicitly, we introduce the bulk-to-boundary propagator from $\sigma_i$ to $\phi$ as
\bea
\cG^{(1)}_{i,~\pm} (k, \eta) &=&\pm i \int_{-\infty}^{0} d\eta' a(\eta')^2  \[ G_{ij,~\pm\pm}^{(1)}(k,\eta, \eta') (\mp ik_j)  K_{\pm}(c_s k, \eta') \right.\nn\\
&& ~~~~~~~~~~~~~~~~~~~~~~~~ \left.- G_{ij,~\pm\mp}^{(1)}(k,\eta, \eta') (\pm ik_j)  K_{\mp}(c_s k, \eta')\] 
\eea
Using \eqref{s1b2b}, $\cG^{(1)}_{i,~\pm} $ can be expressed as
\bea \label{s1mix0}
\cG^{(1)}_{i,~\pm} &=& - \epsilon_i |N_1|^2 k^2
  U^{(1)}_{\eta} \int_{-\infty}^{0} d\eta' a(\eta')^2    \[ G_{\pm\pm}^\s    \partial_{\eta'}K_{\pm} 
 +G_{\pm\mp}^\s   \partial_{\eta'} K_{\mp} \] \mp  \frac{ik}{m^2} \epsilon_i K_{\pm}(k, \eta)~,
\eea
where we have applied integration by parts in $\eta'$ to reduce the $ U^{(1)}_{\eta'}$ operator in the integrand. This expression can be written in terms of the scalar mixed propagators
\bea \label{s1mix}
\cG^{(1)}_{i,~\pm} (k, \eta) =
\pm i \epsilon_i \frac{H^2|N_1|^2}{c_s^3k^3\eta} 
\mathcal{U}^{(1)}_{k} \hat\cG^{(1)}_\pm (k\eta;c_s) \mp  \frac{ik}{m^2} \epsilon_i  K_{\pm}(c_s k, \eta)
\eea
where $\hat\cG^{(1)}_\pm $ is the dimensionless building block of the mixed propagator in \eqref{mix-ncs} with $n=1$.
For the operator $\mathcal{U}^{(1)}_{k}$, we have used the fact that $\hat\cG^{(1)}$ is a function of the combination $k\eta$, such that we are able to trade $\eta$-derivatives with $k$-derivatives. This gives us  
\be
U^{(1)}_{\eta} = \frac{1}{k} \(\partial_\eta-\frac{2}{\eta}\)  ~~~ \rightarrow ~~~\frac{1}{k\eta} \mathcal{U}^{(1)}_{k}~, ~~~~{\rm with} ~ \mathcal{U}^{(1)}_{k} \equiv k\(\partial_k-\frac{2}{k}\) ~,
\ee
which is the spin-raising operator for $s=1$. Using this differential operator, we can derive the spin-1 mixed propagator  from the scalar one.
Notice that the piece which comes from the $\delta$-function term in \eqref{s1b2b} is the free  propagator $K_{\pm}$ of a massless scalar in \eqref{s1mix}. This term leads to a standard contact interaction in the exchange diagram, whose contribution to the bispectrum is the same with the ones from single field inflation. In the following, we shall drop this contribution, and focus on the effect of the first term of the spin-1 mixed propagator in \eqref{s1mix}.

\vskip4pt
It is straightforward to extend the analysis to the quadratic interactions with higher derivatives. 
We can move all the derivatives to the scalar field $\phi$ via integration by parts, then we get interactions
like $\partial_t^n\partial_i\phi\s_i$. These additional derivatives on $\phi$ change the form of the integrand in \eqref{s1mix0}. As we have seen in Section \ref{sec:moremix}, the integral can always be written as a linear combination of the building blocks $\hat{\cG}^{(n)}$, while the spin-raising operator remains unaffected.

\subsubsection{Spin-1 Exchange Bispectra}

With the spin-1 mixed propagator, we now   bootstrap the three-point functions from the single exchange diagram.
We start with  the bispectrum of two conformally coupled scalars $\vp$ with a massless scalar $\phi$.
As one leg of the cubic vertex should be attached to the mixed propagator with the $\s_i$ longitudinal mode,
the lowest derivative interaction is given by $\vp\partial_i\vp\s_i$. 
As a result, the bispectrum is given by
\be 
\langle \vp_{{\bf k}_1}\vp_{{\bf k}_2} \phi_{{\bf k}_3} \rangle' = i   \int_{-\infty}^0 d\eta a^2 \[  K_+^\vp(k_1,\eta)(-ik_2^i) K_+^\vp(k_2,\eta) \cG^{(1)}_{i,~+} (k_3, \eta) 
-c.c.\]+{\rm perm.} 
\ee
By  using the mixed propagator \eqref{s1mix} from the $\partial_i\phi\s_i$ mixing, we can rewrite the bulk integral above in the form
\bea \label{conbi-s1}
\langle \vp_{{\bf k}_1}\vp_{{\bf k}_2} \phi_{{\bf k}_3} \rangle' &=& i \frac{A^\vp_1c_s}{k_1k_3^3} (\hat{\bf k}_2\cdot \hat{\bf k}_3) \mathcal{U}^{(1)}_{k_3} \int_{-\infty}^0 \frac{d\eta}{\eta} \[e^{ic_s k_{12}\eta} \hat\cG^{(1)}_+ (k_3\eta)- c.c.\] +{\rm perm.}\nn\\
&=& \frac{A^\vp_1}{k_1k_3^3} (\hat{\bf k}_2\cdot \hat{\bf k}_3)~ \mathcal{U}^{(1)}_{k_3}~ \[k_3\partial_{k_{12}} \mathcal{\hat I}^{(1)} \(u;c_s\)\]+{\rm perm.}~,
\eea
where $A^\vp_1 = H^4|N_1|^2\eta_0^2/4c_s^6$ and $\mathcal{\hat I}^{(1)} $ is the generalized scalar seeds in \eqref{seedcs} with $n=1$.
Therefore, we derive the spin-1 exchange bispectrum from the seed functions by using the spin-raising operator $\mathcal{U}^{(1)}_{k_3}$.
This example demonstrates the generic structure of the three-point correlator from spinning exchange in our approach. In addition to the spin-raising operator and the scalar seed, the $\hat{\bf k}_2\cdot \hat{\bf k}_3 =\cos\theta$ factor can be written as the Legendre polynomial $P_1(\cos\theta)$, which encodes the angular dependent signal of the spinning particle; the $k_3\partial_{k_{12}}$ operator is associated with the form of the cubic interaction.

\vskip4pt
Next, we consider the inflaton bispectrum with three massless scalars. 
From the EFT of inflation with spinning fields \cite{Lee:2016vti}, the lowest derivative cubic interaction with two inflatons and one massive spin-1 field is given by $\dot\phi\partial_i\phi\s_i$, as shown in \eqref{Lmix1}. This vertex only arises in boost-breaking theories.\footnote{In theories with full de Sitter isometries, the cubic interaction appears as $(\nabla_\mu\phi\nabla^\nu\nabla^\mu\phi-\nabla^\nu\nabla_\mu\phi\nabla^\mu\phi)\sigma_\nu$, which leads to vanishing bispectra after permutation \cite{Arkani-Hamed:2015bza, Arkani-Hamed:2018kmz}.}
Using the mixed propagator $\cG^{(1)}_{i,~\pm} (k, \eta)$, we find 
\bea \label{s1bispec}
\langle \phi_{{\bf k}_1}\phi_{{\bf k}_2} \phi_{{\bf k}_3} \rangle' & =& i   \int_{-\infty}^0  d\eta a(\eta) \[ \partial_\eta K_+ (k_1,\eta)(-ik_2^i)  K_+ (k_2,\eta) \cG^{(1)}_{i,~+} (k_3, \eta) - c.c.
\] +{\rm perm.}\nn\\
&=&  \frac{A_1}{k_1^2k_2^2k_3^3} (\hat{\bf k}_2\cdot \hat{\bf k}_3)~\mathcal{U}^{(1)}_{k_3} ~\mathcal{W}^{(1)} \[ k_3\mathcal{\hat I}^{(1)}(u) \]+{\rm perm.}~,
\eea
where the coefficient $A_1=-{H^5|N_1|^2}/{4c_s^8}$, and we have used the following weight-shifting operator  
\be
\mathcal{W}_{12}  \equiv  (1- k_2  \partial_{k_{2}})k_1\partial_{k_{1}}~.
\ee
This result provides the analytical shape of the inflationary bispectrum from the boostless spin-1 exchange. We will discuss its phenomenological implications in Section \ref{sec:pheno}. 
Here let us simply look at its behaviour in the squeezed limit\footnote{There is a minor difference with the results in \cite{Lee:2016vti}. In Eq. (C.23)  of \cite{Lee:2016vti} and the  discussions below, for odd spins, the squeezed limit has $i\cosh\pi\mu_1$ instead of $(1+i\sinh\pi\mu_1)$. We expect this is because the relative signs of the sum in Eq.(C.2) of \cite{Lee:2016vti} should become different for odd spin exchanges.} 
\bea
\lim_{k_3\rightarrow0}\langle \phi_{{\bf k}_1}\phi_{{\bf k}_2} \phi_{{\bf k}_3} \rangle' &=& 
- \frac{H^3\sqrt{\pi}}{32c_s^5k_1^3k_3^3}\frac{\Xi_1(\mu_1,c_s)}{\cosh\pi\mu_1}
(\hat{\bf k}_1\cdot \hat{\bf k}_3+\hat{\bf k}_2\cdot \hat{\bf k}_3)
\nn\\&& \times\[(7+2i\mu_1)(1+i\sinh\pi\mu_1)\frac{\Gamma(-i\mu_1)}{\Gamma(\frac{1}{2}-i\mu_1)} \(\frac{k_3}{4c_sk_1}\)^{\frac{3}{2}+i\mu_1}+c.c.\]~,
\eea
where the function $\Xi_1$ is defined in \eqref{Xixi}.
Since from momentum conservation we have
\be
\lim_{k_3\rightarrow0}(\hat{\bf k}_1\cdot \hat{\bf k}_3+\hat{\bf k}_2\cdot \hat{\bf k}_3) = -\frac{k_3}{k_1}\[1-(\hat{\bf k}_1\cdot \hat{\bf k}_3)^2\] +\mathcal{O}(k_3^2/k_1^2)~,
\ee
the squeezed limit scales as $(k_3/k_1)^{5/2}$, which is more suppressed than the equilateral shape of the bispectrum.
This behaviour is generic for odd spin exchange, but not for even spins, as we show below.

\vskip4pt
To summarize, we have shown from this simplest example of spinning exchange that the spinning propagators are related to the scalar ones by applying differential operators. This is because only the helicity-$0$ longitudinal mode of the massive field contributes to the linear mixing. Therefore, spin-exchange bispectra can be mapped from scalar-exchange correlators in a simple way. In the following we will follow the same strategy to bootstrap the bispectra of higher spin exchange.

\subsection{Spinning Propagators}
\label{sec:mix-s}
Now we consider correlators involving bosonic particles $\s_{\mu_1...\mu_s}$ of higher spin $s$ and arbitrary mass $m$. 
We present a detailed review of its free theory in de Sitter space in  Appendix \ref{app:spin} (see also Appendix A of \cite{Lee:2016vti}). Again, like in the analysis of the massive scalar and spin-1 fields , we have absorbed the sound speed by rescaling $c_\s k \rightarrow k$.

The first step is to derive the spinning mixed propagators. 
For this purpose, we discuss only the longitudinal mode $\lambda=0$ and consider its projection on the spatial slicing $\s_{i_1...i_n\eta...\eta}$. 
In the polarization basis, it is given by
\be
\sigma_{i_1...i_n\eta...\eta} = \sigma^{(n)}_s \epsilon_{i_1...i_n} ~.
\ee
The polarization tensor  $\epsilon_{i_1...i_n}$  satisfies 
\be 
\hat{k}_{i_1}...\hat{k}_{i_n}\epsilon_{i_1...i_n}({\bf k})=1~,~~~~~~
\hat{q}_{i_1}...\hat{q}_{i_n}\epsilon_{i_1...i_n}({\bf k})=P_n(\hat{\bf q}\cdot \hat{\bf k})~,
\ee
where $P_n(\hat{\bf q}\cdot \hat{\bf k})$ is the Legendre polynomial of order $n$.
We focus on the mode functions $\sigma^{(n)}_s$. Like in the spin-1 case, we use the upper index $n\leq s$ to denote the number of polarization directions, and the lower index for the spin.
In particular, the $n=0$ mode $\sigma_s^{(0)}$ 
can be expressed in terms of the scalar mode function \eqref{sigmak}, with $\mu=\mu_s$
\bea \label{s0-scalar}
 \sigma_s^{(0)}(k,\eta) 
 =   N_s k^s \sigma_k(\eta)~,
\eea
where $N_s$ is a normalization constant, defined in \eqref{Ns}, and
\be
\mu_s=\sqrt{\frac{m^2}{H^2}-\(s-\frac{1}{2}\)^2}~.
\ee
For cosmological collider bispectra, we are interested in the $n=s$ mode, $\sigma_s^{(s)}$, whose equation of motion  becomes rather complicated. However, its solution can be obtained iteratively from the transverse condition $\nabla^{\mu_1}\s_{\mu_1...\mu_s}=0$, which in Fourier space becomes 
\be \label{ss-s0}
\sigma_s^{(s)} = 
\(-i\)^s U^{(s)}_\eta \sigma_s^{(0)}  .
\ee
Here we have introduced the differential operator 
\be
U^{(s)}_\eta \equiv \sum_{m=0}^{s} \frac{a^{(s)}_m}{k^m}  \(\partial_\eta -\frac{2}{\eta} \)^m ,
\ee
with $a^{(s)}_s=1$, $a^{(s)}_{s-1}=a^{(s)}_{s-3}=...=0$ and $a^{(s)}_{s-2n}$ are real constants fixed by the transverse condition.~\footnote{See \eqref{sigmans} for the explicit formulae.}
For example, the $s=1$ operator is given in \eqref{s1-s0}, while the one for spin-2 fields is 
\be \label{s2u2}
U^{(2)}_\eta =  \frac{1}{k^2} \(\partial_\eta -\frac{2}{\eta} \)^2 + \frac{1}{3}
\ee
We leave more examples and further details of higher spins to Appendix \ref{app:spin}.
Using \eqref{s0-scalar} and \eqref{ss-s0}, 
we can establish the connection between the $\sigma_s^{(s)} $ mode and the massive scalar mode function in general.
In particular, we are interested in the bulk-to-bulk propagators of the $\sigma_{i_1i_2...i_s}$  mode. They are expressed in terms of the $G^\sigma$-propagators as
\bea \label{Gspm}
G_{i_1...i_sj_1...j_s,\pm\mp}^{(s)}(k,\eta, \eta') &=&  |N_s|^2 k^{2s}
U^{(s)}_\eta {U}^{(s)}_{\eta'}  G_{\pm\mp}^\sigma(k,\eta, \eta')\epsilon_{i_1...i_s}\epsilon_{j_1...j_s}
~, \\ \label{Gspp}
G_{i_1...i_sj_1...j_s,\pm\pm}^{(s)}(k,\eta, \eta') 
&=& |N_s|^2 k^{2s}
U^{(s)}_\eta {U}^{(s)}_{\eta'}  G_{\pm\pm}^\sigma(k,\eta, \eta')\epsilon_{i_1...i_s}\epsilon_{j_1...j_s} +\cdots 
\eea
where the $\cdots$ represent terms with $\delta(\eta-\eta')$.
As we have seen in the spin-1 case, these extra pieces lead to contact terms in the final bispectra. Since they are degenerate with the non-Gaussian shapes from single field inflation, we drop them in the following.
 
\vskip4pt
Now we derive the mixed propagator with a spin-$s$ massive field. 
From the EFT analysis in \eqref{spinsLmix}, the lowest derivative quadratic interaction between the inflaton and a spin-$s$ field is  
\be \label{quadra-s}
  \partial_{i_1...i_s}\phi \sigma_{i_1...i_s} ~ ,
\ee
which is allowed to have large couplings in boost-breaking theories.
From this two-point vertex, only the helicity-$0$ longitudinal mode with maximal number of polarization directions ($n=s$) gives a nonzero contribution, and we find the spin-$s$ mixed propagator~\footnote{For simplicity, we focus on  $\cG^{(s)}_{i_1...i_s,+}$, while $\cG^{(s)}_{i_1...i_s,-}$ is given by complex conjugation.}
\bea \label{mix-spins}
\cG^{(s)}_{i_1...i_s,+} (k,\eta;c_s) &=& i (-i)^s k_{j_1}...k_{j_s} \int_{-\infty}^{0} d\eta' a(\eta')^{4-2s}  \[ G_{i_1...i_sj_1...j_s,++}^{(s)}(k,\eta, \eta')   K_{+}(k, \eta';c_s)\right.\nn\\
&& ~~~~~~~~~~~~~~~~~~~~~~~~~~~~~~ \left.-(-1)^sG_{i_1...i_sj_1...j_s,+-}^{(s)}(k,\eta, \eta')    K_{-}(k, \eta';c_s)\] \nn\\
&=&  i(-i)^s  |N_s|^2 k^{3s} \epsilon_{i_1...i_s}  {U}^{(s)}_\eta \int_{-\infty}^{0} d\eta' a(\eta')^{4-2s}  \[  {U}^{(s)}_{\eta'} G_{++}^{\sigma}(k,\eta, \eta')   K_{+}(k, \eta';c_s) \right.\nn\\
&& ~~~~~~~~~~~~~~~~~~~~~~~~~~~~~~ \left.
-(-1)^s  {U}^{(s)}_{\eta'}G_{+-}^{\sigma}(k,\eta, \eta')    K_{-}(k, \eta';c_s)\] + \cdots
\eea
where in the second equality we used \eqref{Gspm} and \eqref{Gspp}.
The ellipses represent terms proportional to the conventional bulk-to-boundary propagator $K_{\pm}(k,\eta;c_s)$.

Notice that the constant factors and the $U^{(s)}_\eta$ operator  in \eqref{Gspm} and \eqref{Gspp}  have been moved outside of the integral, while we still need to address the $U^{(s)}_{\eta'}$ operator inside in order to rewrite the spinning mixed propagators in terms of $ \hat\cG^{(n)}$.
For the spin-1 case, $U^{(1)}_{\eta'}$ has only one derivative and can be simplified via integration by parts. For particles with arbitrary spin, it becomes more complicated. In general, the higher order time derivatives on the massive field   (i.e. the $U^{(s)}_{\eta'}$ operator on the $G^\s$ propagators here) can be reduced to terms with at most one derivative by using the equation of motion of the $\sigma$ field.
Let us take the spin-2 fields  as an explicit example whose $U^{(s)}_{\eta'}$ operator is given in \eqref{s2u2}. 
Applying the scalar mode equation of motion, the $ \sigma_2^{(2)}$ mode is 
 \be
 \sigma_2^{(2)}(k,\eta') =- N_2 k^2  U^{(2)}_{\eta'} \s_k(\eta') = - N_2 k^2   \(\frac{2}{k^2\eta'}\partial_{\eta'}  + \frac{\mu_2^2-{15}/{4}}{k^2\eta'^2}+ \frac{2}{3}\) \s_k(\eta')~.
\ee
The same approach also works for higher spin modes $\sigma_s^{(s)}$, where the $ U^{(s)}_{\eta'}$ operator reduces to the one with at most one derivative.
Then integrating by parts in \eqref{mix-spins}, we move this time derivative onto the inflaton propagator $K$. Now we are able to rewrite the mixed propagator as a linear combination of $ \hat\cG^{(n)}$, which can  be further simplified by using their recursive relations.
Note that we have neglected all the free field propagators $K$  generated in this procedure.

\vskip4pt
In the end, we are able to bring the spin-$s$ mixed propagator to the form
\be \label{mix-ss}
\cG^{(s)}_{i_1...i_s,+} (k,\eta) = i^s  \frac{|N_s|^2H^{2s}}{c_s^3k^{3-s}} \epsilon_{i_1...i_s}  {U}^{(s)}_\eta \[f^{(s)}  \hat\cG^{(s)}_+(k\eta) + \frac{iH}{c_s} g^{(s)}  \hat\cG^{(s-1)}_+(k\eta)  \]
 + ...
\ee
where $f^{(s)} $ and $g^{(s)}$ are constants coming from manipulating the $U^{(s)}_{\eta'}$ operator. 
We present them later.
In the bracket we have a linear combination of two building blocks $\cG^{(s)}$ and $\cG^{(s-1)}$. Since they are functions of $k\eta$,
the operator ${U}^{(s)}_\eta$ can be transformed to a $k$-differential operator by trading time and momentum derivatives
\be \label{Us2Uk}
U^{(s)}_\eta    ~~~ \rightarrow ~~~ \sum_{m=0}^{s} \frac{a^{(s)}_m}{k^m\eta^m}  \mathcal{U}^{(m)}_{k}~, ~~~~{\rm with} ~~ \mathcal{U}^{(m)}_{k} \equiv  k^m\(\partial_k-\frac{2}{k}\)^m ~ .
\ee
Like in the spin-1 case, this is related to the spin-raising operators with arbitrary $s$, as we will show when we bootstrap the three-point correlators.
With $s=2$, we explicitly get $(k\eta)^{-2}\mathcal{U}^{(2)}_{k}+1/3$ from \eqref{s2u2}.

\vskip4pt
Now let us look at the two constants $f^{(s)} $ and $g^{(s)}$.
For spin-1, we find $f^{(s)}=1$ and $g^{(s)}=0$ as shown in \eqref{s1mix}. Their expressions become a bit more subtle for higher spins.
Depending on the recursive relations of  $ \hat\cG^{(n)}$ with $c_s=1$ and $c_s\neq 1$, we express the final results of the spinning mixed propagators into two different cases.

\begin{itemize}
\item $c_s=1$: In this case, by using the recursive relation we can rewrite the mixed propagator in terms of one building block $\hat\cG^{(s)} $. Thus we have $g^{(s)}(c_s=1)=0$, while $f^{(s)}(c_s=1)$ becomes an overall factor fixed by $s$ and $\mu_s$. For $s=2$, it is given by
\be
f^{(2)}(c_s=1) = \frac{985-644\mu_2^2+16\mu_2^4}{36(4\mu_2^2+1)} ~.
\ee

\item $c_s\neq1$:  From the recursive relation with $c_s\neq1$, two building blocks $\hat\cG^{(n)}$ are needed for a general form of the mixed propagator. 
Therefore both $f^{(s)}(c_s)$ and $g^{(s)}(c_s) $ should be present in \eqref{mix-ss}. 
For $s=2$, we find their explicit expressions
\bea
f^{(2)}(c_s,\mu_s) &=& \frac{2(1-3c_s^2)}{3c_s^2} - \frac{8}{3} \frac{1}{1-c_s^2} - \frac{1-c_s^2}{c_s^2}\frac{\mu_2^2-\frac{7}{4}}{\mu_2^2+\frac{1}{4}}~,\nn\\
g^{(2)}(c_s,\mu_s) &=& -\frac{2}{3} \frac{1}{1-c_s^2} \(\mu_2^2+\frac{9}{4} \) + \(\mu_2^2-\frac{7}{4} \) \(\frac{2}{\mu_2^2+\frac{1}{4}} +1\)~.
\eea
In the extremal case   $c_s\rightarrow 0$, the spin-2 mixed propagator is given by
\be
\lim_{c_s\rightarrow0}\cG^{(2)}_{ij,+} (k,\eta) = -  |N_2|^2\frac{H^4}{c_s^3k} \epsilon_{ij} U^{(2)}_k \[ \frac{1}{c_s^2} \frac{\frac{\mu_2^2}{3}-\frac{23}{12}}{\mu_2^2+\frac{1}{4}}
 \hat\cG^{(2)}_+(k\eta)  \]
 + ...
\ee 
\end{itemize}

To summarize, we have applied the ${U}^{(s)}_\eta$ and ${U}^{(s)}_{\eta'}$ operators to map the massive scalar propagators into the spinning ones. In the mixed propagators, the ${U}^{(s)}_\eta$ opertator can be moved outside of the integral, and transformed into the spin-raising operators by trading derivatives as shown in \eqref{Us2Uk}. 
To deal with the ${U}^{(s)}_{\eta'}$ operator, we make the use of the equation of motion, integration by parts, and recursive relations of $\hat\cG^{(n)}$. Although the explicit computation can become cumbersome, this operator in the end leads to an overall prefactor which does not change the form of the mixed propagator. 

\vskip4pt
So far our analysis on the mixed propagator is based on the lowest derivative interaction in \eqref{quadra-s}.
It is straightforward to extend the results above to the linear mixings with more derivatives. 
Through integration by parts and equation of motion, the higher derivative interactions can always be reduced to a combination of lower derivative ones. 
As a result, the general form in \eqref{mix-ss} remains unaffected, while the two constant coefficients $f^{(s)}(c_s)$ and $g^{(s)}(c_s) $
may become different.

\subsection{Spin-$s$ Exchange Bispectra}
\label{sec:spin-s}

Having described the spinning mixed propagators, we now use them to compute the three-point correlators from spin-$s$ exchange. 
We first focus on the lowest-derivative cubic vertices with one massive spinning field, which break the de Sitter boosts and may have large interactions.
Then we consider the generalization to arbitrary interactions.  For de Sitter invariant theories, our computation reproduces the de Sitter bootstrap results.

\paragraph{The $\langle \varphi\varphi\phi \rangle$ bispectrum}
We get started with the simplest three-point function with two conformally coupled scalars as a warmup. For general spin, the cubic interaction here is given by $\varphi\partial_{i_1...i_s}\varphi\s_{i_1...i_s}$, and we find the bispectrum
\be
\langle \vp_{{\bf k}_1}\vp_{{\bf k}_2} \phi_{{\bf k}_3} \rangle' = i   \int_{-\infty}^0 d\eta a^{4-2s} \[  K_+^\vp(k_1,\eta)(-i)^sk_2^{i_1}...k_2^{i_s} K_+^\vp(k_2,\eta) \cG^{(s)}_{i_1...i_s,~+} (k_3, \eta) 
-c.c.\]+{\rm perm.} 
\ee
Using the spin-$s$ mixed propagator \eqref{mix-ss}, we rewrite this bispectrum in terms of the generalized scalar seeds
\be
\langle \vp_{{\bf k}_1}\vp_{{\bf k}_2} \phi_{{\bf k}_3} \rangle' =A_s^\vp
\frac{ k_2^s k_3^{-s} }{k_1 k_2 k_3^{2}}   P_s(\hat{k}_2\cdot\hat{k}_3) ~\mathcal{D}_{23}^{(s)} ~ k_3\partial_{k_{12}}\left[ f^{(s)} \hat{\mathcal{I}}^{(s)}+ \frac{iH}{c_s} g^{(s)} \hat{\mathcal{I}}^{(s-1)}\right]
+{\rm perms.}~,
\ee
where $A_s^\vp = i(-i)^s |N_s|^2 H^{4s} \eta_0^2 /4c_s^{5+s} $, and
 we introduce the {\it spin-raising operator}
\begin{eBox}
\be \label{spinraising}
\mathcal{D}_{23}^{(s)} \equiv    \sum_{m=0}^s 
(ic_s)^{m-s} k_3^{2s-m-1}
a_m^{(s)}\mathcal{U}_{k_3}^{(m)} \partial_{k_{2}}^{2s-m-1} ~.
\ee
\end{eBox}
This is a dimensionless differential operator that maps the scalar seed functions to the $\langle \vp\vp\phi \rangle$ bispectrum with spin-$s$ exchange.
Note that for $s=1$, we simply have $\mathcal{D}_{23}^{(1)} =\mathcal{U}_{k_3}^{(1)} $, and reproduce \eqref{conbi-s1}.
As another example, the spin-2 operator is explicitly given by
\be
\mathcal{D}_{23}^{(2)} \equiv  k_3^3 \[ \( \partial_{k_3}-\frac{2}{k_3}\)^2 \partial_{k_2} -\frac{1}{3c_s^2} \partial_{k_2}^3 \]
~.
\ee
From the bulk intuition, the spin-raising operator is originated from the constraint $\nabla^{\mu_1}\s_{\mu_1...\mu_s}=0$, which we used to establish the connection between $\s_{i_1...i_s} $ and the massive scalar $\s$. Thus these operators exist, since only the helicity-0 longitudinal modes with the maximum spatial polarization directions  contribute in the exchange bispectra.

\paragraph{The  inflaton bispectra}
Now we bootstrap the inflationary three-point functions from spinning exchanges. 
The strategy is to use the weight-shifting procedure to map the $\langle\vp\vp\phi\rangle$ correlator to the one with three external massless fields, as we have done in the scalar exchange cases.
Let us first consider the lowest derivative interactions in boost-breaking theories, which could generate large bispectra.
From the EFT of inflation with spinning fields \cite{Lee:2016vti}, the leading order cubic vertex with one spin-$s$ field is given in \eqref{spinsLmix}
\be \label{cubic-spins}
 \dot\phi \partial_{i_1...i_s}\phi \sigma_{i_1...i_s}~,
\ee
for which the bispectrum is
\begin{small}
\be
\langle \phi_{{\bf k}_1}\phi_{{\bf k}_2} \phi_{{\bf k}_3} \rangle' =i \int d\eta a(\eta)^{3-2s} \[ \partial_\eta K_+ (k_1,\eta)(-i)^sk_2^{i_1}...k_2^{i_s}  K_+ (k_2,\eta) \cG^{(s)}_{i_1...i_s,+} (k_3, \eta) -c.c.\]+{\rm perm.}
\ee
\end{small}With the spin-$s$ mixed propagator \eqref{mix-ss} and generalized scalar seeds,
we find that the bispectrum can be written as
\begin{eBox}
\be \label{spins-bispectr}
\langle \phi_{{\bf k}_1}\phi_{{\bf k}_2} \phi_{{\bf k}_3} \rangle' = \frac{ A_s  k_2^s k_3^{-s}}{ k_1^2k_2^{3}k_3^{2}} P_s(\hat k_2 \cdot \hat k_3) ~ \mathcal{W}_{12}   ~  \mathcal{D}^{(s)}_{23} ~ k_3\Big[ f^{(s)} \hat{\mathcal{I}}^{(s)}+ \frac{iH}{c_s} g^{(s)} \hat{\mathcal{I}}^{(s-1)}\Big]
+{\rm perms.}~,
\ee
\end{eBox}
with $A_s= (-i)^{s+1}H^{4s+1}|N_s|^2/4c_s^{7+s}$. A new weight-shifting operator ${\mathcal{W}}_{12} $ is derived by noticing the relation \eqref{dnK}. For the cubic vertex \eqref{cubic-spins} from the EFT, it becomes
\be
\mathcal{W}_{12}^{\rm EFT}= k_1\partial_{k_1} (1-k_2\partial_{k_2})~.
\ee
The result \eqref{spins-bispectr} provides the analytical expression of inflationary bispectra from the exchange of massive spin-$s$ particles.
It is easy to check the squeezed limit behaviour of this bispectrum. Notice that in the sum of the spin-raising operator, the   $m=s$ term provides the lowest order contribution in the $k_3$ expansion, and thus only this term is relevant when we take $k_3\rightarrow0$. We find
\bea
\lim_{k_3\rightarrow0}\langle \phi_{{\bf k}_1}\phi_{{\bf k}_2} \phi_{{\bf k}_3} \rangle' & \propto & \frac{P_s(\hat k_2 \cdot \hat k_3)}{k_1^{3}k_3^{3}}  \bigg[ (1-i\sinh\pi\mu_s)(5+2s-2i\mu_s) \nn\\
&&   ~~~~~~~~~~~~~~~~ \times\frac{\Gamma(i\mu_s)}{\Gamma(\frac{1}{2}+i\mu_s)} \(\frac{k_3}{4c_sk_1}\)^{\frac{3}{2}-i\mu_s} + c.c.\bigg]~,
\eea
which agrees with Eq.(C.20) in \cite{Lee:2016vti}.
We comment on phenomenological implications of those shapes in the next section.

\vskip4pt
It is also straightforward to consider cubic interactions with more derivatives than in \eqref{cubic-spins}. In analogy with the derivation of the weight-shifting operators for scalar-exchanges in Section \ref{sec:bbws}, here we consider the generic cubic interaction with a spin-$s$ field
\be 
\partial_i^{n_s} \( \partial_t^{n_1} \phi  \partial_t^{n_2} \partial_{i_1...i_s}\phi \)\sigma_{i_1...i_s}~,
\ee
where $n_1$, $n_2$ and $n_s$ are the numbers of extra derivatives in addition to the ones with spinning indices.
We find the general form of the weight-shifting operator for spinning exchanges
\be \label{bbws-s}
\mathcal{W}_{12}=-c_s^{-n_s}({\bf k}_a\cdot {\bf k}_b)^{n_s/2}k_1^{n_1-1}k_2^{n_2}\(1-n_1-k_1\partial_{k_1}\)\(1-n_2-k_2\partial_{k_2}\)\partial_{k_{12}}^{\tilde{n}_T-1}~,
\ee
with $\tilde{n}_T= n_1+n_2+n_s$.
In \eqref{spins-bispectr}, using this general form of the $\mathcal{W}_{12}$ operator, we obtain the spin-s exchange bispectra with arbitrary boost-breaking interactions.
In this generic expression, the spin-raising operator $\mathcal{D}^{(s)}_{23}$ uplifts the spin of the exchanged particle, and the weight-shifting operator incorporates any possible cubic vertices with  two massless  scalars.
Therefore, using this approach, we obtain a complete set of inflationary three-point functions from spin exchange diagrams.

\vskip4pt
As a nontrivial example, here we reproduce the results from interactions with de Sitter symmetries.
The dS-invariant cubic vertex with two inflatons and one spinning field (for even $s$) is given by
\be
\nabla_{\nu}\phi \nabla^{\nu}\nabla_{\mu_1...\mu_s} \phi \sigma^{\mu_1...\mu_s} \rightarrow a^{2-2s} \( \phi' \partial_{i_1...i_s} \phi' - \partial_j\phi \partial_{i_1...i_s} \partial_j\phi\)\sigma_{i_1...i_s}~,
\ee
which leads to a new form of the weight-shifting operator
\be
\mathcal{W}_{12}^{\rm dS} =   \frac{1}{2} k_2 (k_{12}^2-k_3^2) \partial^{3}_{k_{12}} + \frac{1}{2k_1} \(k_{12}^2 -k_3^2-2k_1k_2\)\(  \partial_{k_{12}} -k_{12} \partial^{2}_{k_{12}}\)  .
\ee
Then the bispectrum \eqref{spins-bispectr} with the $\mathcal{W}_{\rm dS}$ operator generates the result from the de Sitter invariant interaction.
Again, let us look at the squeezed limit of this bispectrum
\bea
\lim_{k_3\rightarrow0}\langle \phi_{{\bf k}_1}\phi_{{\bf k}_2} \phi_{{\bf k}_3} \rangle'_{\rm dS} &\propto & \frac{ P_s(\hat k_2 \cdot \hat k_3)  }{k_s^{3}k_3^{3}}  \bigg[ (1-i\sinh\pi\mu_s)\frac{\frac{5}{2}+s-i\mu_s}{\frac{3}{2}-s+i\mu_s} \nn\\
&& ~~~~~~~~~~~~~~~~ \times \frac{\Gamma(i\mu_s)}{\Gamma(\frac{1}{2}+i\mu_s)} \(\frac{k_3}{4k_s}\)^{\frac{3}{2}-i\mu_s} + c.c.\bigg] .
\eea
We find agreement with Eq. (6.142) in \cite{Arkani-Hamed:2015bza} and Eq. (6.20) in \cite{Arkani-Hamed:2018kmz}. 
Therefore we recover the results from the (slow-roll) de Sitter bootstrap, bypassing the need to compute four-point functions.
In this case the size of this bispectrum is required to be slow-roll suppressed, as we are only allowed to consider the mild breaking of conformal symmetries in the de Sitter bootstrap. 
There is more freedom in the boost-breaking scenario, for both the size and shape of cosmological colliders with spinning particles.

\section{Phenomenology}
\label{sec:pheno}

One major advantage of the boostless bootstrap is that the resulting signals of non-Gaussianity can be potentially large, and thus testable in near-future observations. In the de Sitter bootstrap, the conformal symmetry is only weakly broken, and thus the inflationary bispectra are always slow-roll suppressed, and beyond current reach. 
In this section, we will present the phenomenological consequences of the boost-breaking shapes of non-Gaussianity, contrasting them with the de Sitter invariant shapes.

\vskip8pt
\noindent{\underline{\it Convention}}~~~ In previous sections we used  ``$c_s$'' for the ratio of the sound speeds of the inflaton and  massive field. To avoid confusion to the reader that came straight to this section, we reintroduce the individual sound speeds $c_s$ and $c_\s$, referring to their ratio explicitly i.e. $c_s/c_\s$. This implies that $u\equiv c_\s k_3 /(c_s k_{12})$.

\vskip 6pt
Following the standard convention, the bispectrum of the primordial curvature perturbation $\zeta=(H/\dot\Phi )\phi$ is given by
\be
\langle \zeta_{{\bf k}_1} \zeta_{{\bf k}_2} \zeta_{{\bf k}_3}  \rangle  =  
(2\pi)^3 \delta({\bf k}_1+{\bf k}_2+{\bf k}_3) \frac{18}{5}\fnl \frac{S(k_1,k_2,k_3)}{k_1^2 k_2^2 k_3^2} P_\zeta^2~,
\ee
where $ P_\zeta$ is the power spectrum of $\zeta$ and $\fnl$ represents the size of the non-Gaussian signal. 
For boost-breaking theories, such as the ones with a small sound speed of the inflaton fluctuations, large non-Gaussianity with $\fnl>1$ is naturally allowed \cite{Lee:2016vti}.
The main focus of this work is the shape function  of the primordial bispectrum $S(k_1,k_2,k_3)$. 
In the previous two sections, we have derived a complete set of bispectra from the exchange of one heavy particle during inflation.
For the scalar exchanges, the general form of the shape function can be simply written as
\be
S^{(0)}(k_1,k_2,k_3) = \mathcal{W}_{12} \mathcal{\hat{I}}^{(0)} + {\rm perms.}~,
\ee
where $\mathcal{W}_{12}$ is the weight-shifting operator with the expression \eqref{bbws} for generic boost-breaking interactions.
The spin-$s$ exchanges in general lead to 
\be
S^{(s)}(k_1,k_2,k_3) =   P_s(\hat k_2 \cdot \hat k_3) {k_2^{s-1} k_3^{-s}} ~ \mathcal{W}_{12}  ~  \mathcal{D}^{(s)}_{23} ~ k_3\left[ f^{(s)} \hat{\mathcal{I}}^{(s)}+ \frac{iH}{c_s} g^{(s)} \hat{\mathcal{I}}^{(s-1)}\right]
+{\rm perms.}~,
\ee
with the spin-raising operator in \eqref{spinraising} and  another type of the weight-shifting operator \eqref{bbws-s}.
Thus starting with the generalized scalar seeds $\hat{\mathcal{I}}^{(n)}$ and applying the weight-shifting and spin-raising operators, we produce the analytical results of full bispectrum shapes, which are theoretically well-motivated targets in the data analysis of future cosmological surveys. 
Next, we examine these shape functions in detail and compare them with the ones from de Sitter bootstrap.

\vskip4pt
Two novelties arise for cosmological colliders in boost-breaking theories: {\it i}) the cubic interactions are extended to boostless forms;
 {\it ii})  the sound speeds of the inflaton and massive field can differ from each other.
Both of these two effects lead to modifications on the collider signals, as we shall discuss respectively.

\begin{figure}[t!]
   \centering
      \includegraphics[height =4.9cm]{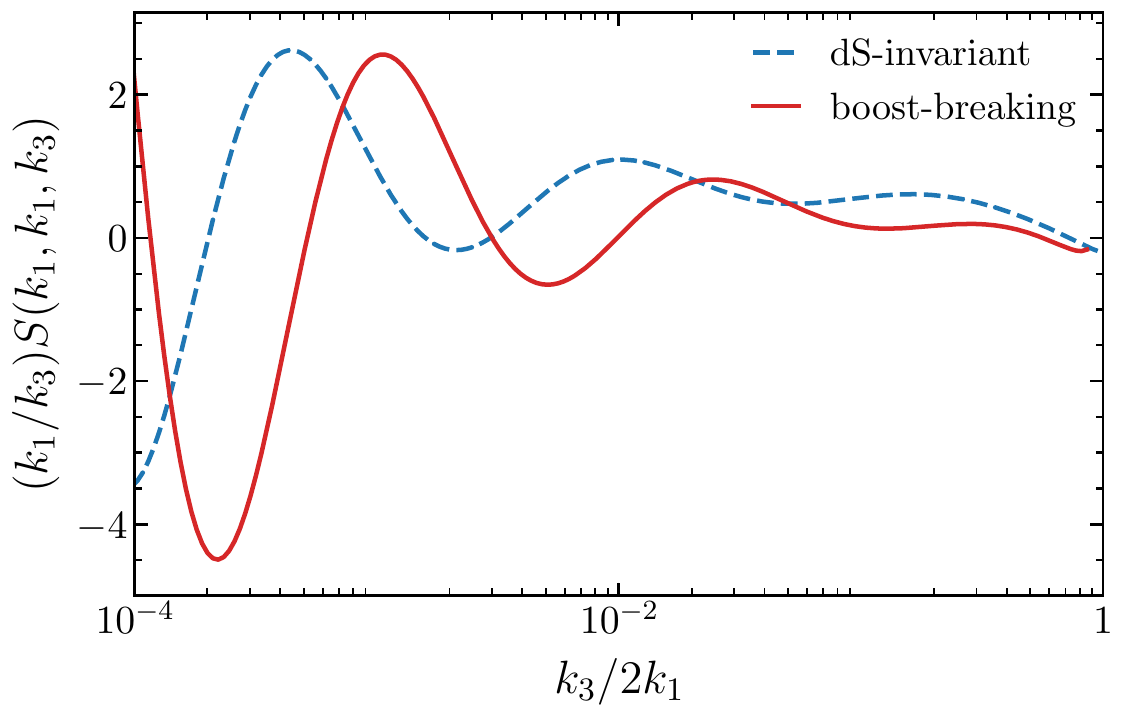}  \hspace{0.1cm}
\includegraphics[height =4.9cm]{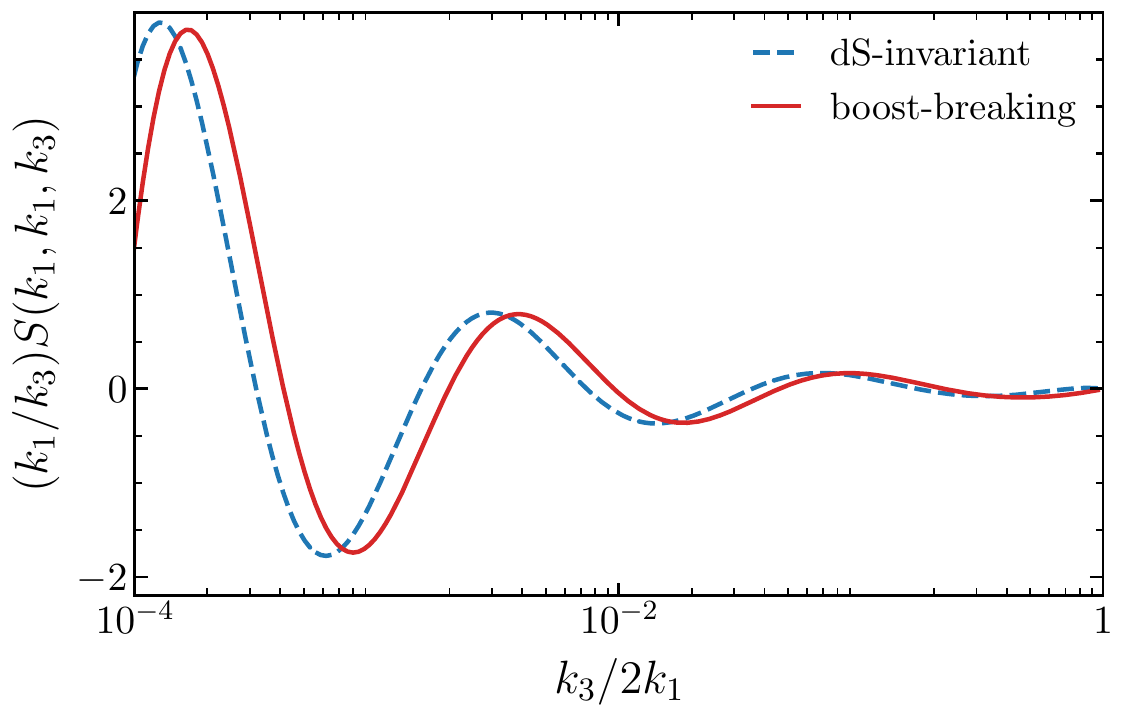} \\
 \includegraphics[height =4.9cm]{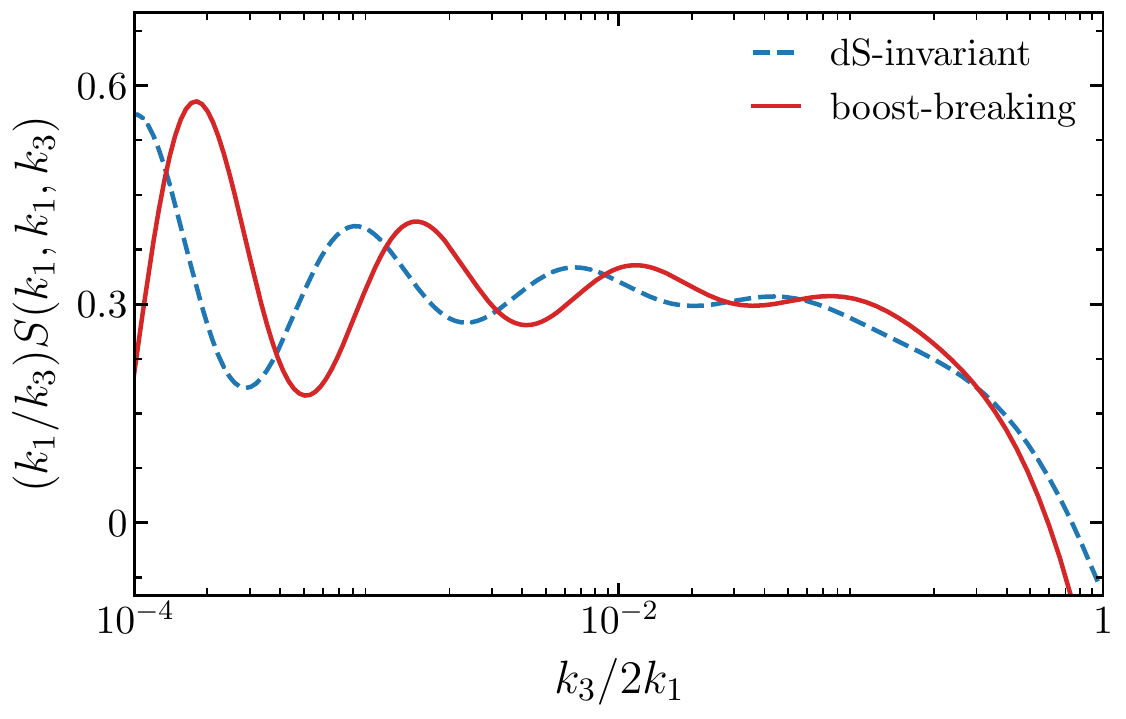}  \hspace{0.1cm}
\includegraphics[height =4.9cm]{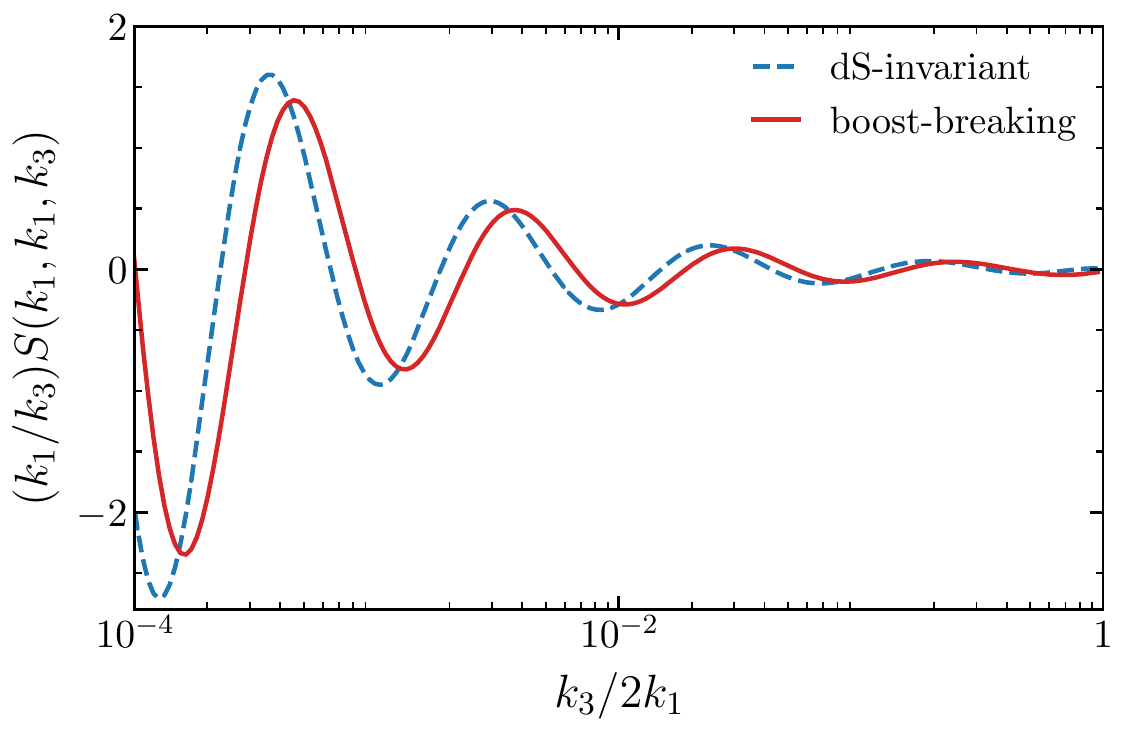}  \\
 \includegraphics[height =4.9cm]{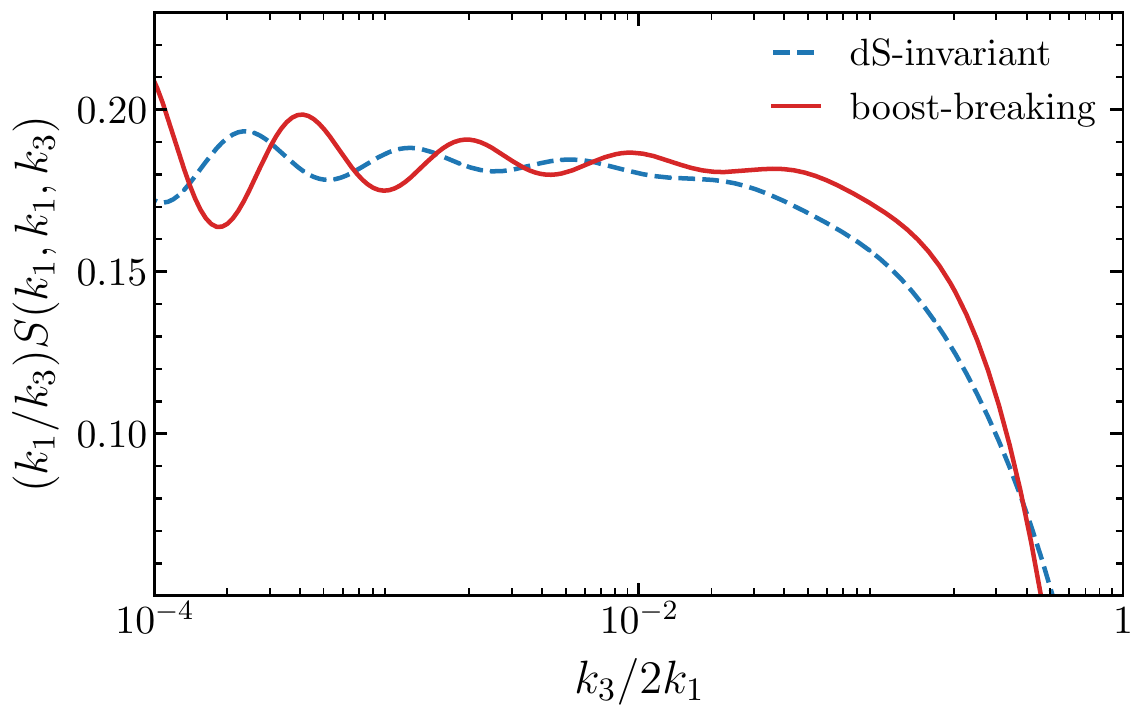}  \hspace{0.1cm}
\includegraphics[height =4.9cm]{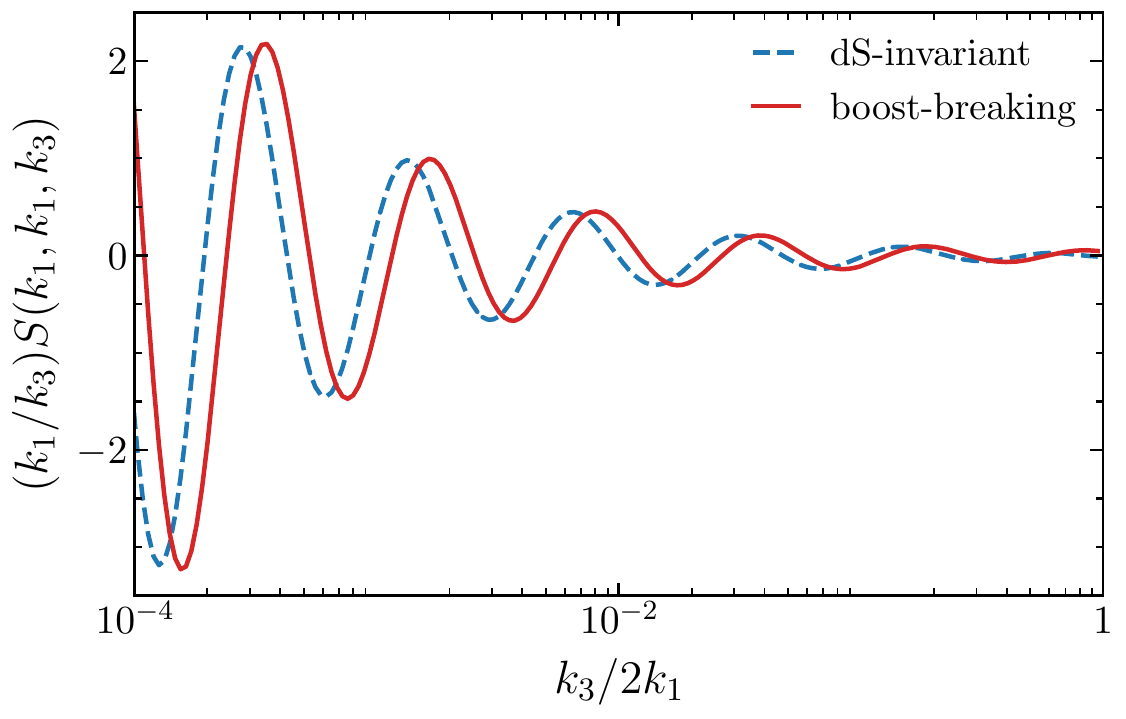} 
      \caption{Shape functions of the cosmological collider bispectra in the isosceles-triangle configuration with $k_1=k_2$ and $c_s/c_\s=1$. {\it Left panel:}  shapes from the massive scalar exchanges with $\mu=2$ (top), $\mu=3$ (middle) and $\mu=4$ (bottom). {\it Right panel:} shapes from the massive spin-2 exchanges with $\mu=2$ (top), $\mu=3$ (middle) and $\mu=4$ (bottom). For the boost-breaking results we have chosen one particular combination of interactions for demonstration.} 
      \label{fig:shapes}
\end{figure}

\paragraph{Phase of cosmological colliders}
Let us first take a look at the consequences of boost-breaking interactions in the bispectrum shapes. 
For this part of the analysis we may assume $c_s/c_\s=1$ for simplicity, such that the comparison with the de Sitter bootstrap is easier.
From the EFT analysis, interactions with lowest derivatives give the dominant contribution to the three-point function. 
Let us take the scalar exchange as an example. There the two leading boost-breaking cubic vertices are given by $\dot\phi^2\s$ and $(\partial_i\phi)^2\s$, while the dS-invariant one is a combination of these two $(\partial_\mu\phi)^2\s$.
Thus in de Sitter bootstrap the weight-shifting operator which is the one given in \eqref{ws-dS}, has been uniquely fixed by the conformal symmetry. The boost-breaking interactions are not constrained by the symmetry, and we are free to consider arbitrary combinations of the two boost-breaking weight-shifting operators from the $\dot\phi^2\s$ and $(\partial_i\phi)^2\s$ vertices.
This generalization modifies the phase in the oscillatory signals of cosmological colliders.
It is convenient to look at the squeezed limit of the shape function, which in general can be written as
\be \label{shapesq0}
\lim_{k_3\rightarrow0}S^{(0)}(k_1,k_2,k_3) \sim \(\frac{k_3}{k_1}\)^{1/2}\cos\[\mu\log\(\frac{c_\s k_3}{4c_sk_1}\)+\delta(\mu)\]~.
\ee
The phase $\delta$ is a function of the mass parameter $\mu$.
Interestingly the explicit expression of this function is determined by the form of the cubic interactions (or equivalently the form of the weight-shifting operators).
For the one with de Sitter symmetries, the squeezed bispectrum is given in \eqref{sq3pt-0-dS}, which fixes the phase to be
\be \label{dSphase}
\delta^{\rm dS}(\mu) =\arg \[i\frac{\Gamma\(\frac{7}{2}+i\mu\)}{\Gamma\(1+i\mu\)}\frac{(1+i \sinh\pi\mu )}{\frac{1}{2}+i\mu}\]~.
\ee
Meanwhile the squeezed bispectra from the two boost-breaking interactions are given in \eqref{sq3pt-0-bb} and \eqref{sq3pt-0-bb2}. In general, their combinations could lead to arbitrary phases of the cosmological collider signals. This is shown in the left panel of Figure \ref{fig:shapes}. 
For demonstration, we show only the shape functions for the dS and boostless bootstrap. Of course, the resulting signals are potentially larger in the boost-breaking case. While in the dS-invariant case the shape function is completely fixed once we know the mass of the new particle, we are still free to shift the phase of the oscillatory signals in boost-breaking theories.
This indicates that any deviation from the phase  \eqref{dSphase} can be seen as a signature for the breaking of the de Sitter boosts.

The same analysis applies to the spinning exchanges. The squeezed limit of the shape function from an internal massive spin-s field is generally given by
\be \label{shapesqs}
\lim_{k_3\rightarrow0}S^{(s)}(k_1,k_2,k_3) \sim P_s(\hat{k}_1\cdot\hat{k}_3) \(\frac{k_3}{k_1}\)^{1/2}\cos\[\mu_s\log\(\frac{c_\s k_3}{4c_sk_1}\)+\delta_s(\mu_s)\]~.
\ee
For the bispectrum from dS-invariant interactions, we find the phase is fully determined by
\be
\delta_s^{\rm dS}(\mu_s) =\arg \left[ (1-i\sinh\pi\mu_s)\frac{\frac{5}{2}+s-i\mu_s}{\frac{3}{2}-s+i\mu_s}\frac{\Gamma(i\mu_s)}{\Gamma(\frac{1}{2}+i\mu_s)}\]~,
\ee
while this phase can be arbitrarily shifted in the boost-breaking theories, as shown in the right panel of Figure \ref{fig:shapes}.
Therefore, the breaking of the de Sitter boosts is manifested in
the phases of cosmological colliders.

\begin{figure} 
   \centering
      \includegraphics[height =5.4cm]{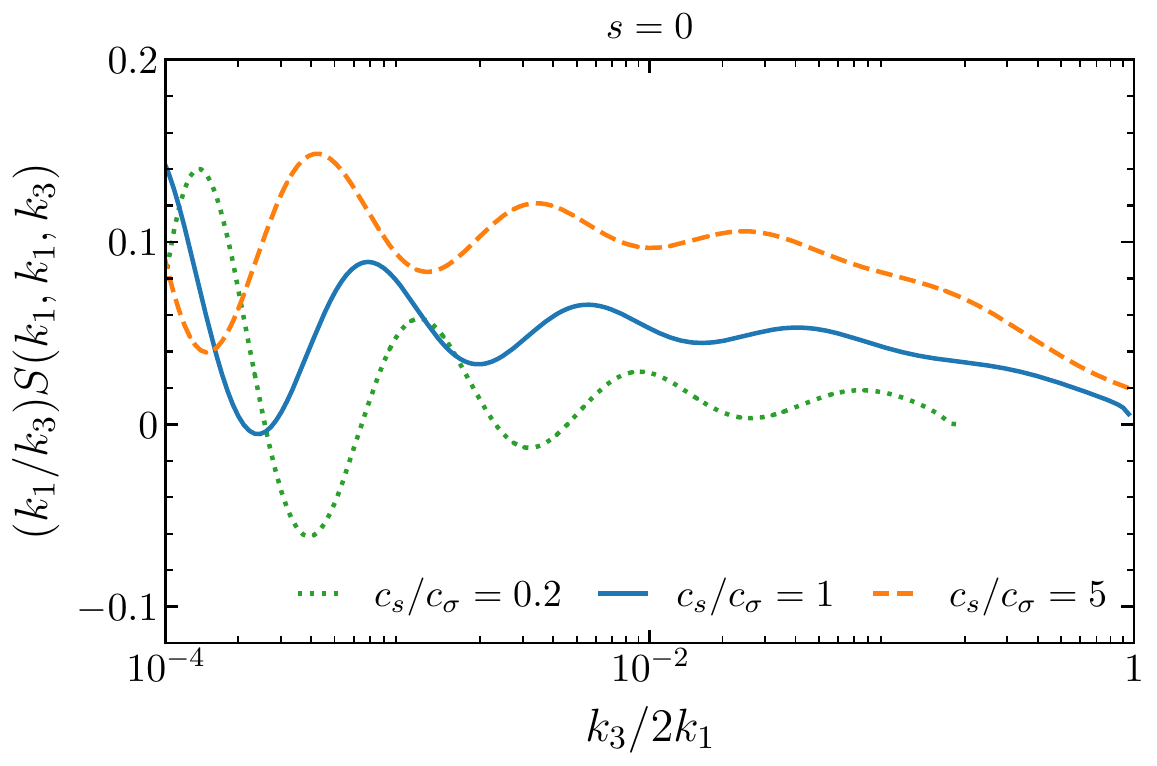}  \hspace{0.1cm}
\includegraphics[height =5.4cm]{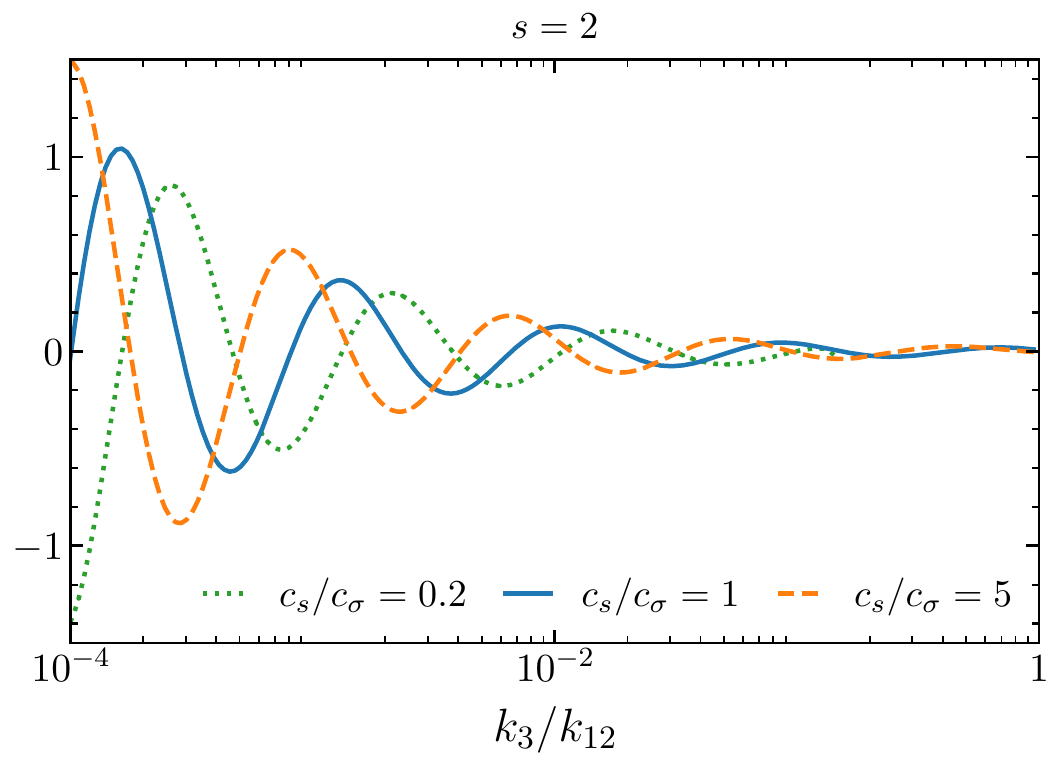} 
      \caption{Shape functions from boost-breaking interactions with nontrivial sound speeds, for scalar exchange ({\it left}) and  spin-2 exchange ({\it right}) diagrams with $\mu=3$. For demonstration we have normalized the amplitudes of the oscillations to be the same size and take the  isosceles-triangle configuration with $k_1=k_2$. As the seed functions are valid for $u\in[0,1]$, we stop the plots at $k_3/k_{12}=c_s/c_\s$ when $c_s/c_\s<1$.} 
      \label{fig:shapecs}
\end{figure}

\paragraph{Collider signals in the equilateral limit}
Now we present the effects of nontrivial sound speeds in the cosmological collider.
New features arise when the sound speeds of two fields $c_s$ and $c_\s$ become different.
As the squeezed limit is given by taking $u\equiv c_\s k_3/(c_sk_{12})\rightarrow0$ and the oscillations there are in terms of $u$, when $c_s\neq c_\s$ we see in \eqref{shapesq0} and \eqref{shapesqs} the phases of collider signals are further shifted. 
For $c_s<c_\s$,  the oscillations are shifted to the left with smaller momentum ratios. This means that one has to look into more squeezed configuration to identify the collider signals.
In the extremal case with $c_s/c_\s\rightarrow0$, the oscillatory behaviour would be manifested only when we consider $k_3/k_{12}\ll c_s/c_\s$. Although the size of the signal can be amplified in this small sound speed limit, it becomes more difficult to probe in observations, as the cosmic variance becomes more significant in the super squeezed limit of the bispectra. 
Meanwhile we find the opposite behaviour when the $\s$ sound speed is the smaller one (i.e.
 $c_s>c_\s$).
The oscillations are shifted to the right with larger momentum ratios, and thus we may be able to identify the signals even in the less squeezed triangle configurations. 
Figure \ref{fig:shapecs} shows how phases change when the two sound speeds differ from each other.

\begin{figure} 
   \centering
      \includegraphics[height =4.8cm]{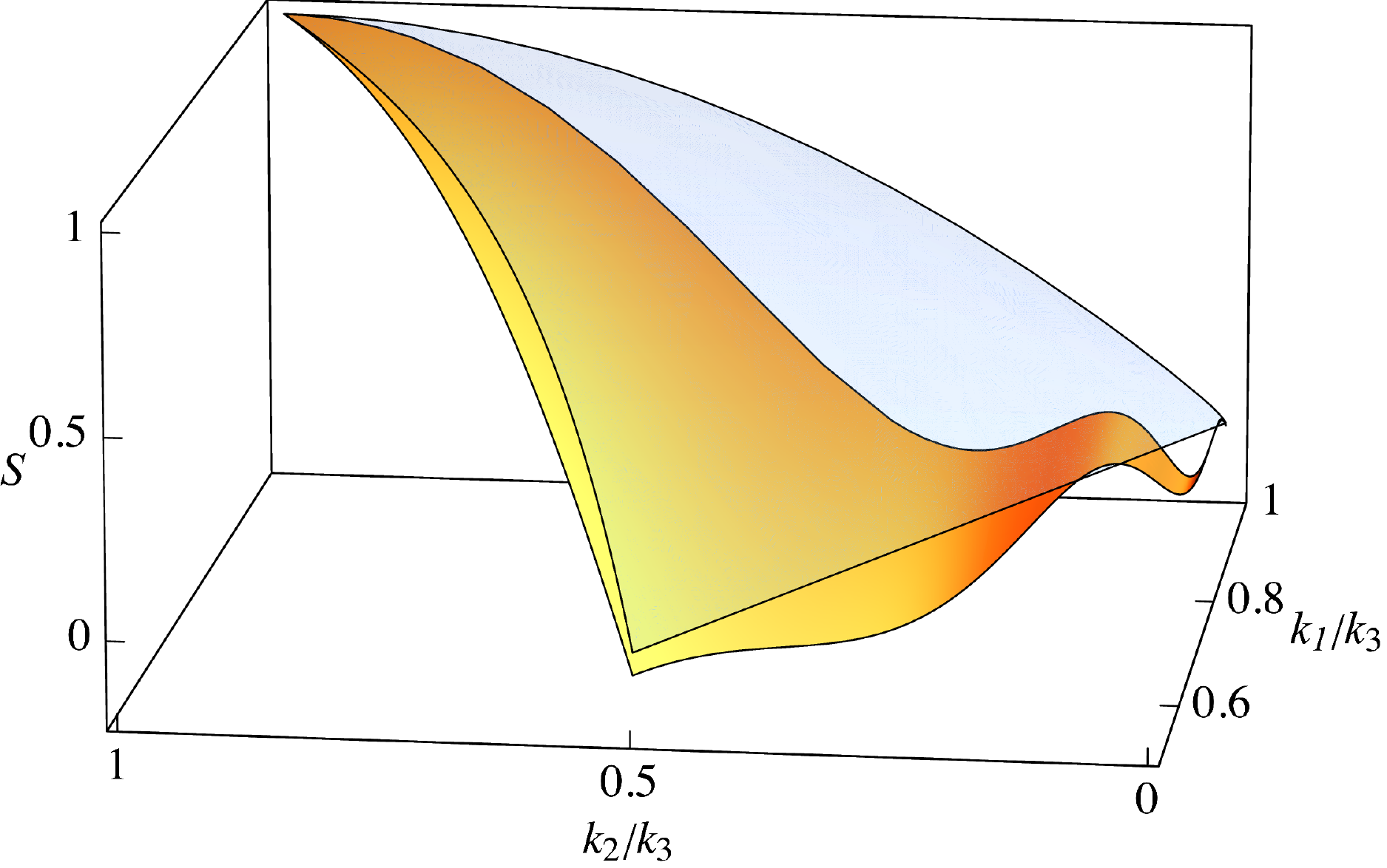}  \hspace{0.2cm}
\includegraphics[height =4.8cm]{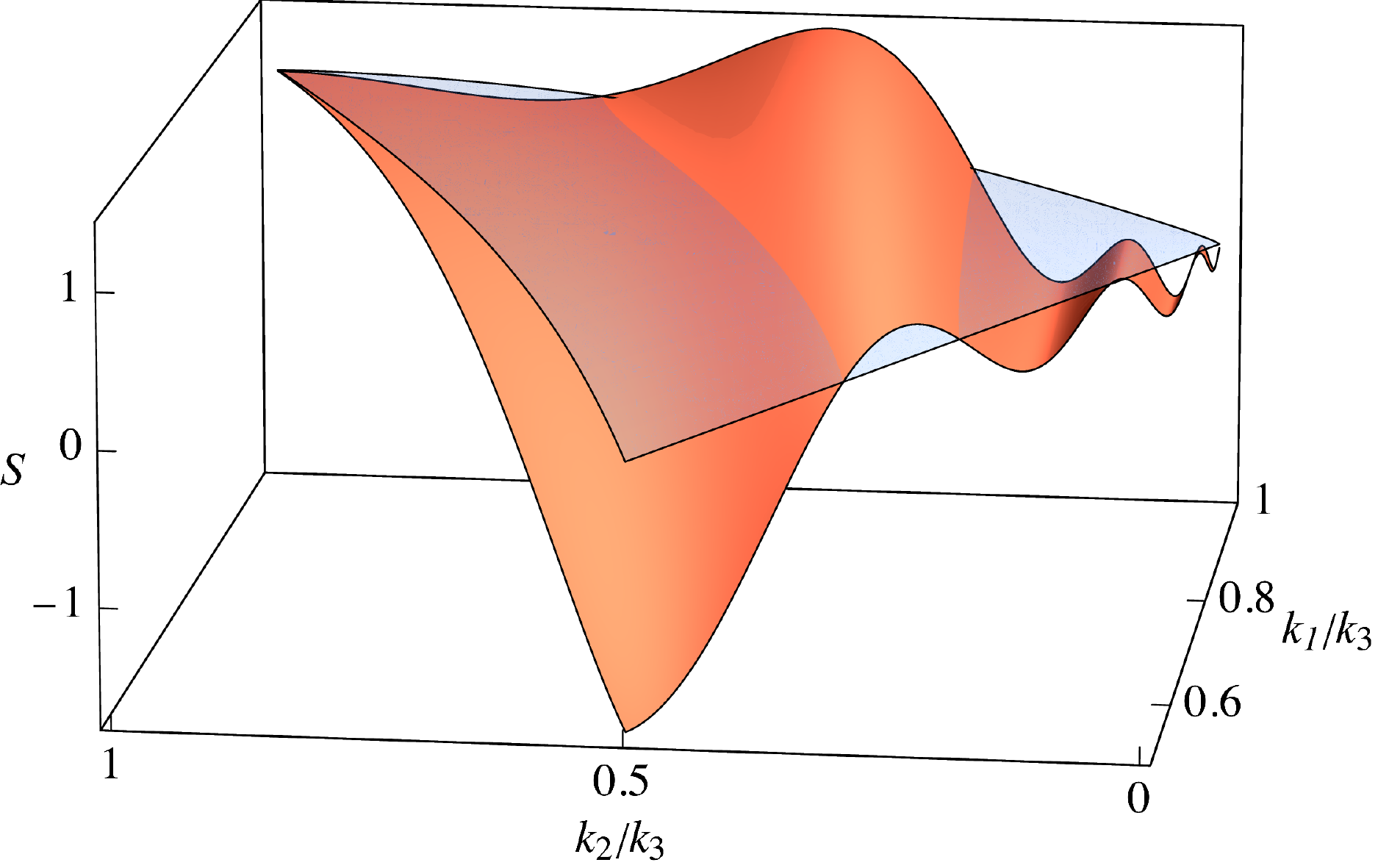} 
      \caption{The equilateral collider shapes $ S^{\rm eq.col.}(k_1,k_2,k_3)$ with $c_s/c_\s=10$, $\mu=3$ ({\it left}) and $c_s/c_\s=20$, $\mu=5$ ({\it right}). We have also plotted the standard equilateral shape ({\it blue transparent surfaces}) for comparison. The shapes are normalized to be $1$ at the equilateral limit $k_1=k_2=k_3=1$.} 
      \label{fig:eqcol}
\end{figure}
 
One particularly interesting case is when $c_s\gg c_\s$, i.e. the sound speed of the massive field is much smaller than the one of the inflaton. 
Since $0\leq k_3/k_{12}\leq1$, in this case we always have $u\ll1$ in the shape functions, and thus can use the squeezed limits \eqref{shapesq0} and \eqref{shapesqs} as good approximations.
As a result, the oscillatory collider signals get shifted outside of the squeezed limit, and can be present even around the equilateral configuration with $k_1\simeq k_2 \simeq k_3$.
We dub this interesting phenomenon  the {\it equilateral collider}. As an example, the shape function from the $\dot\phi^2\s$ scalar exchange is given by
\be 
 S^{\rm eq.col.}(k_1,k_2,k_3) = \frac{k_1k_2}{(k_1+k_2)^2} \(\frac{k_3}{k_1+k_2}\)^{1/2}\cos\[\mu\log\(\frac{c_\s k_3}{2c_s(k_1+k_2)}\)+\delta \] + {\rm perms.}~.
\ee
Figure \ref{fig:eqcol} demonstrates two examples of this new shape. 
In our conventional understanding, the equilateral configurations are dominated by the equilateral and orthogonal shapes which come from the self-interactions of the inflaton in single field models. 
As we can see here, the equilateral collider shape contains oscillations which qualitatively differ from these single field shapes.\footnote{Notice that the equilateral collider oscillations are {\it scale-invariant}, which differs from the inflationary features with scale-dependent oscillations in the equilateral configuration. In feature models, the de Sitter dilation is explicitly broken (for instance, by time-dependent background parameters), which leads to violations of scale-invariance in both the power spectrum and higher-point correlators. See \cite{Chen:2006xjb,Achucarro:2012fd} for specific examples, and \cite{Chluba:2015bqa, Slosar:2019gvt} for reviews.}
Considering that cosmological observations are usually more sensitive to signals with large momenta,
 the non-squeezed configurations of the bispectra may be easier to probe in cosmological data. Therefore we expect this large deviation from the equilateral and orthogonal shapes to provide new templates for the observational search of cosmological collider signals.

\section{Conclusions and Outlook}
\label{sec:concl}

In this paper, we classified inflationary three-point functions in boost-breaking theories, with an emphasis on the correlators coming from the exchange of a massive, scalar/spinning particle.
Our approach is to solve boundary differential equations in the external momenta, thus seeing ``time without time'', and using weight-shifting and spin-raising operators to obtain the most general results. Our findings provide analytical shapes to study cosmological collider physics in scenarios of phenomenological interest, as the resulting signals might be detectable in ongoing and upcoming cosmological surveys.  

\vskip4pt
First, a short recap. Our analysis focuses on the inflaton three-point correlators from the single-exchange diagrams with both scalar and spinning particles. 
Despite the absence of boost symmetry, there are still boundary differential equations. We established them by exploiting the bulk evolution of massive fields. 
As a stepping stone, we first studied the ``mixed propagators'' which are generated by the quadratic interactions between the inflaton and an additional scalar.
From those, we computed the three-point scalar seeds, a correlation function of two conformally coupled scalars and one massless scalar carrying the linear mixing.
From these building blocks, the inflaton bispectra were systematically derived by introducing the boost-breaking versions of the weight-shifting and spin-raising operators. 
Since all the boost-breaking interactions are captured in our approach, their sizes are not slow-roll suppressed, and within the range of validity of the EFT of inflation, the signals can be large enough to be detectable. Moreover, the resulting shapes have interesting new phenomenology compared to the de Sitter invariant cases. 
In particular, two novel features arise in the bispectra shapes of the boostless scenarios.
First, the phases of the cosmological collider signals are generally shifted away from the de Sitter invariant results.
Second, when the massive field has a sound speed much smaller than the one of the inflaton, we find the oscillations appear around the equilateral configuration of the momentum triangle.

\vskip4pt
To wrap up, we mention a few new avenues for future exploration of this rich topic: 

\begin{itemize}

\item Since we cannot leverage all de Sitter symmetries, our approach, although ``bootstrappy'' in nature, largely relies on manipulating the time integrals from bulk evolution. Proceeding that way, we derived the differential equations satisfied by boundary correlators. This was the case for the scalar seeds, as well as weight-shifting and spin-raising operators.
It would be more satisfying to find a systematic derivation of the boostless bootstrap from a purely boundary perspective, which may require a deeper understanding of the symmetries at play in this scenario.

\item Linear mixing is a general consequence of the time-dependent background in cosmology, thus having no analogue within flat space scattering amplitudes. 
Here we have demonstrated how the resulting mixed propagators can be applied in the cosmological bootstrap. 
At a practical level, by considering their differential equations, we managed to directly bootstrap the boost-breaking bispectra, bypassing the computations of four-point functions. 
It would be nice to further explore  implications of the mixed propagators in cosmology, which may reveal deeper connection/distinction 
 between physics in flat and curved spacetime.

\item While in this work we focused on the scalar bispectrum from inflation, it would be interesting to extend this analysis to other types of cosmological correlators. 
For example, we expect the same bootstrap approach to be applicable to correlators of gravitons. Another example is the higher-point functions of cosmological colliders with large boost-breaking interactions.
These correlators may also be phenomenologically interesting. 

\item 
Our analysis focused on the single-exchange diagrams with one internal massive field (i.e. one mixed propagator). Meanwhile, there are also 
double-exchange and triple-exchange diagrams, where two and three internal fields are present. From the bulk perspective, they usually lead to more nested time integrals, which become rather complicated to analyze. 
As we have seen in this work, one major advantage of the mixed propagators is that, by using them, the exchange diagrams can be simplified to ``contact'' ones. 
Therefore we expect the mixed propagators to help with the analysis of multiple-exchange diagrams. We make a few comments about these diagrams in Appendix \ref{app:double}. It would be nice to systematically study these diagrams, to complete the classification of bispectra in this scenario.

\item 
Our results  provide physically motivated templates for future observational surveys.
 It would be very interesting to assess the detectability of the  novel features present in the shape functions,  in particular the oscillations around the equilateral limit. 
As these observational signatures with possibly large sizes are imprinted at the hot Big Bang, they should be robust to late time evolution, and present not only in the CMB bispectrum, but also in LSS probes of primordial non-Gaussianity 
\cite{Meerburg:2016zdz, MoradinezhadDizgah:2017szk, MoradinezhadDizgah:2018ssw}.

\end{itemize}

\vspace{0.5cm}
\paragraph{Acknowledgements} 

We are grateful to Ana Ach\'ucarro, Daniel Baumann, Carlos Duaso Pueyo, Dra\v{z}en Glavan,  Harry Goodhew, Tanguy Grall, Aaron Hillman, Austin Joyce, Barak Kol, Hayden Lee, Mang Hei Gordon Lee, Scott Melville, Enrico Pajer, David Stefanyszyn, and Zhong-Zhi Xianyu for helpful discussions. We also thank Sadra Jazayeri and S\'ebastien Renaux-Petel for discussions, and sharing their upcoming work \cite{Jazayeri:2022kjy} on closely related topics. GLP thanks the Weizmann Institute of Science, and in particular Ofer Aharony and Kfir Blum, for their hospitality during the development of this project.
DGW thanks the Lorentz Institute of Leiden University and the Amsterdam Cosmology Group for hospitality while parts of this work were being
completed.
GLP is supported by the Netherlands Organisation for Scientific Research (NWO/OCW)~and through the Delta-ITP consortium of NWO/OCW.
DGW is supported by the Netherlands Organisation for Scientific Research (NWO) through the research program VIDI with Project
No. 680-47-535, and a Rubicon Postdoctoral Fellowship.


\newpage
\appendix

\section{More on Mixed Propagators}\label{app:mix-p}

In this Appendix, we shall provide more details about the  mixed propagators introduced in Section \ref{sec:moremix}.
In particular the asymptotic behaviours of the mixed propagators will be carefully examined.
We focus our analysis on the dimensionless building blocks \eqref{mix-ncs}, and the results of the simplest mixed propagator \eqref{mix0} can be retrieved by  taking $n=0$ and $c_s=1$.  We mainly use $\hat\cG_+^{(n)} $ for demonstration, while the results of $\hat\cG_-^{(n)} $ are given by its complex conjugate.

Explicitly, \eqref{mix-ncs} can be further written as
\be \label{mix-ncs2}
\hat\cG_+^{(n)} (k \eta ; c_s) 
= -\frac{i c_s^2 k^{2+n}}{2H^{2}}  \left[
\s_k(\eta) D^{(n)*}(\eta;c_s)
+\s_k^*(\eta) \tilde{D}^{(n)}(\eta;c_s)
-(-1)^n\s_k^*(\eta) D^{(n)}(0;c_s)
\right]
~.
\ee
where we have introduced two types of integrals as
\bea
D^{(n)}(\eta;c_s)  &\equiv & \int^\eta_{-\infty} d\eta' \frac{1}{(-\eta')^{2-n}} \sigma_k(\eta') e^{-ic_s k\eta'}~,\\
\tilde{D}^{(n)}(\eta;c_s) &\equiv & \int^0_{\eta} d\eta' \frac{1}{(-\eta')^{2-n}} \sigma_k(\eta') e^{ic_s k\eta'}~.
\eea

The asymptotic behaviours of the mixed propagators can be analyzed by looking into the early-time and late-time limits of the two integrals above.
First of all, the soft behaviour corresponds to the late-time limit $\eta\rightarrow0$, where the second term in \eqref{mix-ncs2} vanishes and thus
\bea
\lim_{k\rightarrow 0} \hat\cG_+^{(n)} (k \eta; c_s) &=& 
-i\frac{c_s^2k^{2+n}}{2 H^{2}} \left[ \sigma_k(\eta) D^{(n)*}(0;c_s)  -(-1)^n \sigma^*_k(\eta) D^{(n)}(0;c_s)  \right]~.
\eea
The $D^{(n)}$ integral can be solved analytically for $\eta\rightarrow0$ \cite{Arkani-Hamed:2018kmz}
\be  \label{Dn0}
D^{(n)}(0;c_s)  = \frac{-i H}{\sqrt{2k}}\(\frac{i}{2k}\)^n \Xi_n(\mu, c_s)~,
\ee
where we have introduced a new function
\be \label{Xixi}
\Xi_n(\mu, x)  \equiv  \frac{\Gamma \({1\over 2} +n -i\mu\)\Gamma \({1\over 2} +n +i\mu\)}{\Gamma(n+1)}
{}_{2}F_1 
\[ {1\over 2} +n -i\mu, {1\over 2} +n +i\mu; 1 +n ; \frac{1-x}{2} \] ~.
\ee
In the end the soft limit is given by
\bea \label{softKncs}
\lim_{k\rightarrow 0} \hat\cG_+^{(n)} (k \eta; c_s) &=& 
 \(-\frac{i}{2}\)^{n} \sum_\pm A^{(n)}_\pm \(\frac{-k\eta}{2}\)^{\frac{3}{2}\pm i\mu}  ~, 
\eea
with
\be
A^{(n)}_\pm =    \frac{c_s^2\sqrt{\pi} }{2\sinh(\pi\mu)} \Xi_n(\mu, c_s)   \frac{e^{\frac{\pi\mu}{2}\mp\frac{i\pi}{4}}(1\mp ie^{-\pi\mu})}{\Gamma(1\pm i\mu)}~.
\ee
For $c_s=1$ and $n=0$, we obtain \eqref{softK} for the simplest mixed propagator. As we can see, the effect of the nontrivial sound speed in this limit is contained in the overal prefactor $c_s^2 \Xi_n$.

Next we consider the early-time behaviour of $ \hat\cG_+^{(n)}$, with $\eta\rightarrow-\infty$. In this limit the first term in \eqref{mix-ncs2} vanishes, and we have 
\be
\lim_{\eta\rightarrow-\infty}\hat\cG_+^{(n)} (k \eta ; c_s) 
= -\frac{i c_s^2 k^{2+n}}{2H^{2}}  \left[
 \tilde{D}^{(n)}(\eta\rightarrow-\infty;c_s)
-(-1)^n D^{(n)}(0;c_s)
\right]\s_k^*(\eta)
~.
\ee
If the two integrals in the bracket are regular in this limit, the early-time behaviour of the mixed propagators is reflected by the $\s$ mode function. As we have seen in \eqref{Dn0}, the $D^{(n)}(0;c_s)$ integral just produces a time-independent prefactor. But for the $\tilde{D}^{(n)}(\eta;c_s)$ integral, we need more analysis for its $\eta\rightarrow-\infty$ limit. 
When $c_s\neq 1$, this integral is also well-behaved
\be
 \lim_{\eta\rightarrow-\infty} \tilde{D}^{(n)}(\eta;c_s\neq 1)  = \frac{-i H}{\sqrt{2k}}\(\frac{i}{2k}\)^n \Xi_n(\mu, -c_s)~,
\ee
where $\Xi_n$ is defined in \eqref{Xixi}.
Thus the two $\tilde{D}^{(n)}$ and ${D}^{(n)}$ integrals in the bracket provide a normalization factor, and the early-time limit of the mixed propagator is determined by the $\s$ field.

When $c_s=1$ we see the hypergeometric function in $\Xi_n(\mu,-1)$  diverges.
This singularity is expected from the bulk perspective. In the $c_s\rightarrow1$ limit the  bulk integral probes the early-time behaviour of two fields, and becomes
\baa
 \lim_{c_s\rightarrow1}
 \frac{-iH}{\sqrt{2k}} \int^0_{-\infty} d\eta' \frac{1}{(-\eta')^{1-n}}  e^{i(c_s-1) k\eta'} = -\frac{iH}{\sqrt{2k}}\times
 \begin{cases} \displaystyle \frac{\Gamma(n)}{(i(c_s-1)k)^n}  & n> 0\,,\\[12pt] \displaystyle \frac{[-ik(c_s-1)]^{|n|}}{\Gamma(1+|n|)}\log\[(1-c_s)k\] & n \le 0\, .
	\end{cases} 
\end{align}
 To analyze this singularity, we notice that when $c_s=1$ the $\tilde{D}^{(n)}(\eta;c_s=1)$ integral can be explicitly solved as
\be
\tilde{D}^{(n)}(\eta;c_s=1)= \frac{iH e^{\frac{i\pi}{4}-\frac{\pi\mu}{2}} }{2(2k)^{\frac{1}{2}+n}} \Big[ \mathcal{E}_-{\rm csch}\pi\mu - \mathcal{E}_+(1+\coth\pi\mu)
 \Big]~,
\ee
with
\be
\mathcal{E}_{\pm}= (-2k\eta)^{\frac{1}{2}+n\pm i\mu}  {\Gamma(\frac{1}{2}\pm i\mu)} \Gamma(\frac{1}{2}+n\pm i\mu)
{}_2F_2\Bigg[\begin{array}{c} \frac{1}{2}\pm i\mu,\frac{1}{2}+n\pm i\mu 
\\[2pt] \frac{3}{2}+n\pm i\mu, 1\pm 2{i\mu}\end{array}\Bigg|\, 2ik\eta \Bigg]~.
\ee
Then using the asymptotic expansion at $\eta\rightarrow-\infty$, we find 
\begin{align} 
\lim_{\eta\rightarrow-\infty} \tilde{D}^{(n)}(\eta;c_s=1) = -\frac{iH}{\sqrt{2k}}\times
 \begin{cases} \displaystyle \log(-2k\eta)  & ~~~~~~ n= 0\,,\\[12pt] \displaystyle \frac{(-k\eta)^n}{nk^n} & ~~~~~~ n \geq 1\, .
	\end{cases} 
\end{align}

To summarize, for $c_s\neq 1$ the early-time limit of $ \hat\cG_+^{(n)}$ has the same behaviour as the $\s$ field
\be \label{Kncs-early}
\lim_{\eta\rightarrow-\infty}  \hat\cG_+^{(n)} 
= \frac{c_s^2}{2} \(-\frac{i}{2} \)^{n+1} \[ \Xi_n(\mu,c_s)+ (-1)^{n+1} \Xi_n(\mu,-c_s) \]  k\eta e^{ik\eta}~,
\ee
while for $c_s= 1$, this limit is deformed into 
\begin{align} \label{Kn-early}
\lim_{\eta\rightarrow-\infty} \hat\cG_+^{(n)} = 
 \begin{cases} \displaystyle \frac{i}{4}k\eta\log(-2k\eta) e^{ik\eta}  & ~~~~~~ n= 0\,,\\[12pt] \displaystyle -\frac{i}{4n} (-k\eta)^{n+1} e^{ik\eta} & ~~~~~~ n \geq 1\, .
	\end{cases} 
\end{align}
The $n=0$ case gives us the asymptotic behaviour of the simplest mixed propagator dressed with a logarithmic function in \eqref{tauinf}, while for quadratic interactions with higher derivatives ($n\geq1$), the deformation of the early-time limit is of the power-law form.

\newpage

\section{Details of the Generalized Scalar Seeds}\label{app:seeds}

In this Appendix, we present more technical details of the generalized scalar seeds of Section \ref{sec:gen-seeds}. We first derive the explicit expressions for the particular and homogeneous solutions of $\mathcal{\hat I}^{(n)}$ in Section \ref{app:Insol}, and then examine their singularity structure in Section \ref{app:Insin}. The analysis is in agreement with the recursive relations derived in Section \ref{sec:moremix}.

\subsection{Explicit Solutions} \label{app:Insol}
In Section \ref{sec:gen-seeds}, we have derived the differential equation of the generalized scalar seeds as
\be 
\[ \Delta_{u} +\(\mu^2+ \frac{1}{4}\) \] \mathcal{\hat I}^{(n)} 
= \alpha_n \hat{\mathcal{C}}^{(n)}~,
\ee
where $\alpha_n=(-i)^{n-1} n! c_s^2$ and the source is given by the contact term
\be
\hat{\mathcal{C}}^{(n)} = \(\frac{u}{1+c_s u}\)^{n+1}~.
\ee
Like the case of the primary scalar seed, its $n$-th order solution is comprised of the homogeneous part and the particular part
\be
\mathcal{\hat I}^{(n)} = \mathcal{\hat H}^{(n)}  + \mathcal{\hat S}^{(n)} ~,
\ee
where $\mathcal{\hat S}^{(n)}$ is a series expansion that vanishes at $u\rightarrow0$ and the homogeneous solution $\mathcal{\hat H}^{(n)}$ can be expressed as hypergeometric functions.
In the following we shall derive these two solutions for arbitrary non-negative integer $n$ separately.

\paragraph{Homogeneous solutions}
First let us look at $\mathcal{\hat H}^{(n)}$.
As the left-hand side of the differential equation \eqref{seedneqcs} remains the same with the one of the primary scalar seed, the $n$-th order homogeneous solutions are still the ones in \eqref{homo-0} but with different coefficients $C_{\pm}^{(n)}$
\be \label{homo-n}
\mathcal{\hat H}^{(n)}  = -\frac{\alpha_n}{2^{n+1}}\sum_{\pm}
C_{\pm}^{(n)} \( \frac{iu}{2\mu} \)^{\frac{1}{2}\pm i\mu} {}_{2}F_1 
\[ \frac{1}{4}\pm \frac{i\mu}{2}, \frac{3}{4}\pm \frac{i\mu}{2}; 1 \pm {i\mu} ; u^2 \]~.
\ee
This solution is responsible for the non-analytic behaviour of $\mathcal{\hat I}^{(n)}$ around $u=0$
\be \label{softHn}
\lim_{u\rightarrow0} \mathcal{\hat H}^{(n)}  = -\frac{\alpha_n}{2^{n+1}}\sum_{\pm}
C_{\pm}^{(n)} \( \frac{iu}{2\mu} \)^{\frac{1}{2}\pm i\mu}~,
\ee
and thus the two free coefficients can be fully fixed by imposing the boundary condition in this limit. 
To do so, let us get back to the integral expression of the generalized scalar seeds  \eqref{seedcs}. 
The $u\rightarrow0$ limit of the integral can be analytically solved by using the $k_3 \rightarrow 0$ limit of the mixed propagator in \eqref{softKncs}
\be
\lim_{u\rightarrow 0} \mathcal{\hat I}^{(n)} 
= - \frac{\alpha_n}{2^{n+1}}\sum_{\pm}B_\pm^{(n)}\(\frac{u}{2}\)^{\frac{1}{2}\pm i\mu},~~~~{\rm with} ~~ B_\pm^{(n)} = \sqrt{\pi}\Xi_n \(1\mp\frac{i}{\sinh\pi\mu}\)
 \frac{\Gamma(\frac{1}{2}\pm i\mu)}{\Gamma(1\pm i\mu)}~,
\ee
where $\Xi_n$ is introduced in \eqref{Xixi}. Therefore matching the soft limit of $\mathcal{\hat I}^{(n)}$ with the one in \eqref{softHn}, we find $C_\pm^{(n)} = (-i\mu)^{\frac{1}{2}\pm i\mu} B_\pm^{(n)}$, which fully fixes the solution.
It is easy to check that this generalized solution returns to the one of the primary scalar seed for $n=0$ and $c_s=1$ in \eqref{homo-0}. As we can see here, the modifications from $n>0$ and $c_s\neq1$ are presented in an overall prefactor of the homogeneous solutions $\alpha_n\Xi_n(\mu,c_s)/2^n$, while the $u$-dependence of the seed function remains unaffected.
We also notice that, there is a logarithmic singularity in the hypergeometric functions when we take $u^2\rightarrow1$, as the sum of their first two parameters is equal to the third.
The $u\rightarrow-1$ limit gives a partial-energy pole of the three-point function, while the $u\rightarrow1$ singularity is unphysical and will be cancelled by the series solution. 
We leave the detailed discussion in the next subsection.

\paragraph{Particular solutions}
Motivated by the series expansion of the contact term around $u=0$, we propose the $n$-th order particular solution as 
\be \label{series-ncs}
\mathcal{\hat S}^{(n)}(u) = \alpha_n \sum_{m=0}^{\infty} c_m^{(n)} u^{n+m+1}~.
\ee
Substituting this ansatz into the differential equation \eqref{seedneqcs}, we find the following recursive relation of the series coefficients
\be
\[ \(m+n+\frac{5}{2}\)+\mu^2 \]c_{m+2}^{(n)} =  (m+n+1)(m+n+2) c_{m}^{(n)} + \frac{(m+n+2)!}{n!(m+2)!}(-c_s)^{m+2}
\ee
with the first two given by
\be
\[ \(n+\frac{1}{2}\)^2 + \mu^2\] c_0^{(n)}=1~, ~~~~~~ \[ \(n+\frac{3}{2}\)^2 + \mu^2\] c_1^{(n)}=-(n+1)c_s~.
\ee
Solving the above relation, we obtain
\be
c_m^{(n)} = \sum_{l=0}^{\lfloor m/2\rfloor}
\frac{(-c_s)^{m-2l} (m+n)! /((m-2l)!n!)}{\[\(m+n+\frac{1}{2}\)^2 + \mu^2\]\[\(m+n-\frac{3}{2}\)^2 + \mu^2\]...\[\(m+n-2l+\frac{1}{2}\)^2 + \mu^2\]}~.
\ee
Again we can simply check that this solution returns to the one of the primary scalar seed in \eqref{partisol1} by taking $n=0$ and $c_s=1$.
The effects of general $n$ and $c_s$ become nontrivial in the series solution.
We should notice that the above series is convergent for $|u|<1$, but
may diverge if we consider the regime with $|u|\geq 1$, which is possible when $c_s<1$. In such a situation, one may need to consider the analytical continuation of this series, which is discussed in detail by \cite{Jazayeri:2022kjy}.

For the convenience of singularity analysis, we redefine $m\rightarrow 2m+l$ and $m-2l\rightarrow l$, and express the series solution in the following form
\be
\mathcal{\hat S}^{(n)} =(-i)^{n-1} n! c_s^2
\sum_{m,l=0}^{\infty} c^{(n)}_{ml} u^{2m+l+n+1}
\ee
with
\bea \label{cnml}
c^{(n)}_{ml} &=&  
\frac{(-c_s)^{l} (2m+l+n)! / l!n!}{\[\(2m+l+n+\frac{1}{2}\)^2 + \mu^2\]\[\(2m+l+n-\frac{3}{2}\)^2 + \mu^2\]...\[\(l+n+\frac{1}{2}\)^2 + \mu^2\]} \nn\\
&=& \frac{(-c_s)^{l} (l+1)_{2m+n} \Gamma\( \frac{l}{2}+\frac{1}{4}+\frac{n}{2}-\frac{i\mu}{2}\)  \Gamma\( \frac{l}{2}+\frac{1}{4}+\frac{n}{2} +\frac{i\mu}{2}\)}{4^{m+1} n! \Gamma\( m+\frac{l}{2}+\frac{5}{4}+\frac{n}{2}-\frac{i\mu}{2}\)  \Gamma\( m+\frac{l}{2}+\frac{5}{4}+\frac{n}{2} +\frac{i\mu}{2}\)  }~,
\eea
where in the second line we have used the identity
\be
\(a+\frac{1}{2}\)^2 + \mu^2  = 4 \frac{\Gamma\(\frac{a}{2}+\frac{5}{4}-\frac{i\mu}{2}\)\Gamma\(\frac{a}{2}+\frac{5}{4}+\frac{i\mu}{2}\)}{\Gamma\(\frac{a}{2}+\frac{1}{4}-\frac{i\mu}{2}\)\Gamma\(\frac{a}{2}+\frac{1}{4}+\frac{i\mu}{2}\)}~.
\ee
As a consistency check, the above form of the series solution with $n=0$ reproduces (3.27) in \cite{Arkani-Hamed:2018kmz} by setting $v=1/c_s$ there.

\subsection{Singularity Structure}
\label{app:Insin}

For the differential equation \eqref{seedneqcs}, there can be three singularities in its solution at $u\rightarrow0$ and $u\rightarrow\pm 1$. While the non-analytical behaviour in the soft limit $u\rightarrow0$ has been used to fix the homogeneous solution, now we analyze the singularity structure of the solution at $u\rightarrow\pm1$.
We shall focus on the boundary solutions, but since these limits can also be computed from the bulk integration, we shall compare the results there, which provides a consistency check of our derivation.

First, let us look at the homogeneous solution.  It is straightforward to approach its $u\rightarrow\pm 1$ limits   by considering the singular behaviour of the hypergeometric functions
\be \label{hyperg-u1}
\lim_{u^2\rightarrow1} {}_{2}F_1 
\[ \frac{1}{4}\pm \frac{i\mu}{2}, \frac{3}{4}\pm \frac{i\mu}{2}; 1 \pm {i\mu} ; u^2 \] = -\frac{\Gamma\(1 \pm {i\mu} \)}{\Gamma\(  \frac{1}{4}\pm \frac{i\mu}{2}\)\Gamma\(  \frac{3}{4}\pm \frac{i\mu}{2}\)} \log(1-u^2)~.
\ee
With the fixed coefficients $\hat{C}^{(n)}_\pm$, the logarithmic singularity of $\mathcal{\hat H}^{(n)}$ at $u=1$ is given by
\be \label{homo-nu1}
\lim_{u\rightarrow1}\mathcal{\hat H}^{(n)} = \frac{\alpha_n}{2^{n+1}n!} \Xi_n(\mu, c_s)\log(1-u)~.
\ee
When $c_s=1$ this corresponds to a folded pole at $k_3=k_{12}$, but in general it is the limit where the energy of the exchanged particle equals to the total energy of two external fields in the cubic vertex. This singularity should be absent in physical solutions with the standard Bunch-Davies vacuum.
As we shall show soon, it will be cancelled by the $u\rightarrow1$ singularity in the series solution. 
Similarly, the $u\rightarrow-1$ limit is given by
\be \label{HnkT}
\lim_{u\rightarrow-1}\mathcal{\hat H}^{(n)} =-\frac{1}{2^{n+1}n!} \alpha_n\Xi_n(\mu, c_s)\log(1+u)~,
\ee
which is a partial energy singularity at $E_L=k_3+c_sk_{12}\rightarrow0$.

Next, we examine the singular behaviours of the series solution, which are not trivially manifested.
Here we follow the analysis for the four-point scalar seed in \cite{Arkani-Hamed:2018kmz} and extend it to the situation with general $n$ and $c_s$.
To analyze the singular behaviours around $u^2\rightarrow1$, let us  consider the first derivative of the series solution 
\be \label{partialSn}
\partial_u \mathcal{\hat S}^{(n)} 
= \alpha_n \sum_{m,l=0}^{\infty} (2m+l+n+1) c^{(n)}_{ml} u^{2m+l+n}~.
\ee
With the form of the  coefficients in \eqref{cnml}, the sum over $l$ in the above series can be expressed as
\bea
\sum_{l=0}^{\infty} (2m+l+n+1) c^{(n)}_{ml} u^{l}
&=& \sum_{l=0}^{\infty} \frac{ (l+1)_{2m+n+1} \Gamma\( \frac{l}{2}+\frac{1}{4}+\frac{n}{2}-\frac{i\mu}{2}\)  \Gamma\( \frac{l}{2}+\frac{1}{4}+\frac{n}{2} +\frac{i\mu}{2}\)(-c_s u)^{l} }{4^{m+1} n! \Gamma\( m+\frac{l}{2}+\frac{5}{4}+\frac{n}{2}-\frac{i\mu}{2}\)  \Gamma\( m+\frac{l}{2}+\frac{5}{4}+\frac{n}{2} +\frac{i\mu}{2}\) } \nn\\
&=&  F_1^{(n)} - c_s u (F_2^{(n)}+F_3^{(n)} )+c_s^2 u^2 F_4^{(n)}  
\eea
where we have introduced the following   functions
\begin{align}
	F_1^{(n)} & \equiv\frac{\Gamma(2+2m+n)}{4^{1+m}n!(\frac{1}{4}+\frac{n}{2}-\frac{i\mu}{2})_{1+m}(\frac{1}{4}+\frac{n}{2}+\frac{i\mu}{2})_{1+m}} \nn\\[5pt]
	& ~~~~~~~~ \times {}_4F_3\Bigg[\begin{array}{c} \frac{1}{2}+\frac{n}{2}+m,1+\frac{n}{2}+m,\frac{1}{4}+\frac{n}{2}-\frac{i\mu}{2},\frac{1}{4}+\frac{n}{2}+\frac{i\mu}{2}\\[2pt] \frac{1}{2},\frac{5}{4}+\frac{n}{2}+m-\frac{i\mu}{2},\frac{5}{4}+\frac{n}{2}+m+\frac{i\mu}{2} \end{array}\Bigg|\, c_s^2{u^2}\Bigg] \, , \\[5pt]
	F_2^{(n)} &  \equiv\frac{(2)_{2m+n}}{4^{1+m}n!(\frac{3}{4}+\frac{n}{2}-\frac{i\mu}{2})_{1+m}(\frac{3}{4}+\frac{n}{2}+\frac{i\mu}{2})_{1+m}} \nn\\[5pt]
	& ~~~~~~~~ \times	{}_4F_3\Bigg[\begin{array}{c} 1+\frac{n}{2}+m,\frac{3}{2}+\frac{n}{2}+m,\frac{3}{4}+\frac{n}{2}-\frac{i\mu}{2},\frac{3}{4}+\frac{n}{2}+\frac{i\mu}{2}\\[2pt] \frac{1}{2},\frac{7}{4}+\frac{n}{2}+m-\frac{i\mu}{2},\frac{7}{4}+\frac{n}{2}+m+\frac{i\mu}{2} \end{array}\Bigg|\, c_s^2{u^2}\Bigg]\, ,\\[5pt]
	F_3^{(n)} & \equiv\frac{(2m+n+1)(2)_{2m+n}}{4^{1+m}n!(\frac{3}{4}+\frac{n}{2}-\frac{i\mu}{2})_{1+m}(\frac{3}{4}+\frac{n}{2}+\frac{i\mu}{2})_{1+m}} \nn\\[5pt]
	& ~~~~~~~~ \times   {}_4F_3\Bigg[\begin{array}{c} 1+\frac{n}{2}+m,\frac{3}{2}+\frac{n}{2}+m,\frac{3}{4}+\frac{n}{2}-\frac{i\mu}{2},\frac{3}{4}+\frac{n}{2}+\frac{i\mu}{2}\\[2pt] \frac{3}{2},\frac{7}{4}+\frac{n}{2}+m-\frac{i\mu}{2},\frac{7}{4}+\frac{n}{2}+m+\frac{i\mu}{2} \end{array}\Bigg|\, c_s^2{u^2} \Bigg]\, ,\\[5pt]
	F_4^{(n)} & \equiv \frac{\Gamma(3+2m+n)}{4^{1+m}n!(\frac{5}{4}+\frac{n}{2}-\frac{i\mu}{2})_{1+m}(\frac{5}{4}+\frac{n}{2}+\frac{i\mu}{2})_{1+m}} \nn\\[5pt]
	& ~~~~~~~~ \times   {}_4F_3\Bigg[\begin{array}{c} \frac{3}{2}+\frac{n}{2}+m,2+\frac{n}{2}+m,\frac{5}{4}+\frac{n}{2}-\frac{i\mu}{2},\frac{5}{4}+\frac{n}{2}+\frac{i\mu}{2}\\[2pt] \frac{3}{2},\frac{9}{4}+\frac{n}{2}+m-\frac{i\mu}{2},\frac{9}{4}+\frac{n}{2}+m+\frac{i\mu}{2} \end{array}\Bigg|\, c_s^2{u^2}\Bigg]\, .
\end{align}
To check the singular behaviours around $u\rightarrow\pm 1$, we are interested in the large $m$ limit of the $\partial_u \mathcal{\hat S}^{(n)} $ series. In the above expression, only $F_1^{(n)}$ and $F_3^{(n)}$ will contribute when $m\rightarrow\infty$, which leads to 
\bea
&&\lim_{m\rightarrow\infty} \[ F_1^{(n)} - c_s u (F_2^{(n)}+F_3^{(n)} )+c_s^2 u^2 F_4^{(n)} \] \nn\\
&&~~~~ = 2^{n-1}\frac{\Gamma(\frac{1}{4}+\frac{n}{2}-\frac{i\mu}{2})\Gamma(\frac{1}{4}+\frac{n}{2}+\frac{i\mu}{2})}{n!\sqrt{\pi}}{}_2F_1\[ \frac{1}{4}+\frac{n}{2}-\frac{i\mu}{2},  \frac{1}{4}+\frac{n}{2}+\frac{i\mu}{2} ;\frac{1}{2} ; c_s^2 u^2 \] \nn\\
&&~~~~~~~~ 
 -2^{n} c_s u  \frac{\Gamma(\frac{3}{4}+\frac{n}{2}-\frac{i\mu}{2})\Gamma(\frac{3}{4}+\frac{n}{2}+\frac{i\mu}{2})}{n!\sqrt{\pi}}{}_2F_1\[ \frac{3}{4}+\frac{n}{2}-\frac{i\mu}{2},  \frac{3}{4}+\frac{n}{2}+\frac{i\mu}{2} ;\frac{3}{2} ; c_s^2 u^2 \]~.
\eea
The above expression can be further simplified by using an identity of hypergeometric functions
\bea
{}_2F_1\[ 2\alpha, 2\beta; \alpha+\beta +\frac{1}{2} ; \frac{1+x}{2}\] &=& \frac{\Gamma\( \alpha+\beta +\frac{1}{2}\)\Gamma\(\frac{1}{2}\)}{\Gamma\( \alpha +\frac{1}{2}\)\Gamma\( \beta +\frac{1}{2}\)}
{}_2F_1\[ \alpha, \beta; \frac{1}{2} ; x^2\] \nn\\
&&-x\frac{\Gamma\( \alpha+\beta +\frac{1}{2}\)\Gamma\(-\frac{1}{2}\)}{\Gamma\( \alpha\)\Gamma\( \beta \)}
{}_2F_1\[ \alpha +\frac{1}{2}, \beta+\frac{1}{2}; \frac{3}{2} ; x^2\]~,
\eea
which leads us to
\be
\lim_{m\rightarrow\infty} \[ F_1^{(n)} - c_s u (F_2^{(n)}+F_3^{(n)} )+c_s^2 u^2 F_4^{(n)} \] =  \frac{1}{2^{n}n!}  \Xi_n (\mu, c_s u)~,
\ee
with the $\Xi_n$ function defined in \eqref{Xixi}. Therefore, when $u\rightarrow\pm 1$, the  $\partial_u \mathcal{\hat S}^{(n)} $ series is dominated by
\be \label{duS}
\lim_{u\rightarrow\pm 1}\partial_u \mathcal{\hat S}^{(n)} 
\rightarrow  \frac{\alpha_n}{2^{n}n!}  \Xi_n (\mu, c_s u) u^n \sum_{m=0}^{\infty}   u^{2m}=  \frac{\alpha_n}{2^{n}n!}  \Xi_n (\mu, c_s u)  \frac{u^n}{1-u^2}~.
\ee
The singular behaviours of the series solution can be solved as
\be
\lim_{u\rightarrow\pm 1} \mathcal{\hat S}^{(n)} 
=  \frac{\alpha_n u^{n+1}}{2^{n}(n+1)!}  \Xi_n (\mu, c_s u)~{}_2F_1\[1, \frac{n+1}{2};1+\frac{n+1}{2} ;u^2\] 
~,
\ee
where we have assumed that $\Xi_n (\mu, c_s u)$ approaches to constants when $u\rightarrow\pm 1$. This is generally valid except for the case with $c_s=1$ and $u\rightarrow-1$, where $\Xi_n(\mu,-1)$ diverges. 
We leave this special case at the end of the discussion.
As we see, the hypergeometric function has logarithmic divergence when $u^2\rightarrow1$.

Let us look at the singularities around $u=1$ and $u=-1$ respectively.
The $u\rightarrow1$ limit becomes
\be
\lim_{u\rightarrow 1} \mathcal{\hat S}^{(n)} 
=-\frac{\alpha_n}{2^{n+1}n!}  \Xi_n (\mu, c_s)
 \log(1-u)~,
\ee
which has an unphysical logarithmic singularity. This is exactly cancelled by the homogeneous solution in the same limit in \eqref{homo-nu1}, and thus the final solution is regular in this limit as expected.
For the $u\rightarrow-1$ limit with $c_s\neq1$, we get 
\be
\lim_{u\rightarrow -1} \mathcal{\hat S}^{(n)} =-\frac{\alpha_n (-1)^{n+1}}{2^{n+1}n!}  \Xi_n (\mu, -c_s)\log(1+u)~,
\ee
which is a $E_L$-singularity same with the one in the homogeneous solution \eqref{HnkT}. Combining these two, the $u\rightarrow-1$ singularity of the full solution is given by
\be \label{kTpole-cs}
\lim_{u\rightarrow -1}\( \mathcal{\hat H}^{(n)} + \mathcal{\hat S}^{(n)}  \) = -\frac{\alpha_n}{2^{n+1}n!}  \[\Xi_n (\mu, c_s) +(-1)^{n+1} \Xi_n (\mu, -c_s)\] \log(1+u)~.
\ee
We can also derive this singularity form the bulk perspective. 
Since the $u\rightarrow-1$ limit picks up the $\eta\rightarrow-\infty$ contribution of the bulk integral in \eqref{seedcs}, the early-time limit of the mixed propagator in \eqref{Kncs-early} shall be used. Thus the bulk computation of the generalized scalar seeds leads to
\be
\lim_{u\rightarrow -1}  \mathcal{\hat I}^{(n)} =-\frac{\alpha_n}{2^{n+2}n!}  \[\Xi_n (\mu, c_s) +(-1)^{n+1} \Xi_n (\mu, -c_s)\] 
\int_{-\infty}^0\frac{d\eta}{\eta} \( e^{i E_L \eta} +e^{-i E_L\eta} \)
~,
\ee
where the integral gives $2\log E_L$. This agrees with what we find in the boundary solution in \eqref{kTpole-cs}.
Thus with a nontrivial sound speed, the partial energy pole of the three-point function with a mixed propagator differs from its total energy pole. 

Now let us analyze the special case of the $u\rightarrow-1$ limit with $c_s=1$, where the $E_L$- and $k_T$ singularities coincide with each other. 
Before that, we remind ourselves that the singular behaviour of the homogeneous solution in the $u\rightarrow-1$ limit is given by $\log(1+u)$ in \eqref{HnkT}. 
As we shall see now, this singularity is subdominated compared with the one from series solution in the same limit.
We start from the $\partial_u\mathcal{\hat S}^{(n)}$ series in \eqref{duS}, and notice that there the $\Xi_n$ function becomes singular
\begin{align}
\lim_{u\rightarrow-1} \Xi_n (\mu,  u) = 
 \begin{cases} \displaystyle -\log(1+u)  & ~~~~~~ n= 0\,,\\[12pt] \displaystyle \frac{2^n \Gamma(n) }{(1+u)^n}  & ~~~~~~ n \geq 1\, .
	\end{cases} 
\end{align}
Thus for $n=0$, the $u\rightarrow-1$ singularity of the series solution is solved as
\be
\lim_{u\rightarrow-1}  \mathcal{\hat S}^{(0)}(u)  = - \frac{i}{4}  {\log(1+u)}^2~.
\ee
This gives us the $k_T$-pole of the primary scalar seed in \eqref{series0-kT}, and agrees with result from the bulk integration in \eqref{I0-kT}.
For $n>0$, the singular behaviour becomes
\be \label{InkT}
\lim_{u\rightarrow-1}  \mathcal{\hat S}^{(n)}(u)  = \frac{\alpha_n}{2n^2}\frac{(-1)^{n+1}}{(1+u)^n}~,
\ee
which is an $n$-th order $k_T$-pole in the three-point functions.
From the bulk perspective, we solve the integral of the generalized scalar seeds by using the early-time limit of the mixed propagator in \eqref{Kn-early}, and get
\be
\lim_{u\rightarrow -1}  \mathcal{\hat I}^{(n)} = -\frac{i}{4n} 
\int_{-\infty}^0 \frac{d\eta}{k_3\eta^2} (-k_3\eta)^{n+1} \[ e^{ik_T\eta} + (-1)^ne^{ik_T\eta} \] = -\frac{\alpha_n}{2n^2} \(\frac{k_3}{k_T}\)^n
~,
\ee
which is in precise agreement with \eqref{InkT}.
Thus for $c_s=1$ the leading $k_T$-pole is dominated by the contribution from the series solution.

\newpage
\section{Comments on Diagrams beyond Single-Exchange}
\label{app:double}

\begin{figure} [h]
   \centering
            \includegraphics[width=.7\textwidth]{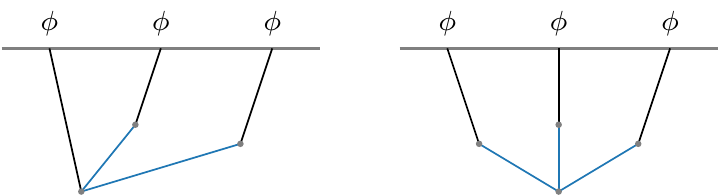}
   \caption{The double-exchange ({\it left}) and triple-exchange ({\it right}) diagrams of the inflaton bispectrum. The blue internal lines are the exchanged massive field $\s$.}
  \label{fig:double}
\end{figure}

In this Appendix, we briefly comment on the double-exchange and triple exchange diagrams with mixed propagators, which can also contribute to the inflationary three-point functions.
Figure \ref{fig:double} shows their Feynman diagrams.
They arise in theories with two or three massive fields in the cubic vertices.

In the bootstrap of the single-exchange diagram, one key observation is that, the line of the exchanged massive field can be ``collapsed" by some differential operation, which leads to the differential equation of the three-point scalar seed. This is an inhomogeneous equation, whose source is the contact three-point function with the ``collapsed" line being removed. The same strategy also works in double-exchange and triple-exchange diagrams.
In the following, we shall take the double-exchange diagram as an example, and derive the differential equations satisfied by the corresponding bispectrum.

For the double-exchange diagram, the leading cubic vertex is given by $\dot\phi\sigma^2$. Using the mixed propagator \eqref{mix0}, we find the bispectrum
\bea
\langle \phi_{\bf k_1} \phi_{\bf k_2} \phi_{\bf k_3} \rangle' 
&\sim & i \int_{-\infty}^{0} d\eta a(\eta)^3\[
\partial_\eta K_+ (k_1, \eta) \cG_+ (k_2, \eta)
\cG_+ (k_3, \eta) -c.c. \]\nn\\
& =&  \frac{-iH}{2 k_2^3 k_3^3} \mathcal{\hat I}_{\rm DE} (k_{1}, k_2, k_3)
\eea
where the double-exchange bulk integral is defined as
\be
\mathcal{\hat I}_{\rm DE} (k_1, k_2, k_3) = \frac{1}{k_1} \int_{-\infty}^{0}  \frac{d\eta}{\eta^2}
\[e^{i k_{1}\eta}\hat\cG_+ (k_2 \eta) 
\hat\cG_+ (k_3 \eta) - c.c.\]~.
\ee
Using the differential equation \eqref{eqGkm}, we can ``collapse" the exchanged massive fields in the two mixed propagators. Let us first do this for the $k_2$ and $k_3$ legs separately, which gives two second order differential equations 
\be \label{dk3}
\frac{1}{k_1}\(k_3^2 \partial^2_{k_3} - 2 k_3 \partial_{k_3} - k_3^2\partial^2_{k_{1}} +\frac{m^2}{H^2}\)  k_1 \mathcal{\hat I}_{\rm DE}   = -\frac{1}{2} \frac{k_3^2}{k_1} \int_{-\infty }^{0} d\eta \[
e^{i k_{13}\eta}\hat\cG_+ (k_2 \eta) -c.c.\]
\ee
\be  \label{dk2}
\frac{1}{k_1}\(k_2^2 \partial^2_{k_2} - 2 k_2 \partial_{k_2} - k_2^2\partial^2_{k_{1}} +\frac{m^2}{H^2}\)  k_1 \mathcal{\hat I}_{\rm DE}   = -\frac{1}{2} \frac{k_2^2}{k_1} \int_{-\infty}^{0} d\eta \[
e^{i k_{12}\eta}\hat\cG_+ (k_3 \eta) -c.c.\]
\ee
Notice that $\mathcal{\hat I}_{\rm DE} $ depends on $k_1,~ k_2,~ k_3$ in the combinations $u\equiv k_3/k_1$ and $v\equiv k_2/k_1$. 
Next, we derive the differential equations in terms of these two new variables.
Let us first take a look at the source terms. These bulk integrals correspond to the single-exchange bispectrum from the $\dot\phi^2\s$ interaction. Basically, since we ``collapse" one internal line using the differential equation, the diagram becomes a single-exchange one. These two integrals can be derived from the primary scalar seed $\mathcal{\hat I}$ in \eqref{singleM} by using the weight-shifting operator
\be
\mathcal{\hat I}_{\rm SE} \(\frac{v}{1+u}\) \equiv {k_2} \int_{-\infty }^{0} d\eta \[
e^{i k_{13}\eta}\hat\cG_+ (k_2 \eta) -c.c.\] = -k_2^2 \partial_{k_{13}}^2\mathcal{\hat I}\(\frac{k_2}{k_{13}}\)
\ee
\be
\mathcal{\hat I}_{\rm SE} \(\frac{u}{1+v}\) \equiv {k_3} \int_{-\infty }^{0} d\eta \[
e^{i k_{12}\eta}\hat\cG_+ (k_3 \eta) -c.c.\] = -k_3^2 \partial_{k_{12}}^2\mathcal{\hat I}\(\frac{k_3}{k_{12}}\)
\ee
Next, in terms of $u$ and $v$, the differential equations in
\eqref{dk3} and \eqref{dk2} become
\begin{small}
\be \label{de2nd-a}
\[ u^2(1-u^2)\partial^2_{u} -2u \partial_{u} +\mu^2+\frac{9}{4}- u^2 \(  v^2\partial_v^2 +2 uv \partial_u \partial_v\)\]  \mathcal{\hat I}_{\rm DE} (u, v) = -\frac{1}{2} \frac{u^2}{v} \mathcal{\hat I}_{\rm SE} \(\frac{v}{1+u}\) ,
\ee
\be \label{de2nd-b}
\[ v^2(1-v^2)\partial^2_{v} -2v \partial_{v} +\mu^2+\frac{9}{4}- v^2 \( u^2\partial_u^2 +2 uv \partial_u \partial_v\)\]  \mathcal{\hat I}_{\rm DE} (u, v) = -\frac{1}{2} \frac{v^2}{u} \mathcal{\hat I}_{\rm SE} \(\frac{u}{1+v}\) .
\ee
\end{small}We may further ``collapse" the other internal line in the diagram by using the equation of mixed propagator  \eqref{eqGkm} once more.
Doing so, we derive one fourth order differential equation of $\mathcal{\hat I}_{\rm DE} $
\begin{small}
\be
\frac{1}{k_1}\(k_2^2 \partial^2_{k_2} - 2 k_2 \partial_{k_2} - k_2^2\partial^2_{k_{1}} +\frac{m^2}{H^2}\) \(k_3^2 \partial^2_{k_3} - 2 k_3 \partial_{k_3} - k_3^2\partial^2_{k_{1}} +\frac{m^2}{H^2}\)  k_1 \mathcal{\hat I}_{\rm DE} =\frac{i}{4}\frac{k_2^2 k_3^2}{k_1 k_T^3}~.
\ee
\end{small}
In terms of $u$ and $v$, this equation has the following compact form
\be \label{de4th}
\Delta_{v,u} \Delta_{u,v}  \mathcal{\hat I}_{\rm DE} (u, v) 
= \frac{i}{4}\frac{u^2 v^2}{(1+u+v)^3}
\ee
where the differential operator $\Delta_{u,v}$ is defined as
\be 
 \Delta_{u,v} \equiv u^2 \partial_u^2 -2 u \partial_u  - u^2 \( u^2\partial_u^2+v^2\partial_v^2 +2 uv \partial_u \partial_v\) +\mu^2+\frac{9}{4} .
\ee
Solving equations \eqref{de2nd-a} and \eqref{de2nd-b}, or equation \eqref{de4th} will help us determine the analytical form of the double-exchange bispectrum. We leave this nontrivial task for future work.

The triple-exchange bispectrum can be analyzed in a similar way. This time, the lowest derivative cubic vertex is the self-interaction $\s^3$ of the massive field, which gives the leading contribution in quasi-single field inflation \cite{Chen:2009zp}.
Since there are three internal lines to be ``collapsed" by the differential operation, in the end we expect a sixth order differential equation for the resulting bispectrum.

\newpage

\section{Spinning Fields in de Sitter}
\label{app:spin}

This Appendix is a brief summary for the free theory of spinning fields in de Sitter space.
In particular, we review the decomposition of spin-$s$ fields and present the results  of the helicity-0 modes,
whose derivation can be found in Appendix A of Ref. \cite{Lee:2016vti}.

For a massive spin-$s$ particle $\s_{\mu_1...\mu_s}$, its equation of motion in de Sitter space is given by
\be \label{spineom}
(\square-m^2_s)\s_{\mu_1...\mu_s} = 0~,
\ee
where $\square\equiv \nabla^\mu \nabla_\mu$ and $m_s^2 = m^2-(s^2-2s-2)H^2$. 
This massive spinning field is allowed to have a reduced sound speed $c_\s$. In \eqref{spineom} and the following analysis, we have absorbed it by rescaling $c_\s\partial_i\rightarrow\partial_i$, or equivalently $c_\s k \rightarrow k$ in Fourier space.
The tensor $\s_{\mu_1...\mu_s}$ should be  totally symmetric  and satisfies the constraints
\be \label{transverse}
\nabla^{\mu_1}\s_{\mu_1...\mu_s}=0~, ~~~~~~
\s^{\mu_1}_{~\mu_1...\mu_s}=0~.
\ee
It is convenient to expand the spinning field into its helicity eigenstates
\be
\sigma_{\mu_1...\mu_s} =\sum_{\lambda=-s}^{s}\sigma^{(\lambda)}_{\mu_1...\mu_s}~.
\ee
Next, we work with $\sigma^{(\lambda)}_{\mu_1...\mu_s}$ and project it onto spatial slices, i.e. $\sigma^{(\lambda)}_{i_1...i_n\eta...\eta} $.
By introducing the helicity-$\lambda$ polarization tensor with $n$ spatial indices $\epsilon_{i_1...i_n}^\lambda$, a helicity-$\lambda$ mode with $n$ polarization directions can be expressed as
\be
\sigma^{(\lambda)}_{i_1...i_n\eta...\eta} = \s^\lambda_{n,s} \epsilon_{i_1...i_n}^\lambda~.
\ee
From the equation of motion in \eqref{spineom}, the mode function of $\s^\lambda_{n,s}$ has different behaviour depending on the helicity $\lambda$ and the number of polarization directions (or the ``spatial spin") $n$.
For $n<|\lambda|$, $\s^\lambda_{n,s}=0$.
For $n=|\lambda|$,  $\s^\lambda_{|\lambda|,s}$ satisfies an equation of motion similar with the one of a massive scalar in \eqref{sigmaeom} and thus has a solution with one Hankel function. 
For $n>|\lambda|$, the equation of motion becomes complicated but the mode function of $\s^\lambda_{n,s}$ can be derived from $\s^\lambda_{|\lambda|,s}$ by using the transverse condition in \eqref{transverse}.

Since only the helicity-0 mode contributes to the cosmological collider signal in the scalar bispectra, next
we are mainly interested in this longitudinal mode with $\lambda=0$. To avoid clutter, we drop $\lambda$ in the indices and  introduce the following notation\footnote{Note that the lower index of $\s^{(n)}_s$ is the spin of the field, which differs from the momentum in the notation of the scalar mode function $\s_k$.}
\be
\s^{(n)}_s = \s^0_{n,s}~,~~~~~~\epsilon_{i_1...i_n}=\epsilon_{i_1...i_n}^0~.
\ee
Let us take the helicity-0 mode of spin-1 field $\s_\mu$ as an explicit example. In our notation, it is decomposed into
\be
\s_\eta =  \s^{(0)}_1~, ~~~~~~ \s_i =  \s^{(0)}_1 \epsilon_i~, ~~~~~~~~~~ {\rm with}~~\epsilon_i(\hat{\bf k})=\hat{k}_i~.
\ee
Similarly, for the spin-2 field $\s_{\mu\nu}$, its  helicity-0 longitudinal mode can be expressed as
\be
\s_{\eta\eta} =  \s^{(0)}_2~, ~~~~~~\s_{i\eta} =  \s^{(1)}_2\epsilon_i ~, ~~~~~~\s_{ij} =  \s^{(2)}_2\epsilon_{ij}~,
\ee
with
\be
\epsilon_{ij}(\hat{\bf k})= \frac{3}{2}\( \hat{k}_i\hat{k}_j-\frac{1}{3}\delta_{ij} \)~.
\ee
For spin-$s$, in general we have $\sigma_{i_1...i_n\eta...\eta} = \s^{(n)}_{s} \epsilon_{i_1...i_n} $ where the polarization tensor satisfies
\be
\hat{q}_{i_1}...\hat{q}_{i_n}\epsilon_{i_1...i_n}(\hat{\bf k}) = P_n(\hat{\bf q}\cdot\hat{\bf k})
\ee
with $P_n(\hat{\bf q}\cdot\hat{\bf k})$ being the Legendre polynomial.

When $n=0$, the $\s^{(0)}_{s} $ mode satisfies the following equation
\be
\(\mathcal{O}_\eta + \mu_s^2 +\frac{9}{4} \) \s^{(0)}_{s}  = 0~, ~~~~~~{\rm with}~~
\mu_s= \sqrt{\frac{m^2}{H^2}-\(s-\frac{1}{2}\)^2}~.
\ee
By assuming the Bunch-Davies initial condition and imposing orthonormality for normalization, this mode function is solved as
\bea
 \sigma_s^{(0)}(k,\eta)  =  -i e^{i\pi/4}e^{-\pi\mu_s/2} N_s k^s (- \eta)^{3/2}H^{(1)}_{i\mu_s}(-k\eta) 
 ~,
\eea
where 
 the normalization factor is given by
\be \label{Ns}
N_s=\frac{i}{H^{s}} \({\frac{s!\Gamma(\frac{1}{2}+i\mu_s)\Gamma(\frac{1}{2}-i\mu_s)}{(2s-1)!! \Gamma(s+\frac{1}{2}+i\mu_s)\Gamma(s+\frac{1}{2}-i\mu_s)} }\)^{1/2}~.
\ee
As explicit examples,
 for spin-1 and spin-2 fields the $N_s$ factor is given by
\be
N_1 = \frac{i}{m}~,~~~~~~
N_2 = i \sqrt{\frac{2}{3}} \frac{1}{H^2} {\[\(\frac{1}{4}+\mu_2^2\)\(\frac{9}{4}+\mu_2^2\)\]^{-1/2}} .
\ee
Notice that the equation for $\sigma_s^{(0)}$ is the same with the one of massive scalars in \eqref{sigmaeom}, and their mode functions differ only in normalization factors. Thus it is convenient to express $ \sigma_s^{(0)}$ as
\bea
 \sigma_s^{(0)}(k,\eta)   
 =   N_s k^s \sigma_k(\eta)~,
\eea
which establishes a connection between the spinning fields and the massive scalars.

For $0<n\leq s$, the equation of motion of $ \sigma_s^{(n)}$ is rather complicated, and the mode function becomes linear combinations of multiple Hankel functions. 
Meanwhile, as we know,  the modes with maximum number of spatial spin, i.e. $\sigma_s^{(s)}$,  are responsible for generating the angular dependent signature in the cosmological collider bispectra.
To describe these modes with $n=s$, one important observation is that,
the  longitudinal modes are related via the transverse condition in \eqref{transverse}\footnote{There is expected to be a typo for the sign of the last term in (A.70) in \cite{Lee:2016vti}.}
\be \label{sigmans}
\sigma^{(n)}_s = -\frac{i}{k}\(\partial_\eta - \frac{2}{\eta}\)\sigma^{(n-1)}_s {+}\sum_{m=0}^{n-1} B_{m,n} \sigma^{(m)}_s
\ee
with
\be
B_{m,n} = \frac{2^n n!}{m!(n-m)!(2n-1)!!} \frac{\Gamma[(1+m+n)/2]}{\Gamma[(1+m-n)/2]} .
\ee
 Note that $B_{m,n}=0$ when $n-m$ is an odd number. 
Therefore, without solving their equations, the mode function of $ \sigma_s^{(s)}$ can be derived iteratively from this recursive relation, which in general yields
\be
\sigma_s^{(s)} = \[ -\frac{i}{k}\(\partial_\eta -\frac{2}{\eta} \)\]^s \sigma_s^{(0)} {+}\sum_{l=0}^{s-1} \[ -\frac{i}{k}\(\partial_\eta -\frac{2}{\eta} \)\]^l \[ \sum_{m=0}^{s-1-l} B_{m,s-l} \sigma_s^{(m)} \] = U^{(s)}_\eta \sigma_s^{(0)}  .
\ee
Here we introduce the differential operator $U^{(s)}_\eta$ as
\be
U^{(s)}_\eta \equiv \sum_{m=0}^{s}  a_m \[ -\frac{i}{k}\(\partial_\eta -\frac{2}{\eta} \)\]^m ,
\ee
where $a_s=1$, $a_{s-1}=a_{s-3}=...=0$ and $a_{s-2n}$ are determined by combinations of $B_{m,n}$. Thus this operator is either real or imaginary.
For illustration, the operators with $s=1,2,3,4$ are given by
\bea
U^{(1)}_\eta & =&-\frac{i}{k}\(\partial_\eta -\frac{2}{\eta} \) \nn\\
U^{(2)}_\eta & =& -\frac{1}{k^2}\(\partial_\eta -\frac{2}{\eta} \)^2 {+}B_{0,2} =  -\frac{1}{k^2} \( \partial_\eta^2 - \frac{4}{\eta}\partial_\eta +\frac{6}{\eta^2}\) - \frac{1}{3} \nn\\
U^{(3)}_\eta & =& \frac{i}{k^3}\(\partial_\eta -\frac{2}{\eta} \)^3 {-}\frac{i}{k}\(\partial_\eta -\frac{2}{\eta} \)(B_{0,2}+B_{1,3}) = \frac{i}{k^3}\(\partial_\eta -\frac{2}{\eta} \)^3 {+}\frac{14}{15}\frac{i}{k}\(\partial_\eta -\frac{2}{\eta} \)\nn\\
U^{(4)}_\eta & =& \frac{1}{k^4}\(\partial_\eta -\frac{2}{\eta} \)^4 {-}\frac{1}{k}\(\partial_\eta -\frac{2}{\eta} \)^2(B_{0,2}+B_{1,3}+B_{2,4}) {+} B_{0,4} +B_{2,4} B_{0,2}
\nn\\ 
&& = \frac{1}{k^4}\(\partial_\eta -\frac{2}{\eta} \)^4 {+}\frac{188}{105}\frac{1}{k^2}\(\partial_\eta -\frac{2}{\eta} \)^2 + \frac{13}{35}
.
\eea
Then by using the relation between  $\sigma_s^{(0)}$ and $\s_k$, we find
\be
\sigma_s^{(s)} = N_s k^s U^{(s)}_\eta \sigma_k ~,
\ee
which maps the massive scalar mode function to the object of interest $\sigma_s^{(s)}$ . This relation plays an important role when we bootstrap the spinning exchange bispectrum from the generalized scalar seeds.

In the end, we notice that for higher spin the number of time derivatives in $U^{(s)}_\eta $ can be reduced by using the equation of motion of $\sigma_s^{(0)}$. By doing this, we are able to express the relation in the following form with at most one derivative
\be
 \sigma_s^{(s)} =  {U}^{(s)}_\eta \sigma_s^{(0)}= \( \frac{\tilde{\alpha}}{k^s\eta^{s-1}} \partial_\eta  + \sum_{m=0}^{s} \frac{\tilde{\alpha}_m}{k^m\eta^m} \) \sigma_s^{(0)} .
\ee
Here $\tilde\alpha_m$ are constants determined by $s$ and $\mu_s$. 
For $s=2$, it takes the following form
\bea
 \sigma_2^{(2)}  &=& \(\frac{2}{k^2\eta}\partial_\eta  + \frac{\mu_2^2-{15}/{4}}{k^2\eta^2}+ \frac{2}{3}\) \sigma_2^{(0)}~.
\eea
We use this expression to simplify the form of mixed propagators with higher spin.

\clearpage
\phantomsection
\addcontentsline{toc}{section}{References}
\bibliographystyle{utphys}
\bibliography{refs}

\end{document}